\documentclass[acmsmall,screen,nonacm]{acmart}

\usepackage{tikz}
\usetikzlibrary{calc, decorations.markings, decorations.shapes, arrows.meta, chains, positioning, shapes.symbols, shapes.geometric}

\usepackage{listings}
\usepackage[normalem]{ulem}  %
\usepackage{lipsum}
\usepackage{wrapfig}
\usepackage{multicol}
\usepackage{xspace}
\usepackage{paralist}  %
\usepackage{titletoc}  %
\usepackage{proof}  %
\usepackage{subcaption}
\usepackage{adjustbox}  %
\usepackage{textcomp}  %
\usepackage{placeins}  %

\usepackage[most]{tcolorbox}  %

\usepackage[scaled=0.77]{FiraMono}
\DisableLigatures{encoding = *, family = tt*}

\input{lib/lstlinebgrd_fixed.tex}

\newcommand{\CodeFigSpaceBefore}{\vspace{-1.3em}}

\newcommand{\SmallerSpaceBeforeNextFig}{\vspace{-1.5em}}
\newcommand{\CancelParBreak}{\vspace{-0.18em}}

\newcommand{\IncludeAppleScreenshot}[1]{%
\makebox[\textwidth][c]{\includegraphics[width=1.09\textwidth]{#1}}%
}
\newcommand{\SmallSpaceBelowAppleScreenshot}{\vspace{-3em}}
\newcommand{\SpaceBelowAppleScreenshot}{\vspace{-2.5em}}
\newcommand{\SmallerSpaceAfterFigure}{\vspace{-0.5em}}

\definecolor{code-color}{RGB}{160, 97, 52}

\definecolor{code-color-lstbg}{RGB}{247, 243, 239}  %

\definecolor{code-label-color}{RGB}{179, 144, 107}

\colorlet{grammar-terminal-color}{code-label-color}

\definecolor{comments-color}{RGB}{197, 170, 141}  %

\definecolor{diff-rm-color}{RGB}{202, 30, 0}
\definecolor{diff-add-color}{RGB}{34, 139, 34}

\lstdefinestyle{delphyne}{
  language=Python,
  basicstyle=\ttfamily,
  showstringspaces=false,
  commentstyle=\color{comments-color},
  columns=flexible,
  keepspaces=true,
  morekeywords={yield,assert,match,case},
  keywordstyle=[2]\uline, %
  deletekeywords={lambda}, %
  deletekeywords=[2]{object,float,int,len,st,list,tuple,all,dict,str,bool,range,enumerate,zip,type}, %
  escapeinside={(@}{@)},
  literate=%
    {->}{{$\rightarrow$}}2%
}

\lstdefinestyle{delphyne-underlined}{%
  morekeywords=[2]{branch,value,ensure,guess,fail,compute}%
}

\newcommand{\ListingDefaultFont}{\linespread{0.88}\ttfamily}
\newcommand{\ListingSmallFont}{\ListingDefaultFont{}\small}

\lstdefinestyle{haskell}{
  language=Haskell,
  commentstyle=\color{comments-color},
  basicstyle=\ListingDefaultFont{},
  showstringspaces=false,
  columns=flexible,
  keepspaces=true,
  numbersep=15pt,
  numbers=none,
  keywordstyle=,
  morekeywords={forall},
  deletekeywords={String, Maybe, return, Bool, Int, Double, msum, map, IO, mzero, mapMaybe, Monad, Just, Nothing, compare, join, Void, true},
  escapeinside={(@}{@)},
  literate=%
    {::}{{$::$}}2%
    {exists}{{$\exists$}}2%
    {->}{{$\rightarrow$}}2%
    {=>}{{$\Rightarrow$}}2%
    {<-}{{$\leftarrow$}}2%
    {:=}{{$\coloneq$}}2%
    {`In`}{{$\in$}}2%
    {~}{{$\sim$}}2%
    {:contRight}{\color{comments-color}{$\hookrightarrow$}}2%
}

\newcommand{\haskellInlineCode}[1]{%
\lstinline[%
  style=haskell,%
  basicstyle=\ttfamily,%
  literate=%
    {::}{{$\!::\!$}}2%
    {->}{{$\!\rightarrow\!$}}2%
    {<-}{{$\!\leftarrow\!$}}2%
    {:=}{{$\coloneq$}}2%
]|#1|
}

\newcommand{\haskellCode}[1]{\inlineCodeBox{\haskellInlineCode{#1}}}

\lstdefinestyle{haskellDiff}{
  style=haskell,
  morecomment=[f][\color{diff-rm-color}]{-\ \ },
  morecomment=[f][\color{diff-add-color}]{+\ \ },
}

\newcommand{\code}[1]{\inlineCodeBox{\texttt{#1}}}

\newenvironment{codeAdjustBox}%
{%
\begin{adjustbox}{valign=t, margin=0 2.0pt 0 0}%
}
{%
\end{adjustbox}%
}

\newtcbox{\inlineCodeBox}[1][]{
  top=1.3pt,
  bottom=1.3pt,
  left=1.2pt,
  right=1.2pt,
  opacityframe=0,
  on line, %
  rounded corners, %
  colback=code-color-lstbg, %
  boxrule=0pt, %
  boxsep=0pt, %
  #1 %
}

\tcbset{
  displayCodeBase/.style={
    enhanced,
    left=2.3mm,
    right=2.3mm,
    before skip=2.2mm, %
    after skip=3.0mm,
    opacityframe=0,
    top=0mm,
    bottom=0mm,
    colback=code-color-lstbg,
    colframe=code-color,
  }
}

\lstdefinestyle{withNumbers}{
  numbers=left,
  numbersep=12pt,
  numberstyle=\small
}

\lstdefinestyle{withSmallFont}{
  basicstyle=\ListingSmallFont{}
}

\newtcolorbox[auto counter, number within=section]{displayCode}[1][]{%
  displayCodeBase,
  overlay={\node[anchor=north east, font=\small, color=code-label-color] at ([xshift=-2mm,yshift=-2mm]frame.north east) {\thetcbcounter};},
  #1
}

\newtcolorbox{displayCode*}[1][]{%
  displayCodeBase,
  #1
}

{%
\begin{displayCode}[#1]%
\begin{center}%
\begin{codeAdjustBox}%
}
{%
\end{codeAdjustBox}%
\end{center}%
\end{displayCode}%
}

\newenvironment{ccodebox*}[1][]%
{%
\begin{displayCode*}[#1]%
\begin{center}%
\begin{codeAdjustBox}%
}
{%
\end{codeAdjustBox}%
\end{center}%
\end{displayCode*}%
}

\newcommand{\DoubleDigitLn}{9mm}

\newenvironment{lcodebox}[1][]%
{%
\begin{displayCode}[#1]%
\begin{codeAdjustBox}%
}
{%
\end{codeAdjustBox}%
\end{displayCode}%
}

\newenvironment{lcodebox*}[1][]%
{%
\begin{displayCode*}[#1]%
\begin{codeAdjustBox}%
}
{%
\end{codeAdjustBox}%
\end{displayCode*}%
}

{%
\begin{displayCode}[#1]%
}
{%
\end{displayCode}%
}

\newenvironment{multicodebox*}[1][]%
{%
\begin{displayCode*}[#1]%
}
{%
\end{displayCode*}%
}

\newenvironment{ccol}%
{%
\begin{codeAdjustBox}%
}
{%
\end{codeAdjustBox}%
}

\newcommand{\codeColRule}{\textcolor{comments-color}{\vrule}}

\newenvironment{itinparaenum}
  {\begin{inparaenum}[\itshape(i)\upshape]}
  {\end{inparaenum}}

\newtheorem{property}{Property}[section]

\newif\ifmuteacmart
\muteacmarttrue %

\ifmuteacmart
\makeatletter
\renewcommand\ClassWarning[2]{}
\renewcommand\ClassWarningNoLine[2]{}
\makeatother
\fi

\setcopyright{acmlicensed}
\copyrightyear{2024}
\acmYear{2024}
\acmDOI{XXXXXXX.XXXXXXX}

\acmConference[PLDI '25]{Programming Language Design and Implementation}{2025}{Seoul}
\acmISBN{978-1-4503-XXXX-X/18/06}

\begin{document}

\title{Oracular Programming}
\subtitle{A Modular Foundation for Building LLM-Enabled Software}

\author{Jonathan Laurent}
\orcid{0000-0002-8477-1560}
\affiliation{%
  \institution{Carnegie Mellon University}
  \city{Pittsburgh}
  \state{Pennsylvania}
  \country{USA}}
  \affiliation{%
  \institution{Karlsruhe Institute of Technology}
  \city{Karlsruhe}
  \country{Germany}}
\email{jonathan.laurent@cs.cmu.edu}

\author{Andr\'e Platzer}
\orcid{0000-0001-7238-5710}
\affiliation{%
  \institution{Karlsruhe Institute of Technology}
  \city{Karlsruhe}
  \country{Germany}}
\email{platzer@kit.edu}

\renewcommand{\shortauthors}{Laurent and Platzer}

\begin{abstract}
Large Language Models (LLMs) can solve previously intractable tasks given only natural-language instructions and a few examples, but they remain difficult to steer precisely and lack a key capability for building reliable software at scale: the modular composition of computations under enforceable contracts. As a result, they are often embedded in larger software pipelines that use domain-specific knowledge to decompose tasks and improve reliability through validation and search. Yet the complexity of writing, tuning, and maintaining such pipelines has so far limited their sophistication. We propose \emph{oracular programming}: a foundational paradigm for integrating traditional, explicit computations with inductive oracles such as LLMs. It rests on two directing principles: the full separation of \emph{core} and \emph{search} logic (allowing the latter to freely evolve without breaking the former), and the treatment of few-shot examples as \emph{grounded} and \emph{evolvable} program components (whose consistency with the rest of the program is enforced through its evolution). Within this paradigm, programmers express high-level problem-solving strategies as programs with unresolved choice points. These choice points are resolved at runtime by LLMs, which generalize from user-provided examples of correct and incorrect decisions. An \emph{oracular program} is composed of three orthogonal components: a \emph{strategy} that consists of a nondeterministic program with choice points that can be reified into a search tree, a \emph{policy} that specifies how to navigate this tree with the help of LLM oracles, and a set of \emph{demonstrations} that describe successful and unsuccessful tree navigation scenarios across diverse problem instances. Each component is expressed in a dedicated programming language. We address the key programming language design challenges of modularly composing oracular programs and enforcing consistency between their components as they evolve.

\end{abstract}

\maketitle

\section{Introduction}\label{sec:introduction}

Large Language Models (LLMs) have introduced a fundamentally new form of inductive computation, leveraging common-sense knowledge and generalization to perform tasks implicitly specified through natural-language instructions and examples. As a consequence, they are significantly widening the range of problems amenable to computational automation, and have led to breakthroughs across a wide range of areas, including reasoning-intensive domains such as program synthesis~\cite{li2022competition}, mathematical problem solving~\cite{lewkowycz2022solving}, and formal theorem proving~\cite{jiang2022draft,first2023baldur}.

Yet, despite these successes, LLMs remain difficult to steer precisely and lack a capability fundamental to building reliable software at scale: the modular composition of computations under enforceable contracts. As a result, LLMs are often deployed as components within larger software pipelines that leverage domain-specific knowledge to decompose complex tasks~\cite{sahoo2024systematic}, provide intermediate feedback~\cite{yao2022react,yao2024tree}, and improve reliability through validation and search~\cite{yao2024tree,lewkowycz2022solving}.

At first glance, developing such pipelines may seem deceptively simple. After all, LLMs can be prompted through straightforward APIs, and developers have access to the full arsenal of existing programming languages and software-engineering tools. Yet, as we argue, building reliable LLM-enabled software raises challenges not properly addressed by existing programming abstractions.
\begin{itemize}
  \item LLMs offer a powerful but inherently unreliable programming primitive, making the ubiquitous use of \emph{search} and \emph{validation} essential for dependability. At the same time, prompting LLMs is expensive, so retries and backtracking incur significant costs. Resolving this tension requires careful tuning of this search logic, which must also adapt to user-specific constraints (e.g., inference budget, latency tolerance). Evolving this logic can require costly and frequent refactorings, especially when it is intertwined with the higher-level logic that governs how large problems are decomposed into smaller subproblems and prompts.
  \item Prompting language models often works best when examples of successfully solving similar task instances are provided (i.e., \emph{few-shot prompting}~\cite{brown2020language}). Such examples are therefore crucial components of LLM-enabled programs but writing them can be time-consuming. In addition, these examples need to be kept synchronized with the rest of the program: any change can silently render some examples obsolete and necessitate new ones. While numerous techniques exist for maintaining correctness and consistency in traditional programs---most notably type systems, contracts and testing---LLM-enabled programs present a unique and largely unaddressed challenge. Yet, these programs stand to gain disproportionately from \emph{fearless refactoring}, since the inherent opacity and unpredictability of LLMs demand frequent iteration and rapid development cycles.
\end{itemize}

Together, these challenges have limited the complexity and robustness of existing LLM-enabled software. This paper tackles them through principled language design, introducing a new paradigm for integrating inductive oracles in traditional programs while enforcing strong modularity, consistency, and evolvability properties. We call this paradigm \emph{oracular programming}. It rests on several interlocking principles and contributions, outlined below.

\begin{description}

\item[Separating core and search logic.]
Existing LLM-enabled programs often conflate two distinct kinds of logic: {\it (i)} the \emph{core logic}, which describes how complex problems can be recursively decomposed into LLM queries and validation contracts; and {\it (ii)} the \emph{search logic}, which specifies how those queries are answered and how the resulting search space is explored (e.g., sequentially or in parallel, using depth-first search or Monte Carlo Tree Search). The former arises naturally from the expression of domain-specific knowledge, whereas the latter typically demands heavier iteration and tuning. A foundational principle of {oracular programming} is the \emph{complete separation} of these two kinds of logic into orthogonal components, which we call \emph{strategies} and \emph{policies} respectively. This separation enables \emph{vastly different} search algorithms and prompting techniques to be explored concurrently, without requiring \emph{any} modification to the program's core logic. Crucially, achieving such separation while supporting a wide range of search algorithms---which no existing framework does---is incompatible with treating LLM queries as ordinary function calls. Instead, we treat them as \emph{nondeterministic assignments}. We define \emph{strategies} as nondeterministic programs that can be reified into search trees (where choice points induce branching nodes and contract violations yield failure leaves; see Figure~\ref{fig:strategy-overview} for a preview) and \emph{policies} as generic functions that manipulate these trees.
\item[A modular and extensible strategy language.]
We define a \emph{strategy language} for expressing high-level problem-solving strategies as nondeterministic programs that can be reified into search trees. Its design is guided by two key requirements of \emph{extensibility} and \emph{modularity}, which drive innovations well beyond traditional formulations of nondeterministic programming~\cite{fischer2011purely}. The \emph{extensibility} requirement arises from the diversity of existing search algorithms, which often exploit specific structure and annotations within search trees to improve efficiency (e.g., independent subgoals, quantitative value estimates). Our proposed strategy language is therefore equipped with an \emph{extensible effect system} that makes it easy to define new tree types. The \emph{modularity} requirement mandates that {\it (i)} heterogeneous strategies producing different types of trees can be composed while keeping their corresponding policies independent, and {\it (ii)} any LLM query can be \emph{locally} and \emph{transparently} refined into a dedicated sub-strategy without impacting its parent strategy or any associated policy. We meet these goals by introducing the concept of \emph{opaque space} that unifies strategies and queries from a policy's perspective, together with a \emph{search stream protocol} that enables arbitrary search algorithms to communicate in a resource-aware manner.

\item[A layered, resource-aware policy language.]
We define a \emph{policy language} for defining functions that navigate strategy-defined trees in pursuit of solutions. This language is \emph{layered}: while policies are typically assembled by composing standard building blocks (in the form of prompting policies, search algorithms, stream transformers, and tree transformers; see Figure~\ref{fig:policy-overview} for a preview), new primitives can also be defined easily using a dedicated language of \emph{search-stream combinators}. In both cases, the proper enforcement of {resource limits} (e.g., LLM inference budget) is guaranteed \emph{by construction}: each policy component can be assigned a global budget limit that is automatically inherited by its sub-components.

\item[Few-shot examples as first-class program components.] %
Few-shot examples are central to LLM-enabled programs, yet are traditionally treated as isolated data, disconnected from the logic that ties prompts together. As a result, they can be difficult to author (even determining \emph{which} questions must be answered to solve a particular problem is often nontrivial in multi-prompt settings), and even harder to maintain as the program evolves. We introduce a \emph{demonstration language} for describing examples while \emph{grounding} them in concrete scenarios of navigating strategy trees. More precisely, a \emph{demonstration} collects a set of query-answer pairs along with a set of unit tests that assemble them into paths within a search tree (see Figure~\ref{fig:demonstration-overview} for a preview). The demonstration language enables a tool-assisted, test-driven workflow in which demonstrations can be created interactively and later repaired when strategy changes cause test failures. The universality of this workflow is established by a strong completeness theorem. Finally, demonstrations are \emph{policy-agnostic}: they can be developed independently of policies and cannot break as a result of policy changes. This independence is achieved by associating each effect definition in the strategy language with a local navigation behavior that extends the semantics of demonstrations.

\end{description}

This paper first motivates and defines the three languages that together constitute the triad of \emph{oracular programming}, highlighting how their designs naturally emerge from the principles outlined above (Section~\ref{sec:triad}). The strategy and policy languages are formally defined via shallow embeddings in Haskell. We next introduce \emph{Delphyne}, an oracular programming framework embedded in Python, with rich tooling support in the form of a dedicated language server (Section~\ref{sec:delphyne}). Python is chosen for its accessibility, its reflection capabilities, and its static type system, which is expressive enough to precisely type the strategy language and the upper layer of the policy language. Finally, we showcase the expressiveness and power of oracular programming in three case studies (Section~\ref{sec:case-studies}). Through concrete user stories, we illustrate the practical benefits of separating strategies from policies and of maintaining few-shot examples in the form of demonstrations.

\paragraph{Summary of Contributions} Overall, this paper makes the following contributions:

\begin{itemize}
  \item It identifies the \emph{separation of core and search logic} and the \emph{grounding of few-shot examples} as key principles for facilitating the writing and evolution of LLM-enabled programs.
  \item It introduces three synergistic languages for writing such programs by composing \emph{strategies}, \emph{policies}, and \emph{demonstrations}, which offer strong modularity and evolvability guarantees.
  \item It presents Delphyne, an open-source implementation of the resulting framework, with extensive documentation, a comprehensive standard library, and rich tooling support.
\end{itemize}

\section{The Oracular Programming Triad}\label{sec:triad}

An \emph{oracular program} is defined by three orthogonal components: a \emph{strategy}, a \emph{policy}, and a set of \emph{demonstrations}. A \emph{strategy} is a nondeterministic program that denotes a high-level plan for solving a particular class of problems. It induces a search tree that can be navigated with LLM guidance, as
\begin{wrapfigure}{r}{0.32\textwidth}
  \vspace{-0.2cm}
  \centering
  \small
  \def\trigSide{1.5cm}
  \def\syncLabel#1{\textsc{#1}}
  \begin{tikzpicture}[
    component/.style={
      fill=code-color-lstbg, rounded corners=3pt,
    }
  ]
      \node[text width=2cm, align=center] (OP) at (0,0) {{Oracular Program}};
      \node[component] (P) at (-150:\trigSide) {Policy};
      \node[component] (D) at (-30:\trigSide) {Demos};
      \node[component] (S) at (90:\trigSide) {Strategy};
      \draw[thick] (S) -- node[midway, sloped, above, inner sep=2mm] {\syncLabel{tests}} (D);
      \draw[thick] (S) -- node[midway, sloped, above, inner sep=2mm] {\syncLabel{types}} (P);
      \draw[thick] (P) -- (D);
  \end{tikzpicture}
  \vspace{-0.2cm}
\end{wrapfigure}
specified by a \emph{policy}. A \emph{demonstration} bundles relevant examples of answering specific LLM requests with unit tests, in such a way as to describe concrete scenarios of successfully or unsuccessfully navigating the search tree for a particular problem instance. 
Consistency between \emph{strategies} and \emph{policies} is enforced through types, while consistency between \emph{strategies} and \emph{demonstrations} is enforced through \emph{navigation tests}. By design, \emph{policies} and \emph{demonstrations} are fully decoupled and so a change in one cannot affect the other. Each of these components can be described in its own dedicated language. The remainder of this section motivates and presents these three languages, which together constitute the triad of \emph{oracular programming}.

We define the \emph{strategy} and \emph{policy languages} via \emph{shallow embeddings} in Haskell, a natural host language due to its purity, its expressive type system and its syntactic facilities for expressing monadic code. Full definitions are available in the appendices, and also as type-checkable Haskell code in the supplementary material. However, our \emph{main implementation}---the Delphyne framework---uses a Python embedding instead, opting for a different trade-off in terms of static type safety, ecosystem integration and accessibility (Section~\ref{sec:delphyne}).

\subsection{The Strategy Language}\label{sec:strategy-language}

We introduce our proposed strategy language in three steps, starting with a naive design based on well-known techniques for defining search trees via monadic programs~\cite{fischer2011purely}, and then successively adding support for \emph{modularity} and \emph{extensibility}.

\subsubsection{Initial Design Attempt}\label{sec:initial-design}

\begin{figure}
\begin{lcodebox*}%
\begin{lstlisting}[style=haskell]
data Tree a = Success a | Failure | forall b. Branch (Query b) (b -> Tree a)
data Query b = Query { prompt :: String, parseAnswer :: String -> Maybe b }
\end{lstlisting}
\end{lcodebox*}
\CodeFigSpaceBefore
\caption{Naive Definition for a Search Tree. A tree producing values of type \code{a} (\code{Tree a}) can be either a \emph{success leaf}, a \emph{failure leaf} that represents a contract violation, or a \emph{branching node} that contains a \emph{query} and one \emph{subtree} for every possible answer to this query. A \emph{query} represents a question being asked to an external oracle and is defined by a prompt along with an answer-parsing function. Note that \code{b} is existentially quantified in the definition of a branching node, meaning that each such node can have children indexed by a different, locally-defined type (perhaps surprisingly, existential types are introduced in Haskell using the \code{forall} keyword. This is because any constructor type $(\exists \alpha \, T(\alpha)) \rightarrow \tau$ is isomorphic to $\forall \alpha \, (T(\alpha) \rightarrow \tau)$).}\label{fig:naive-tree}
\vspace{-1.3em}
\end{figure}
\begin{figure}
\begin{multicodebox*}[label=code:naive-strategy]%
\begin{ccol}%
\begin{lstlisting}[style=haskell]
instance Monad Tree where
  return = Success
  Success a >>= f = f a
  Failure >>= f = Failure
  Branch q k >>= f =
    Branch q (\b -> k b >>= f)
\end{lstlisting}%
\end{ccol}%
\quad%
\codeColRule%
\quad%
\begin{ccol}%
\begin{lstlisting}[style=haskell]
type Strategy = Tree

branch :: Query b -> Strategy b
branch q = Branch q Success
ensure :: Bool -> Strategy ()
ensure b = if b then return () else Failure
\end{lstlisting}
\end{ccol}
\end{multicodebox*}
\CodeFigSpaceBefore
\caption{A Naive Language for Defining Search Trees. The tree type defined in Figure~\ref{fig:naive-tree} can be equipped with a monadic structure, allowing trees to be defined directly and naturally using Haskell's \emph{do-notation} (see Figure~\ref{fig:naive-strategy-example} for an example). For readers unfamiliar with monads, understanding this technical definition is not crucial: what matters more is an intuitive understanding of how a tree (as defined in Figure~\ref{fig:naive-tree}) can be defined via a nondeterministic program (as the one in Figure~\ref{fig:naive-strategy-example}).}\label{fig:naive-strategy}
\vspace{-0.8em}
\end{figure}
\begin{figure}
\begin{multicodebox*}
\begin{ccol}
\begin{lstlisting}[style=haskell]
generateProg :: Spec -> Strategy Prog
generateProg spec = do
  prog <- branch (conjectureProg spec)
  proof <- branch (generateProof spec prog)
  ensure (checkProof spec prog proof)
  return prog
\end{lstlisting}
\end{ccol}
\ \,
\codeColRule{}
\ \,
\begin{ccol}
\begin{lstlisting}[style=haskell]
conjectureProg
  :: Spec -> Query Prog
generateProof
  :: Spec -> Prog -> Query Proof
checkProof
  :: Spec -> Prog -> Proof -> Bool
\end{lstlisting}
\end{ccol}
\end{multicodebox*}
\CodeFigSpaceBefore
\caption{A Minimal Strategy for Program Synthesis. This strategy is expressed using the naive strategy language defined in Figure~\ref{fig:naive-strategy}. To generate a program that provably meets specification \code{spec}, strategy \code{generateProg} first issues a query to conjecture a program (via \code{conjectureProg}, whose type but not definition is provided), then issues a query to obtain a proof that the program indeed meets this specification (via \code{generateProof}), and finally ensures that the proof is correct. For all values of its \code{spec} argument, \code{generateProg} produces a tree with two levels of branching, followed by either success or failure leaves. }\label{fig:naive-strategy-example}
\vspace{-0.8em}
\end{figure}

A minimal strategy language can be defined in just a dozen lines of Haskell, as shown in Figures~\ref{fig:naive-tree}~and~\ref{fig:naive-strategy}. Trees are defined via an algebraic data type and equipped with a monadic structure, allowing them to be constructed using Haskell's \emph{do-notation}, as we demonstrate in Figure~\ref{fig:naive-strategy-example}. Once constructed, trees can be explored by independently definable search algorithms that pattern-match on their structure (see Figure~\ref{fig:naive-dfs} in Appendix~\ref{ap:naive-dfs} for an example).
This initial design has the advantage of simplicity but is severely limited in terms of \emph{extensibility} and \emph{modularity}. The extensibility issue is straightforward: as noted in the introduction, advanced search algorithms leverage a wide variety of specific annotations and structural variations in trees (e.g., value annotations for MCTS~\cite{browne2012survey}), and thus an explicit extension mechanism is needed to accommodate this diversity. The \emph{modularity} issue is more subtle.

Modularity requires every query to be \emph{locally} and \emph{transparently} replaceable by a dedicated strategy. Indeed, a powerful workflow for designing strategies consists in starting with broad queries and then iteratively \emph{refining} them {when} and {where} more control is needed. It might appear
\begin{wrapfigure}{r}{0.54\textwidth}
\vspace{-1.4em}%
\begin{ccodebox*}%
\begin{lstlisting}[style=haskellDiff]
-  conjectureProg :: Spec -> Query Prog
+  conjectureProg :: Spec -> Strategy Prog
...
-  prog <- branch (conjectureProg spec)
+  prog <- conjectureProg spec
\end{lstlisting}
\end{ccodebox*}
\vspace{-1.3em}
\vspace{-0.7em}
\end{wrapfigure}
that our naive design allows this. For example, in the \code{generateProg} strategy from Figure~\ref{fig:naive-strategy-example}, a program is conjectured via a single query. We can update \code{conjectureProg} to use a dedicated strategy instead, as shown on the right.
Doing so is akin to {inlining} the tree associated with \code{conjectureProg} into the tree associated with \code{generateProg}. However, such inlining is hardly transparent and deeply alters the shape of the outer tree, typically requiring changes to policies searching it. For example, assuming the tree induced by \code{generateProg} is explored using depth-first search at depth $d$, such inlining would likely require updating the value of $d$. In addition, a single search algorithm may not necessarily be best adapted to both \code{generateProg} and \code{conjectureProg}: these two strategies should be allowed to produce different types of trees and to leverage different, independent policies.

\subsubsection{A Modular Strategy Language}\label{sec:modular-strategy-language}

\begin{figure}
  \centering
  \includegraphics{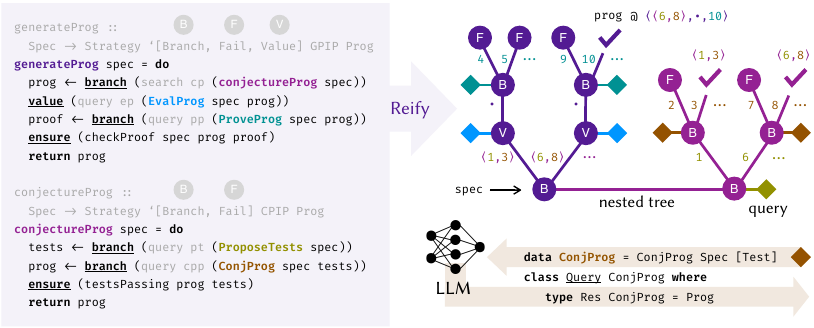}
  \vspace{-0.3em}
  \caption{Example of a Modular Strategy For Program Synthesis. Strategies are nondeterministic programs that can be reified into search trees. They can issue \emph{queries} to be answered by external oracles (queries are represented by diamonds~$\diamond$ and have constructors starting with uppercase letters). \emph{Branching}, \emph{failure}, and \emph{value} nodes are labeled \code{B}, \code{F}, and \code{V} respectively. Horizontal lines denote \emph{nesting} while other lines denote parent-child relations. Child edges and success leaves are decorated with \emph{local value references}, with numbers being used as shortcuts for individual query answers. Grayed-out elements can be ignored on first reading.}\label{fig:strategy-overview}
  \vspace{-0.8em}
\end{figure}

\begin{figure}
\begin{lcodebox*}[left=\DoubleDigitLn]
\begin{lstlisting}[style=haskell,style=withNumbers, style=withSmallFont]
data Strategy s p a  (@\label{line:strategy:type}@)
instance Monad (Strategy s p)  (@\label{line:strategy:monad}@) (@\label{line:strategy:monad:start}@) 
  return :: a -> Strategy s p a
  (>>=) :: Strategy s p a -> (a -> Strategy s p b) -> Strategy s p b  (@\label{line:strategy:monad:end}@) 
class Query q where { type Res q } (@\label{line:strategy:query-def}@) (@\label{line:strategy:query-def:start}@) (@\label{line:strategy:res}@) (@\bigskip@)
fail :: (Fail `In` s) => Strategy s p a  (@\label{line:strategy:fail}@)
branch :: (Branch `In` s) => Opaque p a -> Strategy s p a  (@\label{line:strategy:branch}@)
value :: (Value `In` s) => Opaque p Double -> Strategy s p ()  (@\label{line:strategy:value}@) (@\bigskip@)
query :: (Query q) => (p -> PromptingPolicy q) -> q -> Opaque p (Res q)  (@\label{line:strategy:query-ospace}@)
search :: (p -> SearchPolicy s' p') -> Strategy s' p' a -> Opaque p a  (@\label{line:strategy:search-ospace}@)
\end{lstlisting}
\end{lcodebox*}
\CodeFigSpaceBefore
\caption{A Modular Strategy Language, as a Haskell Monadic DSL. Section~\ref{sec:extensible-strategy-language} demonstrates how to extend this language with new effect types beyond \code{Fail}, \code{Branch}, and \code{Value}. The $\in$ operator expresses membership of a type within a list of types~\cite{swierstra2008data}. An \code{ensure} function can be defined from \code{fail}, as in Figure~\ref{fig:naive-strategy}.}\label{fig:strategy-language-signature}
\end{figure}

We introduce a novel strategy language that addresses the aforementioned challenges of \emph{modularity} and \emph{extensibility}. We begin by tackling \emph{modularity}, proposing a monadic DSL for defining \emph{modular} search trees that may feature \emph{branching}, \emph{failure}, and \emph{value} nodes (value nodes provide quantitative information on how promising a branch is). We later present an extensibility mechanism for defining new kinds of trees (Section~\ref{sec:extensible-strategy-language}).

Figure~\ref{fig:strategy-overview} illustrates a strategy for \emph{program synthesis} expressed in this language, while Figure~\ref{fig:strategy-language-signature} defines its typed syntax in the form of a Haskell module signature. Consider the example strategy from Figure~\ref{fig:strategy-overview}, ignoring the grayed-out elements for now. Given a specification, \code{generateProg} nondeterministically generates a program that satisfies it. It proceeds in four steps: {\it (i)} conjecturing a program, {\it (ii)} independently estimating the likelihood that this program is correct (providing valuable quantitative information to potential search algorithms), {\it (iii)} generating a machine-checkable correctness proof, and {\it (iv)} checking this proof. The first step is itself implemented via a separate strategy named \code{conjectureProg}, which generates executable unit tests before it produces a program and only returns this program if it passes all tests. In contrast to the naive strategy language from Section~\ref{sec:initial-design}, a key innovation already emerges: programs can branch \emph{not only} on the answer to a query, but \emph{also} on the result of another strategy. This is reflected in the structure of the produced tree (Figure~\ref{fig:strategy-overview}), where branching nodes can contain queries (represented as diamonds) but also \emph{nested trees}, in which case children are indexed by the success leaves of this tree.

A \emph{query} is a value that represents a question posed to an oracle.   Queries are stratified by type, and each query type is associated with a \emph{response type}. How exactly a query gets answered is \emph{not} the concern of the strategy language but of the \emph{policy language}, which allows defining \emph{prompting policies} that map queries to streams of possible answers (e.g., by repeatedly sampling and parsing answers from an LLM using a given prompt and a set of examples; see Figure~\ref{fig:few-shot-stream} for a preview). The policy language further allows defining \emph{search policies}, which map strategy-induced trees to streams of values attached to success leaves of these trees. By \emph{modularity}, search policies must remain agnostic to whether the candidate space of a branching node arises from a query or from a nested strategy tree. Exposing this distinction would violate the abstraction barrier: if a search policy had direct access to a query, it would break once that query were refined into a dedicated strategy, contradicting the modularity principle established in Section~\ref{sec:initial-design}. Instead, search algorithms such as \emph{depth-first search} or \emph{MCTS} should operate over a lazy stream of candidates at each branching node, regardless of whether this stream is induced by a query or by a strategy (see Figure~\ref{fig:dfs-stream} for a preview and Section~\ref{sec:search-stream} for details on the exact nature of these \emph{search streams}, which is irrelevant for now). However, by the principle of separation between core and search logic, such streams cannot be directly attached to strategy trees, as they encapsulate policy-specific information. We resolve this tension by introducing the concepts of an \emph{inner policy type} and of an \emph{opaque space}.

\subsubsection{Inner Policy Types and Opaque Spaces}\label{sec:inner-policies-local-spaces}
Every strategy is associated with an \emph{inner policy type}, which appears in its type signature. Values of this type---called \emph{inner policies}---capture all the
\begin{wrapfigure}{r}{0.54\textwidth}%
\vspace{-1.0em}%
\begin{ccodebox*}%
\begin{lstlisting}[style=haskell]
data GPIP = {
  cp :: SearchPolicy '[Branch, Fail] CPIP,
  ep :: PromptingPolicy EvalProg,
  pp :: PromptingPolicy ProveProg }
\end{lstlisting}%
\end{ccodebox*}%
\vspace{-1.6em}%
\caption{Inner Policy Type for \code{generateProg}.}\label{fig:gpip-definition}
\vspace{-0.5em}
\end{wrapfigure}
information required to turn inner queries and sub-strategies into proper streams of elements. For example, Figure~\ref{fig:gpip-definition} defines the inner policy type for the \code{generateProg} strategy from Figure~\ref{fig:strategy-overview}, named \code{GPIP} as an acronym. A value of type \code{GPIP} specifies two prompting policies \code{ep} and \code{pp}, which respectively handle inner instances of the \code{EvalProg} and \code{ProveProg} queries, as well as a search policy \code{cp} for managing the inner call to \code{conjectureProg}. In turn, \code{cp} can be defined by combining a search algorithm such as \emph{depth-first search} (\code{dfs}) with a nested inner policy of proper type (\code{CPIP}). Given that \code{dfs} has type

\CancelParBreak{}%
\begin{ccodebox*}%
\begin{lstlisting}[style=haskell]
dfs :: DFSOptions -> p -> SearchPolicy '[Branch, Fail] p,
\end{lstlisting}%
\end{ccodebox*}%
\CancelParBreak{}%

\noindent a possible policy for \code{generateProg} has shape \haskellCode{bestFirst _ (GPIP (dfs _ (CPIP _ _)) _ _)}, where \code{bestFirst} is a search algorithm capable of handling trees with value nodes, and where \code{\_} holes stand for either search hyperparameters or prompting policies.

An \emph{opaque space} abstracts away a query or a strategy, intentionally hiding its internal structure and exposing only a mapping from the ambient inner policy to a flat stream of elements.  (More information about the internal structure of opaque spaces remains accessible to the demonstration interpreter, but \emph{not} to policies; see Section~\ref{sec:demonstration-language}.) With these intuitions in place, we can now examine the formal signature of our proposed strategy language, as defined in Figure~\ref{fig:strategy-language-signature}.

\subsubsection{Language Signature}\label{sec:language-signature} A \emph{strategy value} is a computation that can be reified into a search tree. We use the more general term \emph{strategy} to denote either a strategy value or a function that returns one, such as \code{generateProg}. A strategy value has type \code{Strategy s p a} (Figure~\ref{fig:strategy-language-signature}, Line~\ref{line:strategy:type}). The first parameter, \code{s}, is its \emph{signature}: the list of effects that it is allowed to invoke---that is, the list of all node types that can occur in the associated tree. The second parameter, \code{p}, is its associated \emph{inner policy type}, and the third parameter, \code{a}, is its \emph{return type} (e.g., \code{Prog} in the case of \code{generateProg}). Strategies are equipped with a monadic structure (Lines~\ref{line:strategy:monad:start}-\ref{line:strategy:monad:end}), allowing them to be composed via the {bind} operator {\small\verb|>>=|} for as long as their signature and inner policy types coincide (only trees with the same type of nodes can be inlined into each other).

Branching occurs over elements of an \emph{opaque space} (Line~\ref{line:strategy:branch}). In turn, an opaque space can be defined via a query (using the \code{query} function, Line~\ref{line:strategy:query-ospace}) but \emph{also} via a strategy of arbitrary signature (using the \code{search} function, Line~\ref{line:strategy:search-ospace}), through pairing with a function that maps a choice of \emph{inner policy} for the surrounding strategy to an adapted \emph{sub-policy}. Often, this mapping consists in accessing a specific field of the inner policy (\code{cp}, \code{ep} or \code{pp} in our example from Figure~\ref{fig:strategy-overview}). Finally, the \code{value} function also takes an opaque space as an argument (Line~\ref{line:strategy:value}), allowing both queries and strategies to be used for computing value estimates. Associated \code{Value} nodes have a unique child (Figure~\ref{fig:strategy-overview}), since value estimates are accessible to the policy but \emph{not} to the strategy continuation (enabling transformations that remove value nodes from a tree; see Section~\ref{sec:assembling-policies}).

\subsubsection{Locality, References, and Traces}\label{sec:locality-references} From the perspective of strategies, branching over an opaque space with element type \code{a} yields a value of type \code{a} (Figure~\ref{fig:strategy-language-signature}, Line~\ref{line:strategy:branch}). However, in the induced tree, the children of the corresponding node are \emph{not} indexed by values of type \code{a} themselves, but by \emph{local references} to such values. When the opaque space is induced by a query, a local reference corresponds to an answer of type \code{a}. When it is induced by a nested tree, it corresponds to a path within that tree leading to a success leaf that contains a value of type \code{a}. Figure~\ref{fig:strategy-overview} shows these references for all child edges and success leaves in an example tree.

This distinction is not a mere technicality: it ensures that every success leaf can be traced back to a concrete sequence of query answers leading to it. This property is essential both for establishing the completeness of our demonstration language (Section~\ref{sec:demo-completeness}) and for extracting learning data from successful runs of an oracular program (Section~\ref{sec:lean-case-study}). When encountering a branching node, policies cannot produce values out of thin air; they must instead select values that are \emph{explicitly} present in the enclosed opaque space, from which they can derive a stream of valid local references.

Whenever a policy explores a search tree, the finite pruned tree containing all visited nodes can be automatically extracted as a \emph{trace}. Traces are serializable, due to references being serializable themselves. They can be visualized using proper tooling (Section~\ref{sec:delphyne}), offering a general debugging mechanism. Moreover, they can be inspected to automatically extract learning data from runs of oracular programs (Section~\ref{sec:lean-case-study}). This ability to reify the outcome of search as a trace, and to let dedicated \emph{teacher programs} analyze it (e.g., by identifying subproblems that required many failed attempts and contrasting them with the ultimately successful attempt) is what we call \emph{search reflection}. Search reflection provides a general foundation for self-improving oracular programs and relies critically on the separation of strategies and policies~(Section~\ref{sec:lean-case-study}).

\subsubsection{Defining New Effects}\label{sec:extensible-strategy-language}

Advanced search algorithms often exploit specific structure and annotations in search trees for efficiency. To accommodate arbitrary algorithms, our strategy language is extensible with new \emph{effects}---each effect introducing a new kind of tree node. We motivate and illustrate this extension mechanism through two examples, and then provide a general definition.

\paragraph{Example: Compare-and-Branch}
Branching nodes provide search policies with a space of candidates to explore, yet no quantitative information is available to compare their quality and prioritize the search. Moreover, branching on syntactically distinct but semantically equivalent candidates (e.g., formulas $x > 0$ and $-x < 0$) is wasteful. The strategy language allows us to remedy this by defining a new \emph{compare-and-branch} effect using the following syntax (not standard Haskell):
\begin{lcodebox}[label=code:cbranch-def]%
\begin{lstlisting}[style=haskell]
effect cbranch :: (CBranch `In` s) =>
  { cands :: Opaque p a, compare :: [a] -> Opaque p [Double] } -> Strategy s p a
\end{lstlisting}
\end{lcodebox}%
\noindent An effect is defined using the \code{effect} keyword, followed by the type of its triggering function, with named arguments (see the types of \code{branch}, \code{fail}, and \code{value} in Figure~\ref{fig:strategy-language-signature} for comparison). From such a declaration, a corresponding node type is automatically derived and can be used in strategy signatures (\code{CBranch} in this case, see Figure~\ref{fig:derived-node-types-examples}).
Compared to \code{branch}, \code{cbranch} takes as an additional argument a comparison function that maps lists of branching candidates to probability distributions over them. This comparison function can be implemented using a query but also via a dedicated substrategy, since it returns an opaque space. In the reified tree, a \code{CBranch} node does not only contain an opaque space of branching candidates (as \code{Branch} does), but \emph{also} a family of opaque spaces indexed by candidate lists containing comparison outcomes. The search policy can use this information in different ways (e.g., generating candidates lazily and performing a comparison for every new candidate, computing a set of candidates first and then only performing comparison once, using a single estimate for each comparison or aggregating several) but, crucially, how it does so is not the concern of the strategy triggering the effect (by our separation principle).

\paragraph{Example: Concurrency in Strategies}
Many search tasks can be decomposed into smaller tasks that can be solved concurrently. For example, in theorem proving, applying a tactic to a goal typically results in multiple subgoals that can each be proved independently. Many proof search algorithms exploit this structure by navigating variants of \emph{conjunction-disjunction} trees~\cite{lample2022hypertree,whalen2016holophrasm,renshaw2011distributed}. We can model this with a \code{Join} effect that is defined as follows:
\begin{lcodebox}[label=code:join-def]
\begin{lstlisting}[style=haskell]
effect join :: (Join `In` s) =>
  { left :: Strategy s p a, right :: Strategy s p b } -> Strategy s p (a, b)
\end{lstlisting}
\end{lcodebox}%
\noindent The \code{join} function takes as its arguments a strategy producing a value of type \code{a} and a strategy producing a value of type \code{b} and returns a strategy producing a value of type \code{(a, b)}. Note that it does \emph{not} take opaque spaces as arguments, but instead \emph{strategies} with identical signatures and inner policy types. Indeed, \code{join} does not abstract away the structure of its arguments, by design. A \code{Join} node in the reified tree contains two nested trees that are directly accessible to the search policy, thus enabling algorithms such as HyperTree Proof Search~\cite{lample2022hypertree} to exploit the resulting bipartite structure. Still, one can combine opaque spaces with \code{join} by first wrapping them with \code{branch}. A \code{Join} node has one child for every pair of success values from its two nested trees.

\subsubsection{Generic Strategy Trees}\label{sec:tree-definition}
We now give an informal definition of generic strategy trees, and then relate it to the examples from the previous section as well as to the grammar of admissible effect declarations (Figure~\ref{fig:effect-grammar}). A formal type definition appears in Figure~\ref{fig:tree-def}, but it can be safely skipped on a first read. Strategy trees are introduced through three mutually recursive definitions, which we present together before discussing them in detail.

{
  \newcommand{\nt}[1]{\langle#1\rangle}
\newcommand{\nts}[1]{\nt{\s{#1}}}
\newcommand{\s}[1]{\text{#1}}
\newcommand{\tid}[1]{\text{#1}}  %
\newcommand{\tsb}[1]{{\color{grammar-terminal-color}\textbf{\textsf{#1}}}}  %
\newcommand{\ts}[1]{{\color{grammar-terminal-color}\textbf{\textsf{#1}}}}
\newcommand{\ntdef}{\Coloneqq}
\newcommand{\GOR}{\ | \ }

  \begin{figure}
    \begin{align*}
      \nt{\s{decl}} \Coloneqq \ &
        \tsb{effect} \ \nt{\s{trigger-name}} \ \ts{::} \ \ts{(}\nt{\s{node-name}} \ \ts{$ \in $ s) $\Rightarrow$} \ \ts{\{}\,\nt{\s{arg}}^*\,\ts{\}} \ \ts{$\rightarrow$} \ \ts{Strategy s p } \nt{\s{res-type}} \\
      \nt{\s{arg}} \Coloneqq \ &
        \nt{\s{name}} \ \ts{::} \ \big(\nt{\s{space}} \ | \ \nt{\s{type}} \ \ts{{$\!\rightarrow\!$}} \ \nt{\s{space}}\big) \\
      \nt{\s{space}} \Coloneqq \ &
        \ts{Opaque p} \ \nt{\s{type}} \ | \ \ts{Strategy s p} \ \nt{\s{type}} \\
    \end{align*}
    \vspace{-3.5em}
    \caption{%
      Grammar of Admissible Effect Declarations.
      Terminal symbols are rendered in brown, while $\nt{\s{res-type}}$ and $\nt{\s{type}}$ denote type expressions with no free variables in $\{\ts{p}, \ts{s}\}$.
      A node type can be automatically derived from such a declaration, as exemplified in Figure~\ref{fig:naive-strategy-example} and formally defined in Figure~\ref{fig:node-type-derivation} (Appendix~\ref{ap:strategy-trees}). The value of $\nt{\s{res-type}}$ determines the node's \emph{action type} (first line), each effect argument defines a member space or parametric space (second line), and spaces can be either opaque spaces or embedded trees (third line).
    }
    \label{fig:effect-grammar}
    \vspace{-0.8em}
  \end{figure}
}

\begin{description}
  \item[Tree]\label{def:tree} A \emph{tree} is either a \emph{success leaf} or an \emph{effect node} (\emph{node} for short). Nodes have types (e.g., \code{Branch}, \code{Join}), and a tree can only feature nodes whose types belong to its signature. Each node is associated with an \emph{action type}, and has child trees indexed by \emph{local values} of this type. It may contain \emph{local spaces}, whose number and nature are determined by its type.
  \item[Local Value]\label{def:local-value} For a given node, a \emph{local value} is either an element of a \emph{local space}, or obtained from other local values through introduction and elimination of lists, tuples (possibly empty), and sum types (e.g., a list of local values is a local value; see Appendix~\ref{ap:references}).
  \item[Local Space]\label{def:local-space} A \emph{local space} is either an \emph{opaque space}, an \emph{embedded tree}, or the result of instantiating a \emph{parametric opaque space} with a local value. An \emph{opaque space} maps inner policies to streams of local values, while an \emph{embedded tree} is a tree that shares the same signature and inner-policy type as its surrounding node, and whose success leaves carry local values.
\end{description}

Let us illustrate these definitions with \code{CBranch} and \code{Join}. From declaration (\ref{code:cbranch-def}), \code{CBranch} nodes contain an opaque space \code{cands} and a parametric opaque space \code{compare}. The associated action type coincides with the element type of the enclosed opaque space. From declaration (\ref{code:join-def}), \code{Join} nodes contain two embedded trees, \code{left} and \code{right}. The associated action type is the product of their result types. More generally, the grammar of admissible effect declarations (Figure~\ref{fig:effect-grammar}) mirrors our definition of generic search trees: each argument of the trigger function defines a (possibly parametric) space, while its return type determines the action type of the corresponding node.

\begin{figure}[t]
\begin{lcodebox*}
\begin{lstlisting}[style=haskell,style=withSmallFont]
data CBranch p t n a where CBranch :: {
    cands :: LocalOpaque p n a,
    compare :: Local n [a] -> LocalOpaque p n [Double] } -> CBranch p t n a  (@\bigskip@)
data Join p t n a where Join :: { left :: t n a, right :: t n b } -> Join p t n (a, b) (@\bigskip@)
getStream :: LocalOpaque p n v -> p -> Stream IO (Local n v)  (@\label{line:get-stream}@)
\end{lstlisting}
\end{lcodebox*}
\vspace{-0.2em}
\CodeFigSpaceBefore
\caption{Examples of Derived Node Types. Each effect declaration induces a GADT node type with four parameters: the node's ambient inner policy type \code{p}, the type \code{t} of embedded trees, a phantom type \code{n} uniquely identifying the node for enforcing locality, and the node's action type \code{a}. The inference rules for deriving node types are defined in Appendix~\ref{ap:strategy-trees} (Figure~\ref{fig:node-type-derivation}) but are better understood through examples, which we provide here for \code{cbranch} (\ref{code:cbranch-def}) and \code{join} (\ref{code:join-def}). \code{Local n a} denotes the type of local values attached to node \code{n}, while \code{LocalOpaque p n a} is the type of a \emph{local} opaque space producing such values (see \code{getStream}).}\label{fig:derived-node-types-examples}
\SmallerSpaceBeforeNextFig
\vspace{0.2em}
\end{figure}
\begin{figure}[t]
\begin{lcodebox*}
\begin{lstlisting}[style=haskell,style=withSmallFont]
data Tree s p n v (@\label{line:tree:start}@)
  = Success (Local n v) (@\label{line:tree:success}@)
  | forall n'. SomeNode (NodeOf s p (Tree s p) n' (Tree s p n v))  (@\label{line:tree:end}@) (@\label{line:tree:some-node}@) (@\bigskip@)
data NodeOf s p t n k where  (@\label{line:node-of:start}@)
  Z :: Node e p t n k -> NodeOf (e ': s) p t n k
  S :: NodeOf s p t n k -> NodeOf (e ': s) p t n k  (@\label{line:node-of:end}@) (@\bigskip@)
data Node e p t n k =
  forall a. (Effect e) => Node { effect :: e p t n a, child :: Local n a -> k }
\end{lstlisting}
\end{lcodebox*}
\CodeFigSpaceBefore
\caption{Definition of a Generic Strategy Tree. Type \code{Tree s p n v} denotes a tree with signature \code{s}, inner policy type \code{p}, return type \code{v}, and contained in a node identified by phantom type \code{n}. Such a tree is either a \emph{success leaf} that contains an \code{n}-local value, or an \emph{effect node} (Figure~\ref{fig:derived-node-types-examples}) whose higher-kinded type \code{e} belongs to \code{s} (\code{NodeOf} expresses such membership, with constructors that encode a Peano integer pointing to an element of \code{s}) and whose children are indexed by local values compatible with the node's action type \code{a}.}\label{fig:tree-def}
\end{figure}

\paragraph{Locality and References} Generalizing our discussion from Section~\ref{sec:locality-references}, the children of a node are indexed by \emph{local values} that cannot be produced by policies out of thin air, and which can only be obtained by combining local space elements. Our formal type definition enforces this property using phantom types (Figure~\ref{fig:tree-def}). Local values carry generalized references that trace their origin.

\paragraph{Reifying Strategies into Trees} Together, Figures~\ref{fig:strategy-language-signature} and~\ref{fig:effect-grammar} define the syntax of our strategy language. Its semantics is given by a \emph{reification} function that maps strategies to trees. The exact definition of this function is largely determined by the types of strategies and trees, and is therefore not particularly insightful. A Haskell implementation is provided in the supplementary material, where strategies are internally represented using free monads~\cite{swierstra2008data}.

\subsection{The Policy Language}\label{sec:policy-language}

\begin{figure}
  \centering
  \begin{subfigure}[t]{0.48\textwidth}
    \centering
    \includegraphics[width=\textwidth]{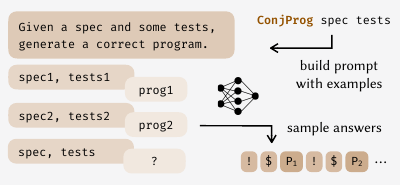}
    \caption{The \emph{few-shot} prompting policy}
    \label{fig:few-shot-stream}
  \end{subfigure}
  \hfill
  \begin{subfigure}[t]{0.48\textwidth}
    \centering
    \includegraphics[width=\textwidth]{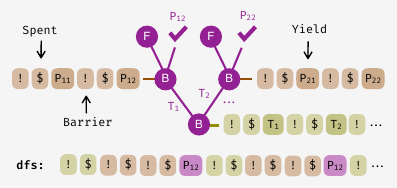}
    \caption{The \emph{depth-first-search} search policy}
    \label{fig:dfs-stream}
  \end{subfigure}
  \vspace{-0.5em}
  \caption{Prompting and search policies map queries and strategies, respectively, to search streams. Section~\ref{sec:dfs} defines \code{dfs}, which, in this example involving the \code{conjectureProg} strategy from Figure~\ref{fig:strategy-overview}, accesses (possibly infinite) streams at each branching node and produces the stream shown at the bottom. }\label{fig:policy-overview}
  \vspace{-1em}
\end{figure}

An oracular program is defined by three components: a \emph{strategy}, a \emph{policy} and a set of \emph{demonstrations}. The previous section explored the first component, defining a language for expressing high-level problem-solving strategies as nondeterministic programs that can be reified into search trees. We now focus on the specification of \emph{policies} that navigate such trees with the assistance of LLM oracles. This separation between \emph{strategies} and \emph{policies}---between \emph{core logic} and \emph{search logic}---provides major practical benefits, as it allows \emph{vastly different} prompting techniques and search algorithms to be explored concurrently without requiring \emph{any} changes in strategies (or in demonstrations). 

We propose a \emph{layered} policy language, which can be used at two levels. At a low level, it offers a series of combinators for building and combining \emph{search streams}, guarding \emph{by construction} against resource misuse (e.g., losing track of LLM API consumption and accidentally overspending). At a higher level, it provides a library of reusable abstractions for assembling policies by combining basic search algorithms and prompting policies with \emph{stream transformers} and \emph{tree transformers}.

\subsubsection{Policies and Search Streams}\label{sec:search-stream}

In the previous section, we defined \emph{prompting} and \emph{search policies} as functions mapping queries and trees, respectively, to \emph{streams} of local values, while remaining intentionally abstract concerning the nature of these streams.

As a first approximation, a \emph{search stream} is simply a (possibly infinite) iterator over local values. When a search policy encounters an opaque space, it can obtain such an iterator from it and use it to extract space elements \emph{on demand}. Already we encounter a problem though, since the search policy may be unwilling to relinquish full control to such an iterator, indefinitely waiting for an element that might never arrive. Instead, it might  want to apply a \emph{budget limit} to that stream (e.g., how many LLM requests or what inference cost it is willing to incur). Doing so requires search streams to be resource-aware. An intuitive view of streams is given in Figure~\ref{fig:policy-overview}: they can be seen as iterators producing messages of three kinds. A \code{Yield} message produces a new value; a \code{Barrier} message (shown as \code{!}) requests authorization to spend an estimated amount of search budget; and a \code{Spent} message (shown as \code{\$}) reports actual resource consumption. Search budgets can be expressed in various metrics, such as number of requests, token usage, or dollar cost for a given LLM API. A consumer of a search stream is therefore capable of monitoring resource usage, and interrupting or resuming the stream whenever it sees fit. 

Supporting concurrency in policies requires a more complex representation of streams. When a policy leverages concurrency internally, consumers of the resulting stream must be able to \emph{selectively} deny spending requests without blocking the entire stream, which in turn calls for bi-directional communication. Details of this representation appear in Appendix~\ref{ap:search-streams}. However, this complexity is abstracted away from users: the policy language hides the underlying representation and lets users build and combine search streams using a combinator language that enforces proper resource management by construction. A representative subset of this language is shown in Figure~\ref{fig:stream-combinators}.

\begin{figure}
\begin{lcodebox*}[left=\DoubleDigitLn]
\begin{lstlisting}[style=haskell,style=withNumbers,style=withSmallFont]
return :: a -> Stream m a  (@\label{line:stream-monad:start}@)
(>>=) :: Stream m a -> (a -> Stream m b) -> Stream m b
lift :: m a -> Stream m a  (@\label{line:stream-monad:end}@)
mzero :: Stream m a
(<|>), parallel :: Stream m a -> Stream m a -> Stream m a (@\bigskip@)
spend :: Budget -> m (a, Budget) -> Stream m a (@\label{line:stream-spend}@)
take :: Int -> Stream m a -> Stream m a
withBudget :: Budget -> Stream m a -> Stream m a
collect :: Stream m a -> m ([a], Budget)  -- only call at top level  (@\label{line:stream:collect}@)
partial :: Stream m a -> Stream m ([a], Budget, Stream m a)  (@\label{line:stream:partial}@)
\end{lstlisting}
\end{lcodebox*}
\CodeFigSpaceBefore
\caption{A Combinator Language for Building Search Streams. The \code{(Monad m)} constraint applies implicitly everywhere, and \code{parallel} additionally requires \code{(MonadIO m)}.}\label{fig:stream-combinators}
\vspace{-0.8em}
\end{figure}

\paragraph{Stream Combinators} Search streams are monad transformers (Figure~\ref{fig:stream-combinators}, Lines~\ref{line:stream-monad:start}-\ref{line:stream-monad:end}) and can therefore be composed through binding (with a behavior analogous to the \code{concatMap} function on lists). Streams can be concatenated with \code{<|>}, for which the empty stream \code{mzero} is the neutral element. The \code{parallel} combinator is a variant, which runs its arguments concurrently, possibly interleaving their values.
Operations that consume resources such as LLM calls can be lifted into streams via \code{spend} (Line~\ref{line:stream-spend}). The \code{spend} function takes as its first argument an estimated over-approximation of the cost that will be incurred. This estimate can be used to proactively abort if not enough resources are available. Its second argument is the resource-consuming operation to be executed, which must return a value along with the \emph{actual} amount of resources that was consumed. The \code{Budget} type can be thought of as representing a multi-dimensional vector of nonnegative real metrics, with possibly infinite values (to represent the absence of a limit). The \code{withBudget} combinator takes a budget limit and a stream as arguments and returns a transformed stream that cannot spend more than the allocated budget, terminating early if needed. Finally, the \code{collect} function (Line~\ref{line:stream:collect}) exhausts a stream, returning all yielded values in a list along with the total spent budget. Note that it is only safe to use at top-level since invoking it \emph{within} another stream loses track of the consumed resources due to the result type not being wrapped into \code{Stream}. A safe variant to use within streams is \code{partial} (Line~\ref{line:stream:partial}), which runs a stream until no more budget is available and then returns the generated elements, the spent budget (for information only, resources are still tracked), and the stream continuation in case one wants to resume it later. In particular, \code{withBudget b (partial s)} gathers as many elements as possible with budget \code{b} and then returns these elements, along with a continuation for \code{s} that is unaffected by this budget limit.

The language presented in Figure~\ref{fig:stream-combinators} for defining search streams offers strong correctness guarantees in terms of resource management (a proof sketch is available in Appendix~\ref{ap:correct-resource-sketch}): 

\begin{property}[Correct Resource Management]\label{prop:correct-resource}
  Let \code{s} be a search stream defined without \code{collect}. Then, collecting the stream \code{withBudget b s} consumes a budget of \code{b} at most, assuming that all calls to \code{spend} provide a correct consumption over-estimate. Failing this assumption, the amount of budget spent over the limit is at most $n \times \delta$, with $n$ the maximum concurrency level of the stream (1 if \code{parallel} is never used) and $\delta$ the maximal estimation error of a call to \code{spend}.
\end{property}

\begin{wrapfigure}{r}{0.521\textwidth}
\vspace{-0.5cm}
\begin{ccodebox*}
\begin{lstlisting}[style=haskell,style=withSmallFont]
dfs :: p -> SearchPolicy '[Branch, Fail] p
dfs p tree = case tree of
  Success x -> return x
  SomeNode (Z (Node (Branch cands) child)) ->
    getStream cands p >>= dfs p . child
  SomeNode (S (Z (Node Fail _))) -> mzero
\end{lstlisting}
\end{ccodebox*}
\vspace{-0.4cm}
\end{wrapfigure}
\subsubsection{Example: Depth-First Search}\label{sec:dfs}

We can now define the \code{dfs} search policy illustrated in Figure~\ref{fig:dfs-stream}. The \code{dfs} function is defined inductively on the structure of a strategy tree whose non-success nodes can have type \code{Branch} and \code{Fail}. The \code{p} argument is the ambient inner policy. As
defined in Figure~\ref{fig:tree-def}, the \code{S} and \code{Z} constructors are used to encode the position of different effects in the \code{[Branch, Fail]} list, thereby encoding extensible sum types in Haskell. Encountering a success node yields an element, encountering a failure node yields no element, while encountering a branching node leads to lazily exploring the \code{cands} space while recursively calling \code{dfs} on each generated element, which we do via the monadic bind combinator. Note that adding a \code{maxBranching} parameter to \code{dfs} is unnecessary, since the prompting policies and search policies in \code{p} can be composed with the \code{take n} stream transformer, as we explain next.

\subsubsection{Assembling Policies}\label{sec:assembling-policies}

Search and prompting policies can be written from scratch using the search stream combinators introduced in Section~\ref{sec:search-stream}. More typically, however, they are assembled by composing standard building blocks (e.g., \code{fewShot}, \code{dfs}, \code{mcts}) with \emph{stream transformers} and \emph{tree transformers}. \emph{Stream transformers} are functions that map streams to streams of the same type. Examples include \code{withBudget b}, \code{take n}, and \code{majorityVote}, which exhausts a stream and yields the most frequently observed element, if any.
A \emph{tree transformer} maps a tree to another tree, often with a different signature so that it can be used with a specific search policy.

For example, the \code{generateProg} strategy from Figure~\ref{fig:strategy-overview} must be paired with a search policy that handles branching, failure, and \emph{value} nodes. This is the case for \code{bestFirst}~\cite{dechter1985generalized} and \code{mcts}~\cite{browne2012survey}, which leverage value information to prioritize branches to explore. Alternatively, \code{dfs} can be used by right-composing it with the \code{dropValues} transformer---which removes value nodes entirely, preventing value computations---or with the \code{threshold} transformer---which converts value nodes whose estimates fall below a given threshold into failure nodes. As another example, the \code{elimJoin} transformer eliminates all \code{Join} nodes in a tree by inlining their subtrees in order (Appendix~\ref{ap:policy-blocks}).

\subsection{The Demonstration Language}\label{sec:demonstration-language}

Although queries can sometimes be answered successfully via \emph{zero-shot} prompting, LLMs generally perform better when given examples of how to answer queries of the same type. Such examples are therefore central to LLM-enabled programs, yet are typically stored as raw query--answer pairs and treated as isolated data, disconnected from the logic that links queries together in multi-prompt pipelines. As a result, they are often difficult to \emph{read}, \emph{write}, and \emph{maintain}. They are difficult to \emph{read} because individual examples are often collected with little or no information about the context in which they arise, and are not grounded in concrete narratives of solving end-to-end problems. They are difficult to \emph{write} because determining \emph{what} questions need to be answered can be nontrivial for queries that occur in the middle of complex pipelines---not to mention the cost of crafting high-quality answers. Finally, they are difficult to \emph{maintain} because changes in the pipeline can render some examples obsolete or incorrect, without clear ways for users to identify which ones.

We introduce a novel language for writing and evolving examples in oracular programs. In this language, related examples are bundled with unit tests into coherent \emph{demonstrations}. Tests themselves can be specified concisely in a dedicated sublanguage. The language is designed to support a rich editor experience through an interactive, test-driven workflow for writing and repairing demonstrations, either manually or with LLM assistance. In addition, demonstrations are policy-agnostic and thus never break as a result of policy changes.

\subsubsection{Overview of Demonstrations}

\begin{figure}
  \centering
  \includegraphics{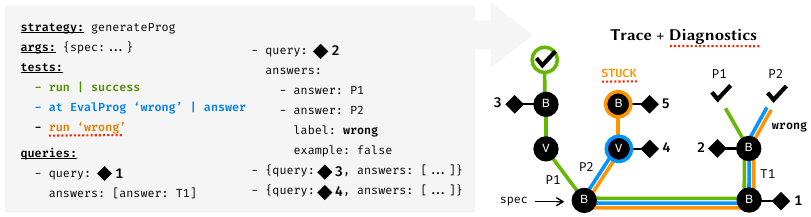}
  \caption{Example of a Demonstration for the Strategy from Figure~\ref{fig:strategy-overview}. A demonstration bundles a set of query answers with a sequence of unit tests that describe concrete scenarios of navigating a specific strategy tree using these answers. Each test describes a path in the tree, highlighted with a matching color.}\label{fig:demonstration-overview}
  \SmallerSpaceAfterFigure{}
\end{figure}

Figure~\ref{fig:demonstration-overview} shows a demonstration for the \code{generateProg} strategy (Figure~\ref{fig:strategy-overview}). A demonstration consists of a \emph{strategy value} (i.e., a strategy function with all its arguments instantiated), a set of \emph{navigation tests}, and a set of \emph{answered queries}. It is typically written in YAML syntax. In our example, the strategy value is specified by the \code{strategy} and \code{args} fields, which together induce a unique search tree. \emph{Answered queries} are listed in the \code{queries} field. An \emph{answered query} is a pair consisting of a query and a list of \emph{answers}. Each answer may optionally carry a \emph{label} (e.g., \code{wrong}) and a flag indicating that it must \emph{not} be used as a few-shot example (we return to this distinction shortly). Figure~\ref{fig:demonstration-overview} depicts queries as diamonds, with identifiers locating them in the search tree, while actual demonstrations store serialized representations of query values.

Crucially, the answered queries are assembled into coherent scenarios of navigating the search tree via \emph{navigation tests}. Three such tests are shown in Figure~\ref{fig:demonstration-overview}. Each test describes a path through the tree, starting at the root and ending at a specific node. Tests can either succeed, as in the first two cases, or fail, as in the third. Also, each test is composed of a sequence of instructions that are chained together using the \code{|} pipe operator. The first test (\code{run \!\!|\!\! success}) ensures that the provided examples are sufficient for solving the particular problem instance that the demonstration is about. The \code{run} instruction can be informally interpreted as ``walk through the tree, using the first listed answer as a response every time a listed query is encountered''; the \code{success} instruction checks that the node at the end of this path is a success leaf. The second test demonstrates how to recover from a suboptimal choice by assigning it a low value. The \code{at EvalProg `wrong`} instruction is similar to \code{run}, except that the answer labeled \code{wrong} must be selected when applicable and navigation must stop upon reaching a node associated with a query of type \code{EvalProg}; the \code{answer} instruction ensures that this query is answered in the demonstration. Since the answer labeled with \code{wrong} is not optimal and only included for the purpose of demonstrating how to reflect on a bad choice, it is marked as unsuitable for use as an example. Finally, the third test aims for a full walk through the tree, selecting the \code{wrong}-labeled answer when appropriate. However, it hits a query that is not listed in the demonstration and thus fails as \emph{stuck}, providing the user with appropriate information on what query must be answered for navigation to proceed. With proper tooling, this mechanism allows writing and repairing demonstrations interactively (Section~\ref{sec:demonstration-workflows}).

\subsubsection{Walking Through the Tree}\label{sec:tests-dsl}

A test consists of a sequence of \emph{instructions}. Each instruction takes as an input a node in the tree (initially the root) and returns a new node. Starting at the current node, the \code{run} instruction uses answered queries to walk through the tree until a leaf node is reached or an answer is missing. But how should this walking behavior be defined? Since our strategy language is extensible, it must accommodate arbitrary effect nodes. Also, it must not depend on a specific policy since demonstrations are meant to be policy-agnostic.

\begin{figure}
\begin{ccodebox*}
\begin{lstlisting}[style=haskell]
navigate (CBranch cands _) = Just (\choose -> choose cands)
navigate Fail = Nothing
navigate (Value _) = Just (\ _ -> return nil)
navigate (Join l r) = Just (\choose ->
    do { vl <- choose l ; vr <- choose r ; return (liftPair (vl, vr)) })
\end{lstlisting}
\end{ccodebox*}
\CodeFigSpaceBefore
\caption{Examples of Standard Navigation Functions. A type signature for the \code{navigate} type class method is provided in Appendix~\ref{ap:effect-def}. The \code{liftPair} function lifts a pair of local values into a local value (Appendix~\ref{ap:references}).
}\label{fig:navigation-function-examples}
\vspace{-0.8em}
\end{figure}

The answer is to have each effect define a canonical \emph{navigation function} that, at any given node, maps a \emph{choice function} capable of selecting elements from local spaces to an action. The type for navigation functions is formally defined in Appendix~\ref{ap:effect-def}. Here, we explain this concept through examples, by looking at the navigation functions of standard effects (Figure~\ref{fig:navigation-function-examples}). Navigating a \code{CBranch} node consists in selecting an element from the \code{cands} space and using it as an action. Failure nodes are leaf nodes and thus cannot be navigated. Value nodes have a unique child, which can be selected without further inspection. Finally, \code{Join} nodes can be navigated by selecting elements from the \code{left} and \code{right} spaces and pairing them up to form an action.

Whenever \code{run} needs to select an element from a space defined by a query, it looks for this query in the demonstration's \code{queries} section and picks the first provided answer. If none is found, it fails at the current node. When \code{run} encounters a space defined by a tree, it recursively navigates this tree. The \code{run} instruction stops when reaching a leaf at the same level of nesting where it started.

\paragraph{Exploring alternative paths with hints}
The \code{run} instruction can be passed a sequence of answer labels as \emph{hints}, so as to specify alternate paths through the tree. Whenever a query is encountered, it is checked whether or not an answer is available whose label matches the first provided hint. If so, this answer is used and the hint is consumed. For example, instruction \haskellCode{run 'foo bar'} can be interpreted as: ``walk through the tree, using answer \code{foo} whenever applicable and then \code{bar}''. Our design allows describing paths concisely, by only specifying the few places in which they differ from a {default} path. This works well for specifying demonstrations, which typically describe a central scenario, on top of which side explorations are made (e.g., showing how a bad decision leads to a low value score---as demonstrated in Figure~\ref{fig:demonstration-overview}---or demonstrating how redundant candidates can be removed at a particular step).

\paragraph{Other instructions}
In most cases, nothing more than \code{run} is needed to reach arbitrary success leaves in the search tree, which we capture as a \emph{strong completeness} property in Section~\ref{sec:demo-completeness}. Other test instructions (Appendix~\ref{ap:demonstration-tests}) are still useful for exploring parts of the search tree that are \emph{not} on any shortest path to a solution (e.g., a sub-strategy that computes value estimates---useful for guiding search but not strictly necessary), or for cases not covered by our completeness result.

\subsubsection{Completeness}\label{sec:demo-completeness}

Our demonstration language offers \emph{two} completeness guarantees. The first (\emph{weak completeness}) states that any node in a search tree can be reached by a demonstration. The second (\emph{strong completeness}) states that, under mild assumptions, any success node can be reached using only the \code{run} instruction. Weak completeness ensures that users cannot get stuck when demonstrating a known solution to a problem due to limitations in the expressiveness of the demonstration language. Strong completeness implies that, \emph{in most cases}, the \code{run} instruction suffices, enabling a maximally simple and interactive demonstration-writing workflow. Proof sketches are provided for both completeness theorems in Appendix~\ref{ap:demonstration-completeness}.

Weak completeness is not hard to obtain in principle, since \emph{Haskell} itself provides a trivially complete testing DSL. The challenge is to achieve it while keeping the DSL simple. The \emph{strong} completeness guarantee is more interesting and subtle. It states that, under a suitable assumption about navigation functions, and for any success leaf of a tree, there exists a demonstration with a single \code{run} test that reaches a success leaf with an \emph{equivalent} value. Two local values (Appendix~\ref{ap:references}) are said to be \emph{equivalent} if they have the same content, despite possibly different references.

The required assumption on navigation functions is \emph{invertibility}. A navigation function is said to be \emph{invertible} if, for any valid action in an associated node, there exists a sequence of local space elements such that the navigation function generates an equivalent action when passed an \emph{inverse} choice function that picks all these elements in sequence.
All effects introduced so far have invertible navigation functions. For example, \code{CBranch} has an invertible navigation function since any valid action is an element from the \code{cands} space that can be picked by the inverse choice function (see technical caveat in Appendix~\ref{ap:cbranch-invertibility}). In addition, the navigation function for \code{Value} is trivially invertible since value nodes have a single child. Our first case study introduces a new \code{Abduction} effect that is \emph{conditionally} invertible (Figure~\ref{fig:abduction-eff}).

\subsubsection{Writing and Repairing Demonstrations}\label{sec:demo-maintenance}\label{sec:demonstration-workflows}

Demonstrations can be written interactively. A typical workflow is to start with an empty \code{queries} section and a single \code{run \!\!|\!\! success} test. Evaluating the demonstration results in the test getting \emph{stuck} at a given node. Editor support allows visualizing this node, along with the attached unanswered query, which can then be added to the demonstration with a single click (Appendix~\ref{ap:writing-repairing-demos}, Figure~\ref{fig:demo-screenshot}). An answer can be written manually, or generated by an LLM and edited, after which the demonstration can be evaluated again. A more advanced workflow is possible, where demonstrations are partially written, with holes automatically filled by running external policies and extracting query answers from successful searches (Appendix~\ref{ap:implicit-answers-hybrid-workflows}).

By design, policy changes cannot break demonstrations. Many strategy changes are also guaranteed to preserve them. For example, adding a call to \code{value} in a strategy cannot cause a previously passing test to fail, nor can introducing or eliminating a call to \code{join}. In the event that a strategy change \emph{does} break a demonstration, the breakage is clearly indicated by one or more failing tests. %

\section{Delphyne: a Python-Based Oracular Programming Framework}\label{sec:delphyne}

We introduce \emph{Delphyne}, an open-source\footnote{Delphyne is available at: \url{https://github.com/jonathan-laurent/delphyne}.}
framework for oracular programming based on Python.\footnote{In Greek mythology, Delphyne was the monstrous serpent, also known as \emph{Python}, that guarded the Delphi oracle.} Beyond being a popular and accessible programming language with a rich ecosystem, Python
\begin{wrapfigure}{r}{0.4\textwidth}
\centering
\vspace{0.1cm}
\includegraphics[width=0.35\textwidth]{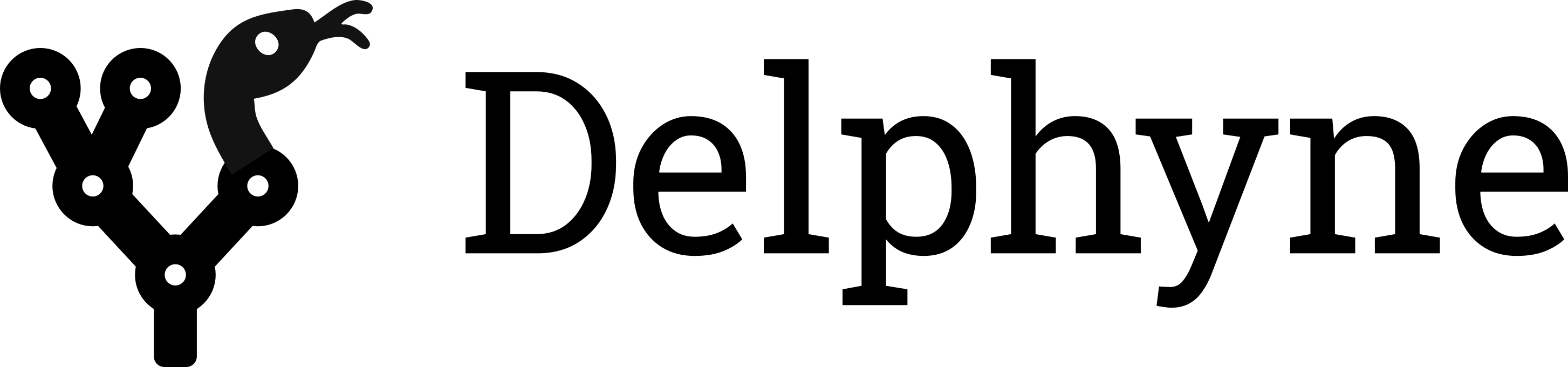}
\vspace{-0.1cm}
\end{wrapfigure}
has several technical properties that make it suitable for implementing an oracular programming framework. First, its \emph{reflection capabilities} allow the definition of new effects and the enrichment of trees with debugging information with minimal boilerplate. Second, its ecosystem includes powerful libraries for structured data serialization and parsing (e.g., Pydantic) and for writing prompts as string templates (e.g., Jinja). Finally, and perhaps surprisingly, its optional static type system is expressive enough to precisely type the strategy language, along with all standard components of the policy language. Gradual typing is only necessary within the internals of primitive policy components (e.g., \code{dfs}), which typical users rarely, if ever, need to implement.
Delphyne offers rich tooling support for writing demonstrations and inspecting search trees, in the form of a VSCode extension (Appendix~\ref{ap:editor-support}). It consists of about 23{,}000 lines of Python and TypeScript (excluding blank lines and comments), and about 45{,}000 words of documentation.

\begin{figure}
\begin{ccodebox*}
\begin{lstlisting}[style=delphyne]
@strategy
def generate_prog(spec: Spec) -> Strategy[Branch | Fail | Value, GPIP, Prog]:
    prog = yield from branch(conjecture_prog(spec).using(lambda p: p.gp))
    yield from value(EvalProg(spec, prog).using(lambda p: p.ep))
    proof = yield from branch(ProveProg(spec, prog).using(lambda p: p.pp))
    yield from ensure(check_proof(spec, prog, proof))
    return prog
\end{lstlisting}
\end{ccodebox*}
\CodeFigSpaceBefore
\caption{An Implementation of \code{generateProg} (Figure~\ref{fig:strategy-overview}) in Delphyne. Strategies are described via coroutines instead of monads (hence \code{yield from}) and the \code{using} method corresponds to \code{query} and \code{search}.}\label{fig:delphyne-strategy-example}
\SmallerSpaceAfterFigure{}
\vspace{-0.3em}
\end{figure}

\paragraph{Embedding the Strategy and Policy Languages}
Delphyne strategies are not expressed with monads (for which Python has poor syntactic support) but with coroutines. An example is provided in Figure~\ref{fig:delphyne-strategy-example}. Since Python coroutines cannot be cloned, reification is implemented via thermometer continuations~\cite{koppel2018capturing}: a tree node is identified by the sequence of actions leading to it, and the strategy is replayed from scratch whenever a child is computed. This does not raise practical performance issues, since expensive computations can be cached (Appendix~\ref{ap:compute-effect}). While strategies must be \emph{pure} and not depend on global mutable state, Delphyne allows \emph{internal} side effects and mutations, which are idiomatic in Python (Appendix~\ref{ap:strategy-mutable-state}). Although defining fully precise types for strategy trees requires advanced type-system features such as GADTs and higher-kinded types (Figure~\ref{fig:tree-def}) that are unavailable in Python, Delphyne offers fully precise static typing for the strategy language (since effect triggers such as \code{branch} can be typed; see Appendix~\ref{ap:delphyne-strategy-type-safety}) and for policy components (i.e., search and prompting policies, stream and tree transformers). Gradual typing (i.e., the use of \code{Any}) is only necessary \emph{inside} the implementation of policy components such as \code{dfs} that manipulate trees (Appendix~\ref{ap:defining-new-search-algos-in-delphyne}). Within these components, locality (Section~\ref{sec:locality-references}) is enforced at runtime.

\section{Case Studies}\label{sec:case-studies}

Having established oracular programming as a \emph{principled} foundation for LLM-enabled software—with strong theoretical guarantees of separation, modularity, consistency, completeness, and evolvability—we empirically demonstrate the power of our framework through three case studies.%

\begin{description}
    \item[Advanced Search for Loop Invariant Discovery (\ref{sec:loop-invariants-case-study}).]
 We showcase Delphyne's ability to leverage advanced search algorithms. An oracular program written in hours is competitive with a ReAct baseline that uses comparable prompts, while achieving a fourteenfold reduction in inference cost by combining smaller models with search. This development speed is enabled by
\begin{itinparaenum}%
    \item the composition of standard-library components, enabled by Delphyne's extensible effect system,
    \item an interactive demonstration-writing workflow allowing examples to be written in minutes, and
    \item concurrently tuning and testing vastly different search algorithms, enabled by the separation of strategies and policies.
\end{itinparaenum}%
    \item[Self-Improvement of a Lean Prover Agent (\ref{sec:lean-case-study}).] We show how oracular programming enables universal yet customizable self-improvement mechanisms by analyzing traces of successful and unsuccessful runs to extract retrievable examples and improve prompts automatically. In addition, we show how it supports the nested integration of both \emph{horizontal} and \emph{vertical} pipelines, in which LLMs orchestrate traditional computations and vice versa.
    \item[Universal Queries (\ref{sec:universal-queries-case-study}).] 
    We show a Delphyne extension that allows writing oracular programs without defining custom queries and prompts, enabling particularly concise programs. 
\end{description}

\noindent Full code for all case studies is available in the supplementary material. %

\subsection{Advanced Search for Loop Invariant Discovery}\label{sec:loop-invariants-case-study}

We showcase Delphyne's ability to leverage advanced search algorithms through the problem of \emph{loop invariant synthesis} in Why3~\cite{filliatre2013why3}. LLM-enabled invariant synthesis is an active area of research~\cite{quokka-wei2025invbench,llama4inv-wu2024llm,loopy-kamath2023finding,lemur-wu2023lemur,autospec-wen2024enchanting,key-isola-beckert2024towards,esbmc-pirzada2024llm} (see the recent survey by Wei et al.~\cite[Introduction]{quokka-wei2025invbench}), and state-of-the-art performance is currently achieved by ReAct-style agents~\cite{yao2022react}, in which large reasoning models interact with a prover within a single feedback loop~\cite{quokka-wei2025invbench}. We implement such an agent in Delphyne, along with an alternative agent that uses similar prompts but leverages smaller models and advanced search (Sections~\ref{sec:inv-react-baseline}--\ref{sec:inv-two-vastly-different-policies}) to drastically reduce inference costs while remaining competitive on the Code2Inv benchmark suite (Section~\ref{sec:code2inv-results}). As we demonstrate, Delphyne is uniquely suited to the rapid development of such advanced pipelines (Section~\ref{sec:inv-case-for-oracular-programming}).

\subsubsection{ReAct Baseline}\label{sec:inv-react-baseline}

A ReAct-style agent can be implemented in a few lines (ignoring custom prompts and demonstrations) using the higher-order \code{interact} strategy from Delphyne’s standard library (Appendix~\ref{ap:interact}). Given an input Why3 program, it repeatedly tries to suggest invariant annotations until the program is successfully proved, receiving feedback from Why3 on failed attempts, and periodically clearing its context at a frequency determined by the policy (Appendix~\ref{ap:invariant-strategies-policies}).

\subsubsection{A Strategy Based on Recursive Abduction}\label{sec:inv-abduction-strategy}

We propose an alternate strategy for invariant synthesis (Appendix~\ref{ap:invariants-case-study}, Figure~\ref{fig:inv-saturation}) that revolves around a query whose role is to suggest auxiliary invariants that may facilitate the proof of a given goal. The query takes as input: a goal (assertion or invariant) that Why3 failed to prove, the associated prover feedback, and a set of already-established invariants. Based on this information, it attempts to generate one or more candidate invariants that could plausibly serve as intermediate lemmas before eventually retrying the original proof obligation. Delphyne defines a dedicated \code{Abduction} effect in its standard library that generalizes this recursive abduction pattern (Figure~\ref{fig:abduction-eff}). Decisions on which abduction candidates to explore, in what order, and assuming which pre-established facts, are left to policies.

\begin{figure}
\begin{lcodebox*}[label=code:abduction-eff]
\begin{lstlisting}[style=haskell]
effect abduction :: forall fact proof feedback. (Abduction `In` s, Eq fact) =>
  { prove :: ([(fact, proof)], Maybe fact) -> Opaque p (Either proof feedback)
  , suggest :: feedback -> Opaque p [fact]
  , searchEquivalent :: ([fact], fact) -> Opaque p (Maybe fact)
  , redundant :: ([fact], fact) -> Opaque p Bool } -> Strategy s p proof
\end{lstlisting}
\end{lcodebox*}
\CodeFigSpaceBefore
\caption{The \code{Abduction} Effect for Recursive Abduction. The \code{prove} function takes as an argument a list of pre-established facts along with a new fact to prove (or \code{Nothing} to refer to the main goal). It returns a proof, or, failing that, some feedback. The \code{suggest} function maps such feedback to a list of suggested auxiliary facts to prove. The \code{searchEquivalent} function determines whether a given fact is equivalent to another one from a given list (useful for removing semantic duplicates), and the \code{redundant} function determines whether a fact is implied by a list of other facts. Actions are proofs of the main goal. Appendix~\ref{ap:abduction-effect} describes the associated navigation function, which is invertible under mild assumptions.
}\label{fig:abduction-eff}
\vspace{-0.5em}
\end{figure}

\subsubsection{Two Vastly Different Policies}\label{sec:inv-two-vastly-different-policies}

We define two policies for our abduction-based strategy. The first (Appendix~\ref{ap:invariants-case-study}, Figure~\ref{fig:inv-par-abduction-policy}) makes short, repeated, and concurrent attempts to generate invariants from scratch, each time exploring a small number of abduction candidates (e.g., two) recursively up to some depth. The second (Appendix~\ref{ap:invariants-case-study}, Figure~\ref{fig:inv-saturation-policy}), used in our Code2Inv experiment, is designed to handle \emph{large} numbers of abduction candidates and uses a saturation-based search algorithm from Delphyne’s standard library.
This search algorithm performs repeated \emph{abduction rollouts} where, starting from the main goal and up to some depth, a large number of suggestions are sampled from an LLM and semantically deduplicated before one of them is recursively examined, based on how many times it was suggested and explored before (in a way reminiscent of the UCT formula~\cite{browne2012survey}). Established facts are accumulated and systematically assumed once proved. The whole process is restarted periodically and runs until some allocated budget is consumed. The invariant suggestion prompt asks for \emph{concise} and \emph{constrained} answers, as lists of pairs that each consist in {\it (i)} a short identifier referring to a relevant piece of advice from the system prompt (Appendix~\ref{ap:inv-sys-prompt}) and {\it (ii)} a proposed invariant. Such concision allows sampling more answers at any given budget, trading precision for quantity and diversity, to be leveraged by search.

\subsubsection{Results on Code2Inv}\label{sec:code2inv-results} We compare our abduction-based oracular program to three ReAct agent baselines, each using a different model. All agents receive demonstrations for the same three example problems, with comparable levels of explanation, and all receive identical advice in their system prompt, including advice on using abduction to identify missing invariants (Appendix~\ref{ap:inv-sys-prompt}). The key difference therefore lies not in how much domain-specific knowledge is shared in prompts, but in how model calls are chained and orchestrated. Each agent is tasked to solve all 124 problems from the Code2Inv benchmark suite~\cite{si2018learning}, with a \$0.20 budget limit on LLM API spending per problem. All baselines are properly tuned (Appendix~\ref{ap:inv:protocol}). Results are shown in Table~\ref{tab:code2inv}. The abduction-based agent solves all problems, with an average spending per problem that is over an order of magnitude lower than baselines. Interestingly, the ReAct agent based on the smaller GPT-4o-mini model outperforms the one based on GPT-4o on all metrics, by virtue of being significantly cheaper and thus being afforded more trials per problem within its budget limit.

These results illustrate the effectiveness of orchestrating well-scoped LLM queries and prover feedback through advanced search. More importantly, Delphyne allows building such pipelines in mere hours, by composing reusable components and writing demonstrations interactively.

\begin{table}
  \begin{tabular}{lccc}
  \toprule
  \textbf{Agent} & \textbf{Problems Solved} & \textbf{Average Price} (\textcent) & \textbf{Median Price} (\textcent) \\
  \midrule
Abduction (gpt-4o-mini) & \textbf{124.0 $\pm$ 0.0} & \textbf{0.12 $\pm$ 0.02} & \textbf{0.05 $\pm$ 0.00} \\
Baseline (gpt-4o) & 113.0 $\pm$ 2.0 & 3.19 $\pm$ 0.31 & 0.80 $\pm$ 0.17 \\
Baseline (gpt-4o-mini) & 120.0 $\pm$ 1.7 & 1.64 $\pm$ 0.24 & 0.10 $\pm$ 0.01 \\
Baseline (o3) & \textbf{124.0 $\pm$ 0.0} & 1.90 $\pm$ 0.03 & 1.58 $\pm$ 0.13 \\
  \bottomrule
  \end{tabular}
\vspace{0.25cm}
  \caption[Experimental results on Code2Inv]{Experimental Results on Code2Inv (124 problems, excluding incorrect ones). Standard deviations are estimated based on three random seeds. A 20\textcent{} budget limit is fixed for each problem.}\label{tab:code2inv}
\vspace{-2em}
\end{table}

\subsubsection{A Case for Oracular Programming}\label{sec:inv-case-for-oracular-programming}

Delphyne allowed us to express the core logic of our abduction-based oracular program in less than fifty lines of strategy code (Appendix~\ref{ap:invariants-case-study}, Figure~\ref{fig:inv-saturation}), which directly express domain-specific knowledge about invariant generation, and which we never had to revisit (in contrast to the search logic, as we discuss next). It allowed us to write three demonstrations in mere minutes, by simply {\it (i)} choosing example programs to verify, {\it (ii)} writing empty demonstrations with single \code{run\,|\,success} tests, and {\it (iii)} following the guidance of the demonstration interpreter to answer relevant queries until all tests passed (the universality of this workflow is guaranteed by the strong completeness property of the demonstration language; Section~\ref{sec:demo-completeness}). Some of these queries were large---featuring Why3 feedback in the form of failing proof obligations---but all were automatically generated and so we only had to \emph{answer} them. Finally, after a strategy update made more detailed Why3 feedback available, the Delphyne VSCode extension enabled us to update demonstrations with almost no effort. It identified breakages---in the form of stuck tests and unused demonstration queries---and proposed query updates in the form of inspectable diffs. By reviewing these diffs, we were able to validate the proposed changes and, in some cases, refine the explanations accompanying our answers to better leverage the new feedback information (see simplified scenario in Appendix~\ref{ap:editor-support}, Figure~\ref{fig:demo-refactor-screenshot}).

By leveraging the standard \code{Abduction} effect, which is definable in its full generality using Delphyne's extensible effect system, our strategy accommodates a wide variety of policies. We iteratively improved our saturation-based policy (Appendix~\ref{ap:invariant-strategies-policies}), introducing additional hyperparameters to better control the computational cost of propagating known facts. This process was most time-intensive, but---crucially---no change was ever needed to the strategy code, and thus to the demonstrations, allowing us to iterate fast and to maintain and optimize multiple policies concurrently. Had our program been written in a traditional way, without a strict separation between core and search logic, frequent refactorings would have been required, and supporting multiple search algorithms at the same time would have introduced significant additional complexity.

\subsection{Self-Improvement of a Lean Prover Agent}\label{sec:lean-case-study}

Every oracular program is automatically equipped with a natural self-improvement mechanism reminiscent of Expert Iteration~\cite{anthony2017thinking}. Each time a problem is solved through search, examples of correct decisions can be extracted along the tree path leading to success. These examples can then be used to improve future executions, for instance through few-shot prompting, retrieval~\cite{gao2023retrieval}, or fine-tuning. Delphyne allows domain-specific knowledge to be leveraged for refining this coarse mechanism---for example, by integrating negative feedback, assigning credit for the successful solving of independent subproblems, or retroactively revisiting answers in strategies that perform iterative repair. It does so by defining a \code{Feedback} effect that strategies can use to declare \emph{hindsight knowledge} about the outcomes of previous choices (Appendix~\ref{ap:self-improvement-and-reflection}).

\begin{wrapfigure}{r}{0.42\textwidth}
\vspace{-1em}
\includegraphics[scale=0.95]{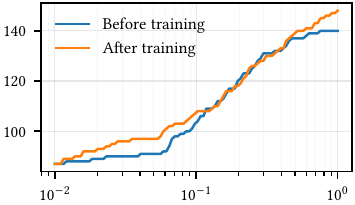}
\vspace{-0.9em}
\caption{Number of MiniF2F test problems solved as a function of the per-problem inference budget (in dollars).}
\label{fig:leandra-scaling}
\vspace{-0.4em}
\end{wrapfigure}

\subsubsection{Self-Improvement Experiment} We implement a strategy for proving theorems in Lean, inspired by Draft-Sketch-Proof~\cite{jiang2022draft}. Given a theorem, a proof sketch is first produced, and the resulting goals are then processed independently by a dedicated substrategy (using the \code{Join} effect). This substrategy has access to a tool for finding relevant theorems in Mathlib, itself implemented as a custom strategy that interacts with Loogle. An example execution is provided in Appendix~\ref{ap:lean-example}. Evaluated on the MiniF2F~\cite{zheng2021minif2f} \emph{test set}, our agent solves 140 problems (57\%). It then undergoes an automatic self-improvement cycle that proceeds as follows. First, it attempts to solve each problem from the MiniF2F \emph{validation set}. Then, examples are extracted from these attempts and made available for few-shot prompting via MMR retrieval~\cite{carbonell1998use}. In addition, a \emph{meta-strategy} examines all search traces (Section~\ref{sec:locality-references}) to identify queries that were ultimately correctly answered, but only at the cost of substantial search effort. For each such query, positive and negative feedback is aggregated and distilled into reusable pieces of advice, to be added back to the agent's prompts (Appendix~\ref{ap:lean-learned-advice}). The improved agent solves 148 problems (60\%) on the MiniF2F test set, while exhibiting a better scaling trend (Figure~\ref{fig:leandra-scaling}).

\subsubsection{A Case for Oracular Programming}

Beyond this quantitative result, the separation of strategies and policies enables generic yet customizable self-improvement mechanisms, by allowing the outcomes of arbitrary runs to be reified into structured traces and then analyzed---a process that we name \emph{search reflection}. In the present case, Delphyne allowed us to implement self-improvement for our Lean prover by merely inserting \emph{four} new lines of strategy code (to collect positive feedback when a subgoal is solved even if the overall proof fails, record prover errors as negative feedback, and link successful proofs obtained after multiple rounds of feedback to all intermediate failed attempts; see Appendix~\ref{ap:lean-strategy-policy-details}) and defining custom prompts for generating and aggregating advice.

In addition, our theorem-proving strategy exploits Delphyne's ability to integrate \emph{horizontal} and \emph{vertical} LLM pipelines, in which LLMs orchestrate traditional computations and vice versa (see Section~\ref{sec:react-comparison} for a comparison): a vertical sketch--prove pipeline delegates subgoal closure to a horizontal agent, giving it access to a lemma-finding tool itself implemented with a custom strategy. This is achieved using Delphyne's standard higher-order \code{interact} strategy (Appendix~\ref{ap:interact}).

Finally, the demonstration language allowed us to write a handful of high-quality examples to bootstrap self-improvement in mere minutes, leveraging a semi-automatic writing workflow in which we provide proof sketches and let our policy try to close as many goals as possible, supplying manual proofs only where it fails, refining the style, and adding explanations (Appendix~\ref{ap:implicit-answers-hybrid-workflows}).

\subsection{Universal Queries}\label{sec:universal-queries-case-study}
\begin{wrapfigure}{r}{0.58\textwidth}
\vspace{-0.9cm}
\begin{ccodebox*}%
\begin{lstlisting}[style=haskell]
x, y, z <- branch (query _ (GenTriple n))
ensure (x^2 + y^2 == z^2 && x + y + z == n)
return (x, y, z)
\end{lstlisting}%
\end{ccodebox*}%
\vspace{-0.6cm}
\end{wrapfigure}
\noindent Consider the strategy computation on the right and imagine being tasked to write a prompt for the \code{GenTriple} query. What would such a prompt say? Most likely, it would only restate the information already present in the surrounding strategy---namely that the oracle must find a Pythagorean triple that sums to \code{n}. In general, what qualifies as an acceptable answer to a query is implicitly specified by the strategy’s \emph{continuation} at the query call site: a good answer is one after which a winning path still exists in the strategy tree. This continuation is fully determined by the strategy’s source code and the current state of its stack, raising the question: can a \emph{universal oracle}~\cite{selsam2020universal} make appropriate decisions directly from this information, eliminating the need to design individual prompts? Delphyne implements this idea in the form of a \code{guess} operator that uses runtime stack inspection to issue instances of a single \emph{universal query}. Appendix~\ref{ap:universal-queries-case-study} demonstrates a short strategy that uses this operator.

\section{Discussion and Limitations}\label{sec:discussion}

\newcommand{\DiscussionNegativeSpace}{}

We situate oracular programming within recent trends in agentic design (Section~\ref{sec:react-comparison}) and discuss its inherent trade-offs and limitations (Section~\ref{sec:limitations}) while identifying opportunities for future work.

\subsection{Comparison with ReAct Agents}\label{sec:react-comparison}

An increasingly popular paradigm for leveraging LLMs is the ReAct-style agent framework~\cite{yao2022react}, exemplified by systems such as Claude Code, in which a reasoning model with an append-only context coordinates calls to external tools. While universal and widely adopted, this framework raises persistent challenges in reliability, steerability, security, and cost. First, long contexts incur substantial inference costs and can degrade model performance through context distraction, confusion, and poisoning~\cite{comanici2025gemini}. Second, composing tool calls through the model’s context is flexible but expensive and exposes fundamental security vulnerabilities via prompt injection~\cite{debenedetti2025defeating} (Appendix~\ref{ap:limitations-tool-composition-through-context}). Third, for improved steerability and predictability, many agents rely on prompts that encode highly specific workflows—essentially algorithms written in natural language (e.g., Numina-Lean-Agent~\cite{liu2026numina}). Yet there is no guarantee that such workflows will be faithfully followed, and none of the standard software tooling exists for exploring or maintaining them.

Having traditional computations orchestrate LLM calls, which oracular programming enables with unprecedented flexibility and generality, offers tight control over semantics and context curation. As demonstrated in our invariant synthesis case study (Section~\ref{sec:loop-invariants-case-study}), it also has the potential to reduce inference cost by leveraging smaller models and search. On the other hand, the dual ReAct paradigm of having LLMs orchestrate traditional computations offers complementary strengths, notably its universality and its ability to tackle broad classes of problems with unanticipated edge cases. We do not call for one paradigm to supplant the other, but for the two to be combined, through \emph{composition} (as demonstrated in our Lean case study; Section~\ref{sec:lean-case-study}) and \emph{staging} (e.g., having ReAct agents generate, analyze, and execute oracular programs, to be explored in future work).

\subsection{Limitations}\label{sec:limitations}

\subsubsection{Physical and Conceptual Overhead} Separating strategies from policies is powerful, but also introduces overhead---both physical and conceptual---that may not always be justified. Physically, fully separating policies introduces overhead through the need to define inner policy types, and redundancy, since the structure of a policy must mirror the structure of the associated strategy (a constraint enforced through types). Conceptually, leveraging oracular programming in its fullest requires users to learn many new abstractions and idioms, which can be a barrier to adoption. Even for advanced users, it can add cognitive overhead, since there are often multiple ways to split a program's logic between strategies and policies with subtle trade-offs (Appendix~\ref{ap:unclear-strategy-policy-boundary}).

Such overhead is mitigated by building abstraction layers on top of oracular programming's foundations that offer a simpler interface for expressing common pipelines (e.g., universal queries, Section~\ref{sec:universal-queries-case-study}). In addition, although our framework allows full separation of strategies and policies, it does not \emph{mandate} it: policies can still be specified inline within strategies. Finally, coding agents could lower the barrier to entry by partially or fully automating the writing of oracular programs. Oracular programs are particularly well suited to this setting because they support strong consistency enforcement mechanisms that provide useful feedback to the agent.

\DiscussionNegativeSpace{}

\subsubsection{An Incomplete Enforcement of Consistency}

Oracular programming provides strong mechanisms for enforcing consistency between strategies, policies, and demonstrations, via types and tests. However, such mechanisms are incomplete and can be further improved.
For example, policies can be composed in ways that are well-typed yet still inconsistent (e.g., using depth-first search on an infinitely branching tree, sequencing two search algorithms when the first is nonterminating, requesting multiple LLM completions with a temperature of zero). Future work may explore ways to detect such inconsistencies through improved type systems, abstract interpretation, or more sophisticated runtime checks. In addition, although demonstration tests ensure that the provided decision examples suffice to solve actual problem instances, they cannot guarantee consistency with the more informal prompting instructions that some policies may provide, especially regarding unparsed chains of thought or reasoning tokens. Future work may investigate using small language models to perform fast sanity checks on the consistency of prompts and examples.

\DiscussionNegativeSpace{}

\subsubsection{Advanced Pipelines Require Advanced Tuning}

Oracular programming particularly shines when building advanced pipelines involving many well-scoped queries and sophisticated search. However, such pipelines require careful tuning: each search hyperparameter or prompt introduces a potential weak link that may cause the entire pipeline to fail if misconfigured. This highlights a key strength of the ReAct paradigm: it introduces few parameters and is relatively forgiving of imperfect prompts, since models operate with access to the full problem context and can disregard prompt inconsistencies. We believe that self-improvement is key for oracular programming to reach its full potential. Our second case study (Section~\ref{sec:lean-case-study}) demonstrated a limited form of self-improvement, where retrievable examples and prompt augmentations are extracted from runs. Beyond those, future work may consider updates to search parameters, policies, and even strategies themselves.

\section{Related Work}\label{sec:related-work}

\subsubsection*{LLM Programming Frameworks} Many existing frameworks aim at bridging prompting and programming and facilitating the development of LLM-enabled software (see survey~\cite{prompting-frameworks-survey}). Ours is unique in leveraging the full power of nondeterministic programming to integrate prompts with general computations, while allowing arbitrarily advanced search logic to be specified independently. It is also the first to identify and address the challenge of treating few-shot examples as \emph{grounded} and \emph{evolvable} program components, deserving of their own language.

Languages such as LMQL~\cite{beurer2023prompting} and SGLang~\cite{zheng2024sglang} enable guiding LLM inference by enforcing strict answer templates, within a single completion. They naturally complement our framework, which allows orchestrating multiple requests. The DSPy framework~\cite{khattab2024dspy} allows building LLM pipelines by composing declarative modules with natural-language typed signatures. Notably, it allows automatically optimizing prompts and examples based on an objective function, drawing a similarity to backpropagation-based learning frameworks such as PyTorch~\cite{paszke2019pytorch}, and inspiring our own self-improvement mechanism. DSPy's separation of LLM computations into \emph{signatures} and \emph{optimizable modules} parallels our own separation of \emph{queries} and \emph{prompting policies}. However, absent a corresponding distinction at the higher level at which such modules are composed (as achieved by our \emph{strategies} and \emph{search policies}), it only supports primitive forms of search, as do other LLM orchestration frameworks such as LangGraph~\cite{langgraph} and Mellea~\cite{mellea_docs} (Appendix~\ref{ap:llm-programming-extra}).

\subsubsection*{Effectful and Nondeterministic Programming}
Our proposed extensible effect system draws inspiration from the Freer monad~\cite{kiselyov2015freer} and interaction trees~\cite{xia2019interaction}. It combines these ideas with open unions for representing signatures and, more importantly, with additional structure imposed on effects in the form of member spaces, a locality constraint over actions, and local navigation.

The idea of defining search spaces via nondeterministic programs is long-standing~\cite{fischer2011purely}, but we push it further by demonstrating the potential of LLMs to serve as universal oracles (an idea pioneered by Selsam~\cite{selsam2020universal}) and by enabling an unprecedented separation between strategies and policies while preserving modularity and type safety via inner policy types. For example, although miniKanren~\cite{minikanren-byrd2009relational} allows some separation between the specification of modeling constraints and search heuristics, the two are still typically interleaved within the same program (e.g., via choices between \code{conde}, \code{conda}, or \code{condi} to express nondeterminism).

\subsubsection*{Probabilistic Programming}%

Our use of nondeterministic programming to separate core and search logic draws parallels with \emph{probabilistic programming}~\cite{probabilistic-programming-book-van2018introduction}. For example, languages such as Pyro~\cite{bingham2019pyro} or Gen~\cite{gen-julia-cusumano2019gen} allow the separate specification of probabilistic programs and inference policies. These policies involve not only global decisions (e.g., whether to use MCMC or variational methods), but also local ones (e.g., which guide distribution to use at each sampling location), making our concept of \emph{inner policy type} relevant for building new, statically type-safe probabilistic programming languages.
Gen~\cite{gen-julia-cusumano2019gen} bears further similarities with Delphyne, by allowing the modular composition of generative functions (i.e., probabilistic programs) through hierarchical address spaces, and offering the full expressiveness of its host language (i.e., Julia) to express them. Still, Gen policies operate on completely different abstractions (choice maps instead of trees), have no resource-awareness, and are not as composable (no direct counterpart to our stream and tree transformers).

\section{Conclusion}\label{sec:conclusion}

LLMs and traditional programs are natural complements: the former provide inductive capabilities that symbolic computation lacks, while the latter provide reliability, modularity, and precise semantics. Yet bringing these complementary strengths together presents unique software-engineering challenges that have constrained the complexity of LLM-enabled programs. We identified the separation of \emph{core logic} and \emph{search} as a key principle for addressing these challenges, along with the elevation of examples into \emph{grounded}, \emph{evolvable} program components. We hope that oracular programming can foster the discovery of new idioms for building intersymbolic AI systems and for weaving explicit and inductive computation in creative ways.

\section*{Data-Availability Statement}

An artifact is provided containing a type-checked embedding of our strategy language in Haskell (Appendix~\ref{ap:orakell}), along with the latest release of Delphyne (Section~\ref{sec:delphyne}) and extensive documentation (quick start guide, tutorials, manual, API reference, and how-tos). Full code for all case studies is included. Quantitative experiments can be replicated without incurring inference cost, since all LLM responses and external tool outputs (from Why3 and Lean) are stored in a human-readable cache that allows oracular program runs to be replayed deterministically. %

\bibliographystyle{ACM-Reference-Format}
\bibliography{main}


\begin{thebibliography}{49}


\ifx \showCODEN    \undefined \def \showCODEN     #1{\unskip}     \fi
\ifx \showISBNx    \undefined \def \showISBNx     #1{\unskip}     \fi
\ifx \showISBNxiii \undefined \def \showISBNxiii  #1{\unskip}     \fi
\ifx \showISSN     \undefined \def \showISSN      #1{\unskip}     \fi
\ifx \showLCCN     \undefined \def \showLCCN      #1{\unskip}     \fi
\ifx \shownote     \undefined \def \shownote      #1{#1}          \fi
\ifx \showarticletitle \undefined \def \showarticletitle #1{#1}   \fi
\ifx \showURL      \undefined \def \showURL       {\relax}        \fi
\providecommand\bibfield[2]{#2}
\providecommand\bibinfo[2]{#2}
\providecommand\natexlab[1]{#1}
\providecommand\showeprint[2][]{arXiv:#2}

\bibitem[AI(2023)]%
        {langgraph}
\bibfield{author}{\bibinfo{person}{LangChain AI}.} \bibinfo{year}{2023}\natexlab{}.
\newblock \bibinfo{title}{LangGraph}.
\newblock \bibinfo{howpublished}{\url{https://github.com/langchain-ai/langgraph}}.
\newblock


\bibitem[Anthony et~al\mbox{.}(2017)]%
        {anthony2017thinking}
\bibfield{author}{\bibinfo{person}{Thomas Anthony}, \bibinfo{person}{Zheng Tian}, {and} \bibinfo{person}{David Barber}.} \bibinfo{year}{2017}\natexlab{}.
\newblock \showarticletitle{Thinking fast and slow with deep learning and tree search}.
\newblock \bibinfo{journal}{\emph{Advances in neural information processing systems}}  \bibinfo{volume}{30} (\bibinfo{year}{2017}).
\newblock


\bibitem[Beckert et~al\mbox{.}(2024)]%
        {key-isola-beckert2024towards}
\bibfield{author}{\bibinfo{person}{Bernhard Beckert}, \bibinfo{person}{Jonas Klamroth}, \bibinfo{person}{Wolfram Pfeifer}, \bibinfo{person}{Patrick R{\"o}per}, {and} \bibinfo{person}{Samuel Teuber}.} \bibinfo{year}{2024}\natexlab{}.
\newblock \showarticletitle{Towards combining the cognitive abilities of large language models with the rigor of deductive program verification}. In \bibinfo{booktitle}{\emph{International Symposium on Leveraging Applications of Formal Methods}}. Springer, \bibinfo{pages}{242--257}.
\newblock


\bibitem[Beurer-Kellner et~al\mbox{.}(2023)]%
        {beurer2023prompting}
\bibfield{author}{\bibinfo{person}{Luca Beurer-Kellner}, \bibinfo{person}{Marc Fischer}, {and} \bibinfo{person}{Martin Vechev}.} \bibinfo{year}{2023}\natexlab{}.
\newblock \showarticletitle{Prompting is programming: A query language for large language models}.
\newblock \bibinfo{journal}{\emph{Proceedings of the ACM on Programming Languages}} \bibinfo{volume}{7}, \bibinfo{number}{PLDI} (\bibinfo{year}{2023}), \bibinfo{pages}{1946--1969}.
\newblock


\bibitem[Bingham et~al\mbox{.}(2019)]%
        {bingham2019pyro}
\bibfield{author}{\bibinfo{person}{Eli Bingham}, \bibinfo{person}{Jonathan~P Chen}, \bibinfo{person}{Martin Jankowiak}, \bibinfo{person}{Fritz Obermeyer}, \bibinfo{person}{Neeraj Pradhan}, \bibinfo{person}{Theofanis Karaletsos}, \bibinfo{person}{Rohit Singh}, \bibinfo{person}{Paul Szerlip}, \bibinfo{person}{Paul Horsfall}, {and} \bibinfo{person}{Noah~D Goodman}.} \bibinfo{year}{2019}\natexlab{}.
\newblock \showarticletitle{Pyro: Deep universal probabilistic programming}.
\newblock \bibinfo{journal}{\emph{Journal of machine learning research}} \bibinfo{volume}{20}, \bibinfo{number}{28} (\bibinfo{year}{2019}), \bibinfo{pages}{1--6}.
\newblock


\bibitem[Brown et~al\mbox{.}(2020)]%
        {brown2020language}
\bibfield{author}{\bibinfo{person}{Tom Brown}, \bibinfo{person}{Benjamin Mann}, \bibinfo{person}{Nick Ryder}, \bibinfo{person}{Melanie Subbiah}, \bibinfo{person}{Jared~D Kaplan}, \bibinfo{person}{Prafulla Dhariwal}, \bibinfo{person}{Arvind Neelakantan}, \bibinfo{person}{Pranav Shyam}, \bibinfo{person}{Girish Sastry}, \bibinfo{person}{Amanda Askell}, {et~al\mbox{.}}} \bibinfo{year}{2020}\natexlab{}.
\newblock \showarticletitle{Language models are few-shot learners}.
\newblock \bibinfo{journal}{\emph{Advances in neural information processing systems}}  \bibinfo{volume}{33} (\bibinfo{year}{2020}), \bibinfo{pages}{1877--1901}.
\newblock


\bibitem[Browne et~al\mbox{.}(2012)]%
        {browne2012survey}
\bibfield{author}{\bibinfo{person}{Cameron~B Browne}, \bibinfo{person}{Edward Powley}, \bibinfo{person}{Daniel Whitehouse}, \bibinfo{person}{Simon~M Lucas}, \bibinfo{person}{Peter~I Cowling}, \bibinfo{person}{Philipp Rohlfshagen}, \bibinfo{person}{Stephen Tavener}, \bibinfo{person}{Diego Perez}, \bibinfo{person}{Spyridon Samothrakis}, {and} \bibinfo{person}{Simon Colton}.} \bibinfo{year}{2012}\natexlab{}.
\newblock \showarticletitle{A survey of {Monte Carlo} tree search methods}.
\newblock \bibinfo{journal}{\emph{IEEE Transactions on Computational Intelligence and AI in games}} \bibinfo{volume}{4}, \bibinfo{number}{1} (\bibinfo{year}{2012}), \bibinfo{pages}{1--43}.
\newblock


\bibitem[Byrd(2009)]%
        {minikanren-byrd2009relational}
\bibfield{author}{\bibinfo{person}{William~E Byrd}.} \bibinfo{year}{2009}\natexlab{}.
\newblock \emph{\bibinfo{title}{Relational programming in miniKanren: techniques, applications, and implementations}}.
\newblock \bibinfo{thesistype}{Ph.\,D. Dissertation}. \bibinfo{school}{Indiana University}.
\newblock


\bibitem[Carbonell and Goldstein(1998)]%
        {carbonell1998use}
\bibfield{author}{\bibinfo{person}{Jaime Carbonell} {and} \bibinfo{person}{Jade Goldstein}.} \bibinfo{year}{1998}\natexlab{}.
\newblock \showarticletitle{The use of MMR, diversity-based reranking for reordering documents and producing summaries}. In \bibinfo{booktitle}{\emph{Proceedings of the 21st annual international ACM SIGIR conference on Research and development in information retrieval}}. \bibinfo{pages}{335--336}.
\newblock


\bibitem[Comanici et~al\mbox{.}(2025)]%
        {comanici2025gemini}
\bibfield{author}{\bibinfo{person}{Comanici} {et~al\mbox{.}}} \bibinfo{year}{2025}\natexlab{}.
\newblock \showarticletitle{Gemini 2.5: Pushing the frontier with advanced reasoning, multimodality, long context, and next generation agentic capabilities}.
\newblock \bibinfo{journal}{\emph{arXiv:2507.06261}} (\bibinfo{year}{2025}).
\newblock


\bibitem[Cusumano-Towner et~al\mbox{.}(2019)]%
        {gen-julia-cusumano2019gen}
\bibfield{author}{\bibinfo{person}{Marco~F Cusumano-Towner}, \bibinfo{person}{Feras~A Saad}, \bibinfo{person}{Alexander~K Lew}, {and} \bibinfo{person}{Vikash~K Mansinghka}.} \bibinfo{year}{2019}\natexlab{}.
\newblock \showarticletitle{Gen: a general-purpose probabilistic programming system with programmable inference}. In \bibinfo{booktitle}{\emph{Proceedings of the 40th acm sigplan conference on programming language design and implementation}}. \bibinfo{pages}{221--236}.
\newblock


\bibitem[Debenedetti et~al\mbox{.}(2025)]%
        {debenedetti2025defeating}
\bibfield{author}{\bibinfo{person}{Edoardo Debenedetti}, \bibinfo{person}{Ilia Shumailov}, \bibinfo{person}{Tianqi Fan}, \bibinfo{person}{Jamie Hayes}, \bibinfo{person}{Nicholas Carlini}, \bibinfo{person}{Daniel Fabian}, \bibinfo{person}{Christoph Kern}, \bibinfo{person}{Chongyang Shi}, \bibinfo{person}{Andreas Terzis}, {and} \bibinfo{person}{Florian Tram{\`e}r}.} \bibinfo{year}{2025}\natexlab{}.
\newblock \showarticletitle{Defeating prompt injections by design}.
\newblock \bibinfo{journal}{\emph{arXiv preprint arXiv:2503.18813}} (\bibinfo{year}{2025}).
\newblock


\bibitem[Dechter and Pearl(1985)]%
        {dechter1985generalized}
\bibfield{author}{\bibinfo{person}{Rina Dechter} {and} \bibinfo{person}{Judea Pearl}.} \bibinfo{year}{1985}\natexlab{}.
\newblock \showarticletitle{Generalized best-first search strategies and the optimality of A}.
\newblock \bibinfo{journal}{\emph{Journal of the ACM (JACM)}} \bibinfo{volume}{32}, \bibinfo{number}{3} (\bibinfo{year}{1985}), \bibinfo{pages}{505--536}.
\newblock


\bibitem[Filli{\^a}tre and Paskevich(2013)]%
        {filliatre2013why3}
\bibfield{author}{\bibinfo{person}{Jean-Christophe Filli{\^a}tre} {and} \bibinfo{person}{Andrei Paskevich}.} \bibinfo{year}{2013}\natexlab{}.
\newblock \showarticletitle{Why3—where programs meet provers}. In \bibinfo{booktitle}{\emph{European symposium on programming}}. Springer, \bibinfo{pages}{125--128}.
\newblock


\bibitem[First et~al\mbox{.}(2023)]%
        {first2023baldur}
\bibfield{author}{\bibinfo{person}{Emily First}, \bibinfo{person}{Markus~N Rabe}, \bibinfo{person}{Talia Ringer}, {and} \bibinfo{person}{Yuriy Brun}.} \bibinfo{year}{2023}\natexlab{}.
\newblock \showarticletitle{Baldur: Whole-proof generation and repair with large language models}. In \bibinfo{booktitle}{\emph{Proceedings of the 31st ACM Joint European Software Engineering Conference and Symposium on the Foundations of Software Engineering}}. \bibinfo{pages}{1229--1241}.
\newblock


\bibitem[Fischer et~al\mbox{.}(2011)]%
        {fischer2011purely}
\bibfield{author}{\bibinfo{person}{Sebastian Fischer}, \bibinfo{person}{Oleg Kiselyov}, {and} \bibinfo{person}{Chung-chieh Shan}.} \bibinfo{year}{2011}\natexlab{}.
\newblock \showarticletitle{Purely functional lazy nondeterministic programming}.
\newblock \bibinfo{journal}{\emph{Journal of Functional programming}} \bibinfo{volume}{21}, \bibinfo{number}{4-5} (\bibinfo{year}{2011}), \bibinfo{pages}{413--465}.
\newblock


\bibitem[Gao et~al\mbox{.}(2023)]%
        {gao2023retrieval}
\bibfield{author}{\bibinfo{person}{Yunfan Gao}, \bibinfo{person}{Yun Xiong}, \bibinfo{person}{Xinyu Gao}, \bibinfo{person}{Kangxiang Jia}, \bibinfo{person}{Jinliu Pan}, \bibinfo{person}{Yuxi Bi}, \bibinfo{person}{Yi Dai}, \bibinfo{person}{Jiawei Sun}, {and} \bibinfo{person}{Haofen Wang}.} \bibinfo{year}{2023}\natexlab{}.
\newblock \showarticletitle{Retrieval-augmented generation for large language models: A survey}.
\newblock \bibinfo{journal}{\emph{arXiv preprint arXiv:2312.10997}} (\bibinfo{year}{2023}).
\newblock


\bibitem[Geng et~al\mbox{.}(2025)]%
        {geng2025generating}
\bibfield{author}{\bibinfo{person}{Saibo Geng}, \bibinfo{person}{Hudson Cooper}, \bibinfo{person}{Micha{\l} Moskal}, \bibinfo{person}{Samuel Jenkins}, \bibinfo{person}{Julian Berman}, \bibinfo{person}{Nathan Ranchin}, \bibinfo{person}{Robert West}, \bibinfo{person}{Eric Horvitz}, {and} \bibinfo{person}{Harsha Nori}.} \bibinfo{year}{2025}\natexlab{}.
\newblock \showarticletitle{Generating structured outputs from language models: Benchmark and studies}.
\newblock \bibinfo{journal}{\emph{arXiv e-prints}} (\bibinfo{year}{2025}), \bibinfo{pages}{arXiv--2501}.
\newblock


\bibitem[{IBM Research}(2024)]%
        {mellea_docs}
\bibfield{author}{\bibinfo{person}{{IBM Research}}.} \bibinfo{year}{2024}\natexlab{}.
\newblock \bibinfo{title}{Mellea Documentation}.
\newblock \bibinfo{howpublished}{\url{https://docs.mellea.ai}}.
\newblock
\newblock
\shownote{Programming framework for reliable generative programs}.


\bibitem[Jiang et~al\mbox{.}(2022)]%
        {jiang2022draft}
\bibfield{author}{\bibinfo{person}{Albert~Q Jiang}, \bibinfo{person}{Sean Welleck}, \bibinfo{person}{Jin~Peng Zhou}, \bibinfo{person}{Wenda Li}, \bibinfo{person}{Jiacheng Liu}, \bibinfo{person}{Mateja Jamnik}, \bibinfo{person}{Timoth{\'e}e Lacroix}, \bibinfo{person}{Yuhuai Wu}, {and} \bibinfo{person}{Guillaume Lample}.} \bibinfo{year}{2022}\natexlab{}.
\newblock \showarticletitle{Draft, sketch, and prove: Guiding formal theorem provers with informal proofs}.
\newblock \bibinfo{journal}{\emph{arXiv preprint arXiv:2210.12283}} (\bibinfo{year}{2022}).
\newblock


\bibitem[Kamath et~al\mbox{.}(2023)]%
        {loopy-kamath2023finding}
\bibfield{author}{\bibinfo{person}{Adharsh Kamath}, \bibinfo{person}{Aditya Senthilnathan}, \bibinfo{person}{Saikat Chakraborty}, \bibinfo{person}{Pantazis Deligiannis}, \bibinfo{person}{Shuvendu~K Lahiri}, \bibinfo{person}{Akash Lal}, \bibinfo{person}{Aseem Rastogi}, \bibinfo{person}{Subhajit Roy}, {and} \bibinfo{person}{Rahul Sharma}.} \bibinfo{year}{2023}\natexlab{}.
\newblock \showarticletitle{Finding inductive loop invariants using large language models}.
\newblock \bibinfo{journal}{\emph{arXiv preprint arXiv:2311.07948}} (\bibinfo{year}{2023}).
\newblock


\bibitem[Khattab et~al\mbox{.}(2024)]%
        {khattab2024dspy}
\bibfield{author}{\bibinfo{person}{Omar Khattab}, \bibinfo{person}{Arnav Singhvi}, \bibinfo{person}{Paridhi Maheshwari}, \bibinfo{person}{Zhiyuan Zhang}, \bibinfo{person}{Keshav Santhanam}, \bibinfo{person}{Sri Vardhamanan}, \bibinfo{person}{Saiful Haq}, \bibinfo{person}{Ashutosh Sharma}, \bibinfo{person}{Thomas~T. Joshi}, \bibinfo{person}{Hanna Moazam}, \bibinfo{person}{Heather Miller}, \bibinfo{person}{Matei Zaharia}, {and} \bibinfo{person}{Christopher Potts}.} \bibinfo{year}{2024}\natexlab{}.
\newblock \showarticletitle{DSPy: Compiling Declarative Language Model Calls into Self-Improving Pipelines}.
\newblock \bibinfo{journal}{\emph{The Twelfth International Conference on Learning Representations}}.
\newblock


\bibitem[Kiselyov and Ishii(2015)]%
        {kiselyov2015freer}
\bibfield{author}{\bibinfo{person}{Oleg Kiselyov} {and} \bibinfo{person}{Hiromi Ishii}.} \bibinfo{year}{2015}\natexlab{}.
\newblock \showarticletitle{Freer monads, more extensible effects}.
\newblock \bibinfo{journal}{\emph{ACM SIGPLAN Notices}} \bibinfo{volume}{50}, \bibinfo{number}{12} (\bibinfo{year}{2015}), \bibinfo{pages}{94--105}.
\newblock


\bibitem[Koppel et~al\mbox{.}(2018)]%
        {koppel2018capturing}
\bibfield{author}{\bibinfo{person}{James Koppel}, \bibinfo{person}{Gabriel Scherer}, {and} \bibinfo{person}{Armando Solar-Lezama}.} \bibinfo{year}{2018}\natexlab{}.
\newblock \showarticletitle{Capturing the future by replaying the past (functional pearl)}.
\newblock \bibinfo{journal}{\emph{Proceedings of the ACM on Programming Languages}} \bibinfo{volume}{2}, \bibinfo{number}{ICFP} (\bibinfo{year}{2018}), \bibinfo{pages}{1--29}.
\newblock


\bibitem[Lample et~al\mbox{.}(2022)]%
        {lample2022hypertree}
\bibfield{author}{\bibinfo{person}{Guillaume Lample}, \bibinfo{person}{Timothee Lacroix}, \bibinfo{person}{Marie-Anne Lachaux}, \bibinfo{person}{Aurelien Rodriguez}, \bibinfo{person}{Amaury Hayat}, \bibinfo{person}{Thibaut Lavril}, \bibinfo{person}{Gabriel Ebner}, {and} \bibinfo{person}{Xavier Martinet}.} \bibinfo{year}{2022}\natexlab{}.
\newblock \showarticletitle{Hypertree proof search for neural theorem proving}.
\newblock \bibinfo{journal}{\emph{Advances in neural information processing systems}}  \bibinfo{volume}{35} (\bibinfo{year}{2022}), \bibinfo{pages}{26337--26349}.
\newblock


\bibitem[Lewkowycz et~al\mbox{.}(2022)]%
        {lewkowycz2022solving}
\bibfield{author}{\bibinfo{person}{Aitor Lewkowycz}, \bibinfo{person}{Anders Andreassen}, \bibinfo{person}{David Dohan}, \bibinfo{person}{Ethan Dyer}, \bibinfo{person}{Henryk Michalewski}, \bibinfo{person}{Vinay Ramasesh}, \bibinfo{person}{Ambrose Slone}, \bibinfo{person}{Cem Anil}, \bibinfo{person}{Imanol Schlag}, \bibinfo{person}{Theo Gutman-Solo}, {et~al\mbox{.}}} \bibinfo{year}{2022}\natexlab{}.
\newblock \showarticletitle{Solving quantitative reasoning problems with language models}.
\newblock \bibinfo{journal}{\emph{Advances in Neural Information Processing Systems}}  \bibinfo{volume}{35} (\bibinfo{year}{2022}), \bibinfo{pages}{3843--3857}.
\newblock


\bibitem[Li et~al\mbox{.}(2022)]%
        {li2022competition}
\bibfield{author}{\bibinfo{person}{Yujia Li}, \bibinfo{person}{David Choi}, \bibinfo{person}{Junyoung Chung}, \bibinfo{person}{Nate Kushman}, \bibinfo{person}{Julian Schrittwieser}, \bibinfo{person}{R{\'e}mi Leblond}, \bibinfo{person}{Tom Eccles}, \bibinfo{person}{James Keeling}, \bibinfo{person}{Felix Gimeno}, \bibinfo{person}{Agustin Dal~Lago}, {et~al\mbox{.}}} \bibinfo{year}{2022}\natexlab{}.
\newblock \showarticletitle{Competition-level code generation with {Alphacode}}.
\newblock \bibinfo{journal}{\emph{Science}} \bibinfo{volume}{378}, \bibinfo{number}{6624} (\bibinfo{year}{2022}), \bibinfo{pages}{1092--1097}.
\newblock


\bibitem[Liu et~al\mbox{.}(2026a)]%
        {liu2026numina}
\bibfield{author}{\bibinfo{person}{Junqi Liu} {et~al\mbox{.}}} \bibinfo{year}{2026}\natexlab{a}.
\newblock \showarticletitle{Numina-Lean-Agent: An Open and General Agentic Reasoning System for Formal Mathematics}.
\newblock \bibinfo{journal}{\emph{arXiv preprint arXiv:2601.14027}} (\bibinfo{year}{2026}).
\newblock


\bibitem[Liu et~al\mbox{.}(2026b)]%
        {prompting-frameworks-survey}
\bibfield{author}{\bibinfo{person}{Xiaoxia Liu}, \bibinfo{person}{Jingyi Wang}, \bibinfo{person}{Xiaohan Yuan}, \bibinfo{person}{Jun Sun}, \bibinfo{person}{Guoliang Dong}, \bibinfo{person}{Peng Di}, \bibinfo{person}{Wenhai Wang}, {and} \bibinfo{person}{Dongxia Wang}.} \bibinfo{year}{2026}\natexlab{b}.
\newblock \showarticletitle{Prompting Frameworks for Large Language Models: A Survey}.
\newblock \bibinfo{journal}{\emph{ACM Comput. Surv.}} (\bibinfo{year}{2026}).
\newblock
\showISSN{0360-0300}
\href{https://doi.org/10.1145/3789253}{doi:\nolinkurl{10.1145/3789253}}


\bibitem[Paszke et~al\mbox{.}(2019)]%
        {paszke2019pytorch}
\bibfield{author}{\bibinfo{person}{Adam Paszke}, \bibinfo{person}{Sam Gross}, \bibinfo{person}{Francisco Massa}, \bibinfo{person}{Adam Lerer}, \bibinfo{person}{James Bradbury}, \bibinfo{person}{Gregory Chanan}, \bibinfo{person}{Trevor Killeen}, \bibinfo{person}{Zeming Lin}, \bibinfo{person}{Natalia Gimelshein}, \bibinfo{person}{Luca Antiga}, {et~al\mbox{.}}} \bibinfo{year}{2019}\natexlab{}.
\newblock \showarticletitle{{PyTorch}: An imperative style, high-performance deep learning library}.
\newblock \bibinfo{journal}{\emph{Advances in neural information processing systems}}  \bibinfo{volume}{32} (\bibinfo{year}{2019}).
\newblock


\bibitem[Pirzada et~al\mbox{.}(2024)]%
        {esbmc-pirzada2024llm}
\bibfield{author}{\bibinfo{person}{Muhammad~AA Pirzada}, \bibinfo{person}{Giles Reger}, \bibinfo{person}{Ahmed Bhayat}, {and} \bibinfo{person}{Lucas~C Cordeiro}.} \bibinfo{year}{2024}\natexlab{}.
\newblock \showarticletitle{Llm-generated invariants for bounded model checking without loop unrolling}. In \bibinfo{booktitle}{\emph{Proceedings of the 39th IEEE/ACM International Conference on Automated Software Engineering}}. \bibinfo{pages}{1395--1407}.
\newblock


\bibitem[Renshaw et~al\mbox{.}(2011)]%
        {renshaw2011distributed}
\bibfield{author}{\bibinfo{person}{David~W Renshaw}, \bibinfo{person}{Sarah~M Loos}, {and} \bibinfo{person}{Andr{\'e} Platzer}.} \bibinfo{year}{2011}\natexlab{}.
\newblock \showarticletitle{Distributed theorem proving for distributed hybrid systems}. In \bibinfo{booktitle}{\emph{International Conference on Formal Engineering Methods}}. Springer, \bibinfo{pages}{356--371}.
\newblock


\bibitem[Sahoo et~al\mbox{.}(2024)]%
        {sahoo2024systematic}
\bibfield{author}{\bibinfo{person}{Pranab Sahoo}, \bibinfo{person}{Ayush~Kumar Singh}, \bibinfo{person}{Sriparna Saha}, \bibinfo{person}{Vinija Jain}, \bibinfo{person}{Samrat Mondal}, {and} \bibinfo{person}{Aman Chadha}.} \bibinfo{year}{2024}\natexlab{}.
\newblock \showarticletitle{A systematic survey of prompt engineering in large language models: Techniques and applications}.
\newblock \bibinfo{journal}{\emph{arXiv preprint arXiv:2402.07927}} (\bibinfo{year}{2024}).
\newblock


\bibitem[Selsam et~al\mbox{.}(2020)]%
        {selsam2020universal}
\bibfield{author}{\bibinfo{person}{Daniel Selsam}, \bibinfo{person}{Jesse~Michael Han}, \bibinfo{person}{Leonardo de Moura}, {and} \bibinfo{person}{Patrice Godefroid}.} \bibinfo{year}{2020}\natexlab{}.
\newblock \showarticletitle{Universal policies for software-defined MDPs}.
\newblock \bibinfo{journal}{\emph{arXiv preprint arXiv:2012.11401}} (\bibinfo{year}{2020}).
\newblock


\bibitem[Si et~al\mbox{.}(2018)]%
        {si2018learning}
\bibfield{author}{\bibinfo{person}{Xujie Si}, \bibinfo{person}{Hanjun Dai}, \bibinfo{person}{Mukund Raghothaman}, \bibinfo{person}{Mayur Naik}, {and} \bibinfo{person}{Le Song}.} \bibinfo{year}{2018}\natexlab{}.
\newblock \showarticletitle{Learning loop invariants for program verification}.
\newblock \bibinfo{journal}{\emph{Advances in Neural Information Processing Systems}}  \bibinfo{volume}{31} (\bibinfo{year}{2018}).
\newblock


\bibitem[Swierstra(2008)]%
        {swierstra2008data}
\bibfield{author}{\bibinfo{person}{Wouter Swierstra}.} \bibinfo{year}{2008}\natexlab{}.
\newblock \showarticletitle{Data types {\`a} la carte}.
\newblock \bibinfo{journal}{\emph{Journal of functional programming}} \bibinfo{volume}{18}, \bibinfo{number}{4} (\bibinfo{year}{2008}), \bibinfo{pages}{423--436}.
\newblock


\bibitem[Van~de Meent et~al\mbox{.}(2018)]%
        {probabilistic-programming-book-van2018introduction}
\bibfield{author}{\bibinfo{person}{Jan-Willem Van~de Meent}, \bibinfo{person}{Brooks Paige}, \bibinfo{person}{Hongseok Yang}, {and} \bibinfo{person}{Frank Wood}.} \bibinfo{year}{2018}\natexlab{}.
\newblock \showarticletitle{An introduction to probabilistic programming}.
\newblock \bibinfo{journal}{\emph{arXiv preprint arXiv:1809.10756}} (\bibinfo{year}{2018}).
\newblock


\bibitem[Wadler(1989)]%
        {wadler1989theorems}
\bibfield{author}{\bibinfo{person}{Philip Wadler}.} \bibinfo{year}{1989}\natexlab{}.
\newblock \showarticletitle{Theorems for free!}. In \bibinfo{booktitle}{\emph{Proceedings of the fourth international conference on Functional programming languages and computer architecture}}. \bibinfo{pages}{347--359}.
\newblock


\bibitem[Wei et~al\mbox{.}(2025)]%
        {quokka-wei2025invbench}
\bibfield{author}{\bibinfo{person}{Anjiang Wei}, \bibinfo{person}{Tarun Suresh}, \bibinfo{person}{Tianran Sun}, \bibinfo{person}{Haoze Wu}, \bibinfo{person}{Ke Wang}, {and} \bibinfo{person}{Alex Aiken}.} \bibinfo{year}{2025}\natexlab{}.
\newblock \showarticletitle{InvBench: Can LLMs Accelerate Program Verification with Invariant Synthesis?}
\newblock \bibinfo{journal}{\emph{arXiv preprint arXiv:2509.21629}} (\bibinfo{year}{2025}).
\newblock


\bibitem[Wen et~al\mbox{.}(2024)]%
        {autospec-wen2024enchanting}
\bibfield{author}{\bibinfo{person}{Cheng Wen}, \bibinfo{person}{Jialun Cao}, \bibinfo{person}{Jie Su}, \bibinfo{person}{Zhiwu Xu}, \bibinfo{person}{Shengchao Qin}, \bibinfo{person}{Mengda He}, \bibinfo{person}{Haokun Li}, \bibinfo{person}{Shing-Chi Cheung}, {and} \bibinfo{person}{Cong Tian}.} \bibinfo{year}{2024}\natexlab{}.
\newblock \showarticletitle{Enchanting program specification synthesis by large language models using static analysis and program verification}. In \bibinfo{booktitle}{\emph{International Conference on Computer Aided Verification}}. Springer, \bibinfo{pages}{302--328}.
\newblock


\bibitem[Whalen(2016)]%
        {whalen2016holophrasm}
\bibfield{author}{\bibinfo{person}{Daniel Whalen}.} \bibinfo{year}{2016}\natexlab{}.
\newblock \showarticletitle{Holophrasm: a neural automated theorem prover for higher-order logic}.
\newblock \bibinfo{journal}{\emph{arXiv preprint arXiv:1608.02644}} (\bibinfo{year}{2016}).
\newblock


\bibitem[Wu et~al\mbox{.}(2024)]%
        {llama4inv-wu2024llm}
\bibfield{author}{\bibinfo{person}{Guangyuan Wu}, \bibinfo{person}{Weining Cao}, \bibinfo{person}{Yuan Yao}, \bibinfo{person}{Hengfeng Wei}, \bibinfo{person}{Taolue Chen}, {and} \bibinfo{person}{Xiaoxing Ma}.} \bibinfo{year}{2024}\natexlab{}.
\newblock \showarticletitle{{LLM} meets bounded model checking: Neuro-symbolic loop invariant inference}. In \bibinfo{booktitle}{\emph{Proceedings of the 39th IEEE/ACM International Conference on Automated Software Engineering}}. \bibinfo{pages}{406--417}.
\newblock


\bibitem[Wu et~al\mbox{.}(2023)]%
        {lemur-wu2023lemur}
\bibfield{author}{\bibinfo{person}{Haoze Wu}, \bibinfo{person}{Clark Barrett}, {and} \bibinfo{person}{Nina Narodytska}.} \bibinfo{year}{2023}\natexlab{}.
\newblock \showarticletitle{Lemur: Integrating large language models in automated program verification}.
\newblock \bibinfo{journal}{\emph{arXiv preprint arXiv:2310.04870}} (\bibinfo{year}{2023}).
\newblock


\bibitem[Xia et~al\mbox{.}(2019)]%
        {xia2019interaction}
\bibfield{author}{\bibinfo{person}{Li-yao Xia}, \bibinfo{person}{Yannick Zakowski}, \bibinfo{person}{Paul He}, \bibinfo{person}{Chung-Kil Hur}, \bibinfo{person}{Gregory Malecha}, \bibinfo{person}{Benjamin~C Pierce}, {and} \bibinfo{person}{Steve Zdancewic}.} \bibinfo{year}{2019}\natexlab{}.
\newblock \showarticletitle{Interaction trees: representing recursive and impure programs in Coq}.
\newblock \bibinfo{journal}{\emph{Proceedings of the ACM on Programming Languages}} \bibinfo{volume}{4}, \bibinfo{number}{POPL} (\bibinfo{year}{2019}), \bibinfo{pages}{1--32}.
\newblock


\bibitem[Yao et~al\mbox{.}(2024)]%
        {yao2024tree}
\bibfield{author}{\bibinfo{person}{Shunyu Yao}, \bibinfo{person}{Dian Yu}, \bibinfo{person}{Jeffrey Zhao}, \bibinfo{person}{Izhak Shafran}, \bibinfo{person}{Tom Griffiths}, \bibinfo{person}{Yuan Cao}, {and} \bibinfo{person}{Karthik Narasimhan}.} \bibinfo{year}{2024}\natexlab{}.
\newblock \showarticletitle{Tree of thoughts: Deliberate problem solving with large language models}.
\newblock \bibinfo{journal}{\emph{Advances in Neural Information Processing Systems}}  \bibinfo{volume}{36} (\bibinfo{year}{2024}).
\newblock


\bibitem[Yao et~al\mbox{.}(2022)]%
        {yao2022react}
\bibfield{author}{\bibinfo{person}{Shunyu Yao}, \bibinfo{person}{Jeffrey Zhao}, \bibinfo{person}{Dian Yu}, \bibinfo{person}{Nan Du}, \bibinfo{person}{Izhak Shafran}, \bibinfo{person}{Karthik Narasimhan}, {and} \bibinfo{person}{Yuan Cao}.} \bibinfo{year}{2022}\natexlab{}.
\newblock \showarticletitle{ReAct: Synergizing Reasoning and Acting in Language Models}.
\newblock \bibinfo{journal}{\emph{arXiv preprint arXiv:2210.03629}} (\bibinfo{year}{2022}).
\newblock


\bibitem[Zhang et~al\mbox{.}(2025)]%
        {zhang2025recursive}
\bibfield{author}{\bibinfo{person}{Alex~L Zhang}, \bibinfo{person}{Tim Kraska}, {and} \bibinfo{person}{Omar Khattab}.} \bibinfo{year}{2025}\natexlab{}.
\newblock \showarticletitle{Recursive language models}.
\newblock \bibinfo{journal}{\emph{arXiv preprint arXiv:2512.24601}} (\bibinfo{year}{2025}).
\newblock


\bibitem[Zheng et~al\mbox{.}(2021)]%
        {zheng2021minif2f}
\bibfield{author}{\bibinfo{person}{Kunhao Zheng}, \bibinfo{person}{Jesse~Michael Han}, {and} \bibinfo{person}{Stanislas Polu}.} \bibinfo{year}{2021}\natexlab{}.
\newblock \showarticletitle{Minif2f: a cross-system benchmark for formal olympiad-level mathematics}.
\newblock \bibinfo{journal}{\emph{arXiv preprint arXiv:2109.00110}} (\bibinfo{year}{2021}).
\newblock


\bibitem[Zheng et~al\mbox{.}(2024)]%
        {zheng2024sglang}
\bibfield{author}{\bibinfo{person}{Lianmin Zheng}, \bibinfo{person}{Liangsheng Yin}, \bibinfo{person}{Zhiqiang Xie}, \bibinfo{person}{Chuyue Sun}, \bibinfo{person}{Jeff Huang}, \bibinfo{person}{Cody~H Yu}, \bibinfo{person}{Shiyi Cao}, \bibinfo{person}{Christos Kozyrakis}, \bibinfo{person}{Ion Stoica}, \bibinfo{person}{Joseph~E Gonzalez}, {et~al\mbox{.}}} \bibinfo{year}{2024}\natexlab{}.
\newblock \showarticletitle{Sglang: Efficient execution of structured language model programs}.
\newblock \bibinfo{journal}{\emph{Advances in neural information processing systems}}  \bibinfo{volume}{37} (\bibinfo{year}{2024}), \bibinfo{pages}{62557--62583}.
\newblock


\end{thebibliography}

\raggedbottom
\newpage

\appendix

\section*{Content of Appendices}

Appendices~\ref{ap:strategy-language}, \ref{ap:policy-language}, and~\ref{ap:demonstration-language} provide details on the three languages of oracular programming. Appendix~\ref{ap:delphyne} describes Delphyne. Appendix~\ref{ap:extra-discussion-items} elaborates on specific items from our Discussion and Related Work sections. Finally, Appendices~\ref{ap:invariants-case-study}, \ref{ap:lean-case-study}, and~\ref{ap:universal-queries-case-study} provide details on the case studies.

\medskip

\startcontents[appendices]
\printcontents[appendices]{}{1}{}

\section{Strategy Language}\label{ap:strategy-language}

This appendix provides more details on our proposed strategy language. Appendix~\ref{ap:orakell} describes our reference, type-checked Haskell embedding (included in the supplementary material), Appendix~\ref{ap:strategy-trees} explains how effect node types are derived from effect declarations, Appendix~\ref{ap:references} formally defines local values and references, Appendix~\ref{ap:effect-def} defines the type class for effects and provides a type signature for navigation functions, and Appendix~\ref{ap:naive-dfs} provides additional details on the naive tree definition presented in Section~\ref{sec:initial-design} for pedagogical purposes.

\subsection{A Type-Checked Embedding in Haskell}\label{ap:orakell}

For reference, we provide a minimal, type-checked Haskell embedding of our proposed strategy language in the supplementary material. All type definitions mentioned in this paper can be found there, sometimes with small, {inconsequential} variations (e.g., additional \code{Typeable} constraints that were omitted from the main text for simplicity). The automatic derivation of node effect types is not implemented, nor are the definitions of \code{spaces} and \code{mapEmbedded}, which are written by hand for several standard effects.

\subsection{Deriving Node Effect Types}\label{ap:strategy-trees}

Figure~\ref{fig:node-type-derivation} provides inference rules for deriving effect node types from effect declarations.

{
\newcommand{\nt}[1]{\langle#1\rangle}
\newcommand{\nts}[1]{\nt{\s{#1}}}
\newcommand{\s}[1]{\text{#1}}
\newcommand{\tid}[1]{\text{#1}}  %
\newcommand{\tsb}[1]{{\color{grammar-terminal-color}\textbf{\textsf{#1}}}}  %
\newcommand{\ts}[1]{{\color{grammar-terminal-color}\textbf{\textsf{#1}}}}
\newcommand{\ntdef}{\Coloneqq}
\newcommand{\GOR}{\ | \ }

\newcommand{\ruleSpace}{\qquad }
\newcommand{\ruleRw}[1]{#1}
\newcommand{\rewrites}{\ \rightsquigarrow\ }
\newcommand{\ruleVSpace}{1ex}

\begin{figure}
\begin{gather*}
\infer{
    \tsb{effect} \ e \ \ts{::} \ \ts{(} E \ \ts{$\in$ s} \ts{) $\Rightarrow$ \{} \, \vec f \, \ts{\} $\rightarrow$ Strategy s p } \alpha  
    \,\rewrites\,
    \tsb{data \,} E \ts{ p t n a } \tsb{\,where\, } E \ts{ :: \{\,} \vec f' \ts{\,\} $\rightarrow$ \ } E \ts{ p t n } \alpha
}{\forall i \ \ f_i \rewrites f'_i}
\\[\ruleVSpace]
\infer{ l \ts{ :: } s \rewrites l \ts{ :: } s' }{s \rewrites s'} \ruleSpace
\infer{ l \ts{ :: } (\tau \ts{ $\rightarrow$ } s) \rewrites l \ts{ :: } (\ts{Local n } \tau \ts{ $\rightarrow$ } s') }{s \rewrites s'} \ruleSpace
\\[\ruleVSpace]
\infer{
\ruleRw{\ts{Opaque p } \tau \rewrites \ts{LocalOpaque p n } \tau }}{}
\ruleSpace
\infer{
\ruleRw{\ts{Strategy s p } \tau \rewrites \ts{t n } \tau}}{}
\end{gather*}
\vspace{-1.2em}
\caption{Inference Rules for Deriving Node Types from Effect Declarations. See Figure~\ref{fig:effect-grammar} for the grammar of effects, and Figure~\ref{fig:derived-node-types-examples} for concrete examples of this transformation. The first line shows how to derive node types as GADTs after translating effect arguments. The second line shows how to translate effect arguments into local spaces and local parametric spaces, respectively. Finally, the third line details the translation of spaces into local spaces.}\label{fig:node-type-derivation}
\end{figure}
}

\subsection{Local Values and References}\label{ap:references}

\begin{figure}
\begin{lcodebox*}[label=code:refs-def,left=\DoubleDigitLn]
\begin{lstlisting}[style=haskell,style=withNumbers]
-- Local value
data Local n a  -- No constructor is exposed   (@\label{line:local-value-def:start}\label{line:local-value}@)
getRef :: Local n a -> ValueRef
dropRef :: Local n a -> a (@\label{line:local-value-def:end}@)

-- Local value reference (relative to a given node)
type ValueRef = Assembly SpaceElementRef  (@\label{line:value-ref}@)
data Assembly a = Atom a | List [Assembly a] | Element Int (Assembly a)  (@\label{line:assembly}@)
data SpaceRef = Main | SpaceRef SpaceName ValueRef  (@\label{line:space-ref}@)
data SpaceElementRef  (@\label{line:space-element-ref:start}@)
  = Answer SpaceRef String
  | Result SpaceRef NodeRef  (@\label{line:space-element-ref:end}@)

-- Node reference, relative to the root of a (possibly nested) tree
data NodeRef = Root | Child NodeRef ValueRef  (@\label{line:node-ref}@)

-- Combining local values
nil :: Local n ()  (@\label{line:local-value-combine:start}@)
liftPair :: (Local n a, Local n b) -> Local n (a, b)   (@\label{line:local-value-pair}@)
liftList :: [Local n a] -> Local n [a]  (@\label{line:local-value-list}@)
liftMaybe :: Maybe (Local n a) -> Local n (Maybe a)
liftEither :: Either (Local n a) (Local n b) -> Local n (Either a b)
unliftPair :: Local n (a, b) -> (Local n a, Local n b)
unliftList :: Local n [a] -> [Local n a]
unliftMaybe :: Local n (Maybe a) -> Maybe (Local n a)
unliftEither :: Local n (Either a b) -> Either (Local n a) (Local n b)
castLocal :: (Typeable a, Typeable b) => Local n a -> Maybe (Local n b)  (@\label{line:local-value-combine:end}@)

-- Recovering a success value from a reference
successValue :: Tree s p n v -> NodeRef -> Maybe (Local n v)  (@\label{line:success-value}@)
\end{lstlisting}
\end{lcodebox*}%
\CodeFigSpaceBefore
\caption{Definition of Local Values and References.}\label{fig:refs-def}
\end{figure}

\noindent Figure~\ref{fig:refs-def} provides type definitions for \emph{local values} and \emph{references} (Sections~\ref{sec:locality-references} and~\ref{sec:tree-definition}). Local values index the children of a tree's nodes, and are obtained by combining elements of local spaces.

\subsubsection{Local Values}
A \emph{local value} (Lines~\ref{line:local-value-def:start}-\ref{line:local-value-def:end}) is a pair consisting of a value and a \emph{value reference} that indicates its origin relative to a tree node identified by a phantom type parameter \code{n}. Local values cannot be constructed directly; instead, they can only be obtained by combining local space elements via a number of combinators (Lines~\ref{line:local-value-combine:start}-\ref{line:local-value-combine:end}).

\subsubsection{References}
A \emph{value reference} (Line~\ref{line:value-ref}) is an \emph{assembly} (Line~\ref{line:assembly}) of local \emph{space element references}---i.e., it refers to a value obtained by combining space elements through the introduction and elimination of lists, pairs, \code{Maybe}, \code{Either}, and \code{()}, all of which are encoded in references using lists.\footnote{For example, a reference to \texttt{(Just \!\_)} is encoded as \texttt{[\_]}, while a reference to \texttt{Nothing} is encoded as \texttt{[]}.} 
A \emph{space element reference} (Lines~\ref{line:space:start}-\ref{line:space:end}) denotes an element of a local parametric space (all spaces are considered parametric, be it over unit type) and is defined by a \emph{space name} (a string denoting a field of the effect type), a \emph{value reference} for the space parameter, an \emph{answer} in the case of a query, and a \emph{node reference} to a success node in the case of a nested tree. 
A \emph{node reference} (Line~\ref{line:node-ref}) consists of a list of \emph{value references} denoting a sequence of actions to follow from the root.

\subsection{Effect Types and Associated Methods}\label{ap:effect-def}

\begin{figure}
\begin{lcodebox*}[left=\DoubleDigitLn]
\begin{lstlisting}[style=haskell,style=withNumbers]
class Effect e where
  spaces :: (Space t, Typeable a) =>  -- auto-generated
    e p t n a -> [(SpaceName, SomeParametricSpace n)]   (@\label{line:spaces}@)
  mapEmbedded :: (Space t, Space t') =>  -- auto-generated
    (forall v. t n v -> t' n v) -> e p t n a -> e p t' n a
  navigate :: (Space t, MonadFail m) =>  -- mandatory (@\label{line:navigate:start}@)
    e p t n a -> Maybe (ChoiceFun m n -> m (Local n a))   (@\label{line:navigate:end}@)
  validAction :: e p t n a -> Local n a -> Bool  -- optional (@\label{line:valid-action}@)
  hasPrimarySpace :: e p t n a -> Bool  -- optional (@\label{line:has-primary-space}@)
  nodeTags :: e p t n a -> [Tag]  -- optional  (@\label{line:node-tags}@)

class Space sp where  -- Instances: `Tree s p` and `LocalOpaque p` (@\label{line:space:start}@)
  source :: sp n v -> SpaceSource n v
  spaceTags :: sp n v -> [Tag]  (@\label{line:space:end}@) (@\label{line:space-tags}@)

data SpaceSource n v  (@\label{line:space-source:start}@)
  = forall s p. SourceTree (Tree s p n v)
  | forall q. (Query q, Res q ~ v) => SourceQuery (AttachedQuery n q)  (@\label{line:space-source:end}@)

data SomeParametricSpace n =  (@\label{line:some-parametric-space:start}@)
  forall sp i v. (Space sp, Typeable i, Typeable v) =>
  SomeParametricSpace (Local n i -> sp n v)  (@\label{line:some-parametric-space:end}@)

type ChoiceFun m n = forall sp v. (Space sp) => sp n v -> m (Local n v)   (@\label{line:choice-fun}@)
\end{lstlisting}
\end{lcodebox*}
\CodeFigSpaceBefore
\caption{Definition of the \code{Effect} Type Class.}\label{fig:eff-space-def}
\end{figure}

\noindent All effect types (derived using the rules defined in Figure~\ref{fig:node-type-derivation}) must implement the \code{Effect} type class defined in Figure~\ref{fig:eff-space-def}. Methods of \code{Effect} are only relevant to the demonstration language (Section~\ref{sec:demonstration-language}), with the exception of \code{mapEmbedded}, which is useful for defining {tree transformers}, since tree transformers must \emph{generically} update all embedded trees in nodes that contain them (e.g., \code{Join}). The \code{spaces} and \code{mapEmbedded} methods can be automatically generated from the effect declaration. All other methods have default implementations, so the only method that must be defined manually in every case is \code{navigate} (Appendix~\ref{ap:navigate-details}).

As a reminder, an {effect} type \code{e} is a type constructor with parameters \code{p} (the {inner policy type} associated with the surrounding tree), \code{t} (the type of {embedded trees}), \code{n} (the phantom type identifying the current node), and \code{a} (the action type of the current node).

\subsubsection{Spaces}
The \code{spaces} function returns a list of \emph{parametric} and \emph{nonparametric spaces}, each indexed by a \emph{local value} (Appendix~\ref{ap:references}) of some type (\code{()} for nonparametric spaces). Spaces can be of two kinds---\emph{opaque spaces} and \emph{embedded trees}---corresponding to the two instances of the \code{Space} type class (Lines~\ref{line:space:start}-\ref{line:space:end}). The \code{spaces} function is useful for manipulating \emph{generic} trees with \emph{unknown} or \emph{arbitrary} effects. Its output features runtime type annotations that can be used to perform safe casting and runtime type checks, hence the \code{Typeable} instances (some \code{Typeable} constraints were omitted in the main text for simplicity; see the full Haskell embedding described in Appendix~\ref{ap:orakell} for details). These features are useful for implementing an interpreter for the demonstration language (Section~\ref{sec:demonstration-language}), which is dynamic in nature. However, specific policies such as \code{dfs} can access nested spaces directly via the fields of specific effects, thereby benefiting from strong static typing. For the purposes of the demonstration language, a \emph{space} (Lines~\ref{line:space:start}-\ref{line:space:end}) must specify a \emph{source}, which is either a \emph{tree} or a \emph{query}. Spaces can also specify \emph{tags}, which by default are derived from the name of the attached strategy or query.

\subsubsection{Navigation Functions}\label{ap:navigate-details}
Navigation functions must be implemented for every effect type via the \code{navigate} method (Lines~\ref{line:navigate:start}-\ref{line:navigate:end}). For leaf nodes such as \code{Fail}, \code{navigate} must return \code{Nothing}. For other nodes, it must return a navigation function that maps a \emph{choice function} (Line~\ref{line:choice-fun}) to an action. A choice function maps a local space to one of its elements. Note that choice functions (and hence \code{navigate}) are allowed to be monadic, since the demonstration interpreter uses choice functions that inspect answered queries from demonstrations and that may fail in the case of missing answers. In addition, navigation functions themselves may fail, hence the \code{MonadFail} constraint on Line~\ref{line:navigate:start} (see the navigation function for the \code{Abduction} effect in Figure~\ref{fig:abduction-navigate} for an example, Line~\ref{line:abduction-nav-assert}).

\subsection{Policy Example for Naive Trees}\label{ap:naive-dfs}

We show an implementation of \emph{depth-first} search for the naive trees from Figure~\ref{fig:naive-tree} in Figure~\ref{fig:naive-dfs}.

\begin{figure}
\begin{ccodebox*}%
\begin{lstlisting}[style=haskell]
dfs :: (String -> IO [String]) -> Tree a -> MaybeT IO a
dfs _ (Success x) = return x
dfs _ Failure = mzero
dfs oracle (Branch (Query prompt parse) k) = do
  answers <- lift (oracle prompt)
  msum (map (dfs oracle . k) (mapMaybe parse answers))
\end{lstlisting}
\end{ccodebox*}
\CodeFigSpaceBefore
\caption{Defining Depth-First Search for Naive Search Trees. See Figure~\ref{fig:naive-tree} for the definition of naive search trees. The \code{dfs} function takes as its first argument an \emph{oracle} in the form of a function that maps a prompt to a list of possible answers. Such an oracle can be implemented by sampling multiple answers from a large language model. The \code{MaybeT} monad transformer adds failure capability to a monad, \code{mzero} denotes a failure, \code{msum} returns the first successful value out of a list of alternatives, and \code{lift} wraps an IO value into \code{MaybeT}.}\label{fig:naive-dfs}
\end{figure}

\section{Policy Language}\label{ap:policy-language}

This appendix provides details on the policy language. Appendix~\ref{ap:search-streams} describes a representation for search streams and defines the \code{withBudget} and \code{parallel} operators on this representation. Appendix~\ref{ap:correct-resource-sketch} provides a proof sketch of resource management correctness (Property~\ref{prop:correct-resource}). Appendix~\ref{ap:policy-blocks} provides details on tree transformers and shows the implementation of \code{elimJoin}.

\subsection{Implementing Search Streams}\label{ap:search-streams}
\begin{figure}
\begin{ccodebox*}
\begin{lstlisting}[style=haskell]
type Skipped = Int
data StreamElem m a =
      Done
    | Yield a (Stream m a)
    | Barrier Budget (Bool -> Stream m a)
    | Spent Budget Skipped (Stream m a)
newtype Stream m a = Stream {runStream :: m (StreamElem m a)}
\end{lstlisting}
\end{ccodebox*}
\CodeFigSpaceBefore
\caption{Internal Representation of Search Streams.}\label{fig:streams-internal}
\end{figure}
\noindent We propose an internal representation for search streams in Figure~\ref{fig:streams-internal}.
A stream is a monadic value producing a \emph{stream element} (search streams are often effectful---e.g., sending LLM requests). There are four kinds of stream elements: \code{Done} indicates that the stream has terminated, \code{Yield} produces a value, and \code{Barrier} and \code{Spent} handle resource management. A \code{Barrier} element requests authorization to spend some resources, up to a provided upper-estimate. The client must answer with a boolean---granting the resources or not---before the stream can resume\footnote{One cannot just interrupt the stream for denying resource requests since other requests for smaller amounts may follow (along with \texttt{Spent} elements that must not be dropped),  especially when \texttt{parallel} is used.}. Each \code{Barrier} element must be later paired with a \code{Spent} element that reports \emph{actual} resource consumption. When using \code{parallel}, a \code{Spent} element is not necessarily paired with the latest \code{Barrier} element in the stream, due to calls to \code{spend} happening concurrently. Thus, \code{Spent} elements carry an integer that indicates how many previous barrier elements are to be skipped before encountering the matching \code{Barrier} element. We now explain how to implement particular stream combinators. More examples can be found in our Haskell implementation (Appendix~\ref{ap:orakell}).

\subsubsection{Implementing ``{withBudget}''}\label{ap:with-budget-impl}
The \code{withBudget b s} stream intercepts and retransmits all elements from \code{s}, while passing back all answers to resource requests. It overrides these answers whenever granting acceptance risks exceeding spending limit \code{b}. To do so, it not only memorizes how much budget has been spent already, but it also maintains an estimate on how much spending is \emph{pending} at any moment in time---that is, what amount of resources is currently frozen by currently executing \code{spend} operations. Every granting of a \code{Barrier} request increases this pending amount by the associated consumption estimate, which is then subtracted upon encountering the matching \code{Spent} element. A full implementation is provided in Figure~\ref{fig:with-budget}.

\subsubsection{Implementing ``{parallel}''} The \code{parallel} function spawns one thread for each of its arguments. Each thread terminates upon encountering the \code{Done} element. \code{Yield} and \code{Spent} elements encountered by any thread are transmitted back (although care must be taken to correctly update the \code{Skipped} argument of \code{Spent} messages to account for interleaving). Handling \code{Barrier} messages is more subtle. When a thread encounters a \code{Barrier} message and the client desires to allow it, the thread does so. However, if the client denies the request, the thread does \emph{not} transmit this answer right away. Rather, if the \emph{other} thread has \code{spend} operations pending (i.e., unmatched barrier elements), it waits until these operations are finished, and then asks the client \emph{again} (whose answer might be different upon seeing an updated estimate of pending expenses).

\begin{figure}
\begin{ccodebox*}
\begin{lstlisting}[style=haskell]
withBudget :: (Monad m) => Budget -> Stream m a -> Stream m a
withBudget lim = aux 0 []
  where
    aux :: (Monad m) =>
      Budget -> [(Skipped, Budget)] -> Stream m a -> Stream m a
    aux spent pending (Stream me) = Stream $ do
      e <- me
      case e of
        Done -> return Done
        Yield x cont -> return (Yield x (aux spent pending cont))
        Barrier b cont ->
          return (Barrier b (\allow ->
            let allow' =
              allow && (spent + b + sum (map snd pending) <= lim) in
            let reserved = if allow' then b else 0 in
            let pending' =
              (0, reserved) : map (\(s, v) -> (s + 1, v)) pending in
            aux spent pending' (cont allow')))
        Spent b skipped cont ->
          let spent' = spent + b in
          let freed = fromJust (lookup skipped pending) in
          let pending' = delete (skipped, freed) pending in
          return (Spent b skipped (aux spent' pending' cont))
\end{lstlisting}
\end{ccodebox*}
\CodeFigSpaceBefore
\caption{Implementation of \code{withBudget}. See Appendix~\ref{ap:with-budget-impl} for explanations.}\label{fig:with-budget}
\end{figure}

\subsection{Proof Sketch of Resource Management Correctness}\label{ap:correct-resource-sketch}

To show \emph{correct resource management} (Property~\ref{prop:correct-resource}), we can prove that all combinators from Figure~\ref{fig:stream-combinators} preserve the following invariants on streams (as defined in Appendix~\ref{ap:search-streams}):
\begin{itemize}
  \item Every \code{Spent} message is associated with a unique \code{Barrier} message and vice versa.
  \item If all spending requests are denied after seeing a given prefix from the stream, then the \emph{total amount spent} after exhausting the stream (i.e., the sum of all amounts in \code{Spent} messages) is at most the sum of {\it(i)} the amount spent in the prefix, {\it(ii)} the pending amount (i.e., the sum of all amounts in \code{Barrier} messages that are not matched to a \code{Spent} message in the prefix), and {\it(iii)} the number of pending \code{Barrier} messages times the maximum resource estimation error $\delta$.
\end{itemize}
The proof can be concluded by noticing that the number of pending \code{Barrier} messages in any stream prefix is at most its concurrency level $n$.

\subsection{Tree Transformers}\label{ap:policy-blocks}

For illustration purposes, we show the implementation of \code{elimJoin} in Figure~\ref{fig:elim-join-def}. It uses the \code{bindTree} function, which is the tree equivalent of the monadic bind operation on strategies. It also uses the \code{elimEffect} utility, which handles the plumbing of recursively applying a given local transformation to the full tree, using \code{mapEmbedded} to also transform embedded trees (Appendix~\ref{ap:effect-def}). See our Haskell implementation for more details (Appendix~\ref{ap:orakell}).

\begin{figure}
\begin{ccodebox*}[label=code:elim-join]
\begin{lstlisting}[style=haskell]
elimJoin :: Tree (Join ': s) p n v -> Tree s p n v
elimJoin = elimEffect (\(Node (Join l r) child) ->
  bindTree (elimJoin l) (\lv ->
    bindTree (elimJoin r) (\rv ->
      elimJoin (child (trackPair (lv, rv))))))
\end{lstlisting}
\end{ccodebox*}
\CodeFigSpaceBefore
\caption{Implementation of the \code{elimJoin} Transformer.}\label{fig:elim-join-def}
\end{figure}

\section{Demonstration Language}\label{ap:demonstration-language}

This appendix provides details on the demonstration language. Appendix~\ref{ap:demonstration-tests} provides a full grammar for demonstration tests and explains the semantics of all instructions. Appendix~\ref{ap:demonstration-completeness} provides proof sketches of weak and strong completeness for the demonstration language. 

\subsection{Syntax and Semantics of Navigation Tests}\label{ap:demonstration-tests}

Figure~\ref{fig:navigation-tests-grammar} provides a full grammar for navigation tests. A test consists of a sequence of \emph{instructions}. Each instruction takes as an input a node in the tree (initially the root) and returns a new node. Instructions can also fail or emit warnings. The semantics of \code{run} is discussed in Section~\ref{sec:tests-dsl}. We motivate and describe the other instructions below.

{
\newcommand{\nt}[1]{\langle#1\rangle}
\newcommand{\nts}[1]{\nt{\s{#1}}}
\newcommand{\s}[1]{\text{#1}}
\newcommand{\tid}[1]{\text{#1}}  %
\newcommand{\tsb}[1]{{\color{grammar-terminal-color}\textbf{\textsf{#1}}}}  %
\newcommand{\ts}[1]{{\color{grammar-terminal-color}\textbf{\textsf{#1}}}}
\newcommand{\ntdef}{\Coloneqq}
\newcommand{\GOR}{\ | \ }

\newcommand{\hintQuote}{\char`\'}
\newcommand{\groupSpacing}{0cm}

\begin{figure}
\begin{align*}
  \nts{test}           \ntdef\ & \nts{instr} \GOR \nts{test} \ \ts{|} \ \nts{test} \\
  \nts{instr}          \ntdef\ & \ts{run} \ \nts{hints}? \GOR
                              \ts{at} \ \nts{node-sel} \ \nts{hints}? \GOR
                              (\ts{go} \GOR \ts{answer}) \ \nts{space-ref} \GOR
                              \\
                       &\ts{take} \ \nts{val-ref} \GOR \ts{success} \GOR \ts{save} \ \s{node-var} \GOR \ts{load} \ \s{node-var} \\
  \nts{hints}          \ntdef\ & \ts{\texttt{`}} \s{hint}^* \ts{\texttt{'}} \\[\groupSpacing]
  \nts{node-sel}       \ntdef\ & \s{node-tag} (\ts{{\#}}\s{int})? \GOR 
                              \nts{space-sel} \ts{\texttt{/}} \nts{node-sel} \\
  \nts{space-sel}      \ntdef\ & \s{space-tag} (\ts{{\#}} \s{int})? \\[\groupSpacing]
  \nts{space-ref}      \ntdef\ & \s{space-id}(\ts{(}\nts{val-ref}\ts{)})? \\
  \nts{elt-ref}        \ntdef\ & \nts{space-ref} \, \ts{\{} \nts{hints} \ts{\}} \GOR \nts{hints} \GOR \ts{\%} \s{node-var} \\
  \nts{val-ref}        \ntdef\ & \nts{elt-def} \GOR \ts{[} \nts{val-ref} \, \ts{,}\, ... \,\ts{,}\, \nts{val-ref} \ts{]} \GOR \nts{val-ref} \ts{[} \s{int} \ts{]}
\end{align*}
\vspace{-0.45cm}
\caption{Grammar of Demonstration Tests.}\label{fig:navigation-tests-grammar}
\end{figure}
}

\subsubsection{Stopping at Particular Nodes}\label{ap:test-at-instruction}
The \code{at} instruction works like \code{run}, except that it allows specifying a node at which the walk must stop. All nodes in a tree are associated with a set of tags. The \code{at} instruction takes as an additional argument a node selector, the simplest form of which denotes a tag to match. For example, in Figure~\ref{fig:demonstration-overview}, instruction \haskellCode{at EvalProg 'wrong'} behaves similarly to \haskellCode{run 'wrong'}, except that it stops when encountering a node tagged with \code{EvalProg}. All \emph{spaces} are tagged with the name of the associated query or strategy, and each node inherits the tags of its \emph{primary space} if it has one (effect nodes can optionally define at most one \emph{primary space}---see the definition of \code{Effect} in Appendix~\ref{ap:effect-def}). Effect nodes can also define custom node tags (Figure~\ref{fig:eff-space-def}, Line~\ref{line:node-tags}).

Importantly, \code{at} can only stop {within} the same tree that it started in and \emph{not} inside a nested tree. In the example from Figure~\ref{fig:demonstration-overview}, \haskellCode{at ConjProg} will error instead of stopping at the unique node that contains a \code{ConjProg} query (query 2). This design choice is mandated, once again, by \emph{modularity}. Indeed, individual strategies can be made responsible for setting unambiguous tags for nodes that they control but cannot be made responsible for ensuring the absence of clashing tags in \emph{other} strategies. In order to stop at the node mentioned earlier, one must use instruction \haskellCode{at conjectureProg/ConjProg} instead. The node selector that is used here refers to the first node with tag \code{ConjProg} \emph{within} the first space with tag \code{conjectureProg}. Finally, \code{at foo\#2/bar\#3} stops at the \emph{third} node with tag \code{bar}, within the \emph{second} space with tag \code{foo}.

\subsubsection{Entering Nested Spaces} The \code{go} instruction allows entering a tree nested within the current node. For example, if the current node is a \code{CBranch} node, \code{go cands} enters the tree that defines the \code{cands} space or errors if \code{cands} is defined by a query. This instruction can be shortened to \code{go}, since \code{cands} is the primary space of \code{CBranch} nodes. More interestingly, suppose that the demonstration already explores two paths within \code{cands} that reach different success leaves and thus correspond to two different branching candidates. As previously discussed, each of these paths can be described through a sequence of \emph{hints} so let us assume that the first candidate is identified by \haskellCode{''} (no hints, or default path) and the second candidate is identified by \haskellCode{'foo'} (use answer \haskellCode{'foo'} when appropriate). Then, instruction \haskellCode{go compare([cands\{''\}, cands\{'foo'\}])} can be used to enter the strategy tree comparing those two candidates. It can be shortened to \haskellCode{go compare(['', 'foo'])} since \code{cands} is a primary space. In general, any element of a local space can be referred to via a (possibly empty) sequence of hints. For spaces defined by queries, at most one hint is expected that indicates which answer to use. For spaces defined by trees, a sequence of hints is expected that leads to a success leaf by calling \code{run} recursively.

The \code{answer} instruction is similar to \code{go}. It takes a space selector as an argument but then expects to find a query instead of a tree when entering this space. It then succeeds if the corresponding query is answered in the demonstration and fails otherwise.

\subsubsection{Visiting a Child} The \code{take} instruction takes a local value as an argument and updates the current node to the corresponding child. For example, at a \code{Join} node, the instruction written \code{take [left\{''\}, right\{'foo bar'\}]} visits the child associated with the pair whose first component is the default element from the \code{left} space and whose second component is the element from the \code{right} space described by hints \code{'foo bar'}.

\subsubsection{Loading and Saving Nodes} The \code{save} and \code{load} instructions allow saving nodes into named variables and later recovering them. For example, \code{save id} saves the current node in a variable named \code{id} while \code{load id} ensures that the \code{id} variable is set (possibly by a previous test) and sets the current node to the corresponding value. When specifying a local space or a local value in the \code{take} and \code{go} commands, \code{\%id} can be used to refer to the value attached to success leaf \code{id} (and errors if \code{id} does not refer to a success leaf).

\subsection{Proof Sketches for Completeness}\label{ap:demonstration-completeness}

We precisely state and prove the completeness properties of our demonstration language.

\subsubsection{Weak Completeness} The \emph{weak completeness} property (Section~\ref{sec:demo-completeness}) can be stated as follows:

\begin{property}[Weak Completeness]
For any node in a search tree, there exists a demonstration with a test (expressed using the grammar defined in Figure~\ref{fig:navigation-tests-grammar}) that reaches this node.
\end{property}

The challenge is to obtain this property while keeping the demonstration language simple, and without allowing arbitrary Haskell values to be explicitly constructed. This is enabled by \emph{locality} (Section~\ref{sec:locality-references}), which guarantees that actions can only result from combining local space elements, which can themselves be identified by answer labels (for spaces induced by queries) or recursive sequences of actions (for spaces induced by trees). Thus, any path in a tree can be described using the \code{take}, \code{go}, \code{load}, and \code{save} instructions (Appendix~\ref{ap:demonstration-tests}).

\subsubsection{Strong Completeness} \emph{Strong completeness} (Section~\ref{sec:demo-completeness}) can be stated as follows:

\begin{property}[Strong Completeness]
For any success node in a tree, and assuming that all involved effects have invertible navigation functions, there exists a demonstration with a test of the form \code{run <hints>?} that reaches a success node carrying an equivalent value.
\end{property}

A proof sketch proceeds as follows. Given a success leaf in the tree, the attached reference specifies a precise path leading to it as a sequence of actions. At every step along this path, and by invertibility, an equivalent action can be produced by \code{navigate}. Whenever an inverse choice function selects an element from a space defined via a query, the corresponding answer is added to the demonstration based on the element’s reference. For spaces based on nested trees, the process is invoked recursively. Passing hints to \code{run} is only useful in the rare cases where the same query must be answered differently at different steps. In such cases, labels can be introduced to distinguish the different answers and provided as hints in the appropriate order.

\subsubsection{On the Invertibility of \code{CBranch}}\label{ap:cbranch-invertibility}
In Section~\ref{sec:demo-completeness}, we argue that \code{CBranch} has an invertible navigation function since any valid action is an element from the \code{cands} space that can be picked by the inverse choice function.
Technically, it is possible for an action at a \code{CBranch} node to be obtained through a nontrivial combination of elements from \code{cands} such as the following, with \code{e} and \code{e'} elements from \code{cands}: \code{fst (unliftPair (liftPair (e, e')))} . However, using a \emph{parametricity} argument~\cite{wadler1989theorems}, any action built by a policy (which does not have knowledge of the concrete action types of nodes, themselves hidden by existentials in the definition of trees; see Figure~\ref{fig:tree-def}) must be equivalent to an element from \code{cands} (\code{e} in our example). Alternatively, one can enforce at runtime that valid \code{CBranch} actions must be elements of \code{cands}, by implementing the optional \code{validAction} method (see Appendix~\ref{ap:effect-def}).

\section{The Delphyne Framework}\label{ap:delphyne}

This appendix provides details on the Delphyne framework (Section~\ref{sec:delphyne}). Appendices~\ref{ap:python-embedding-strategy} and~\ref{ap:python-embedding-policy} discuss the embeddings of the strategy and policy languages, respectively. Appendix~\ref{ap:editor-support} discusses Delphyne's editor support for writing and repairing demonstrations and visualizing traces. Appendix~\ref{ap:compute-effect} explains the \code{Compute} effect for performing impure computations inside strategies, and Appendix~\ref{ap:interact} presents the standard higher-order \code{interact} strategy for building ReAct-style agents.

\subsection{Embedding the Strategy Language}\label{ap:python-embedding-strategy}

Delphyne strategies are not expressed with monads (for which Python has poor syntactic support) but with coroutines instead (hence the use of \code{yield from}). An example is provided in Figure~\ref{fig:delphyne-strategy-example}. Since Python coroutines cannot be cloned, reification is implemented via thermometer continuations~\cite{koppel2018capturing}: a tree node is identified by the sequence of actions leading to it and the strategy is replayed from scratch every time a child is computed. Doing so does not raise practical performance issues since expensive computations can be cached (see \code{Compute} effect in Appendix~\ref{ap:compute-effect}).

\begin{figure}
\begin{lcodebox*}
\begin{lstlisting}[style=delphyne]
@dataclass
class ProveProg(Query[Proof]):
    """
    Given a specification and a program, produce a checkable
    proof that the program meets the specification.
    """
    spec: Spec
    prog: Prog
\end{lstlisting}
\end{lcodebox*}
\CodeFigSpaceBefore
\caption{Defining a Query in Delphyne.}\label{fig:delphyne-query-example}
\end{figure}

\subsubsection{Defining Queries.}\label{ap:queries-in-delphyne} Queries producing answers of type \code{T} can be defined as dataclasses inheriting from \code{Query[T]} (Figure~\ref{fig:delphyne-query-example}). By default, the docstring is used as a system prompt, and instance prompts include a serialized JSON representation of the query object produced by Pydantic. LLMs are tasked with producing \emph{structured output}~\cite{geng2025generating} from an automatically generated schema, with the resulting JSON parsed by Pydantic. All these behaviors can be customized. Python's reflection capabilities are particularly useful here.

\subsubsection{Strategies and Mutable State.}\label{ap:strategy-mutable-state} In general, strategies must be \emph{pure} and not depend on global, mutable state. Otherwise, reifying the same strategy computation twice could yield different trees. Haskell allows enforcing this discipline through types, but Python does not. We explicitly allow \emph{internal} side effects and mutations inside strategies, which are idiomatic in Python, \emph{as if} the associated continuation never yielded control. Thus, the following is acceptable within a strategy, where \code{xs} and \code{ys} are local list variables:

\begin{ccodebox*}
\begin{lstlisting}[style=delphyne]
ys = yield from branch(Query(xs).using(...)) ; xs.append(0) ; ys.append(0)
\end{lstlisting}
\end{ccodebox*}
\noindent To ensure soundness, we restrict the arguments and return values of effects to be either deep-copyable values or pure functions. This is partially enforced by runtime checks, although these cannot distinguish pure functions from impure ones. In our experience, this soundness hole is very difficult to trigger accidentally in idiomatic Python code. Impure, stochastic, or non-replicable computations (e.g., calls to external theorem provers with wall-time timeouts) can be performed inside strategies via the \code{Compute} effect (Appendix~\ref{ap:compute-effect}).

\subsubsection{Static Type Safety.}\label{ap:delphyne-strategy-type-safety} Perhaps surprisingly, the strategy language can be fully typed using Python's static type system. For example, the \code{branch} effect trigger is typed as follows:

\begin{ccodebox*}
\begin{lstlisting}[style=delphyne]
def branch[P, T](cands: Opaque[P, T]) -> Strategy[Branch, P, T]: ...
\end{lstlisting}
\end{ccodebox*}

\noindent Strategy signatures can be represented using \emph{union types} (see Figure~\ref{fig:delphyne-strategy-example}), which is more ergonomic than our Haskell encoding based on \emph{type lists}, since unions are automatically commutative. The \code{Strategy} type is covariant with respect to its first and third arguments, and so \code{Strategy[N, P, T]} is a subtype of \code{Strategy[N | M, P, T]}. Overall, writing strategies in Delphyne with type checkers such as Pyright enabled in strict mode delivers an experience comparable to that of a statically typed language such as Haskell. Strategies can also be written without type annotations, providing the feel of a dynamic language.

One limitation of current Python type checkers is that they often struggle to infer the argument types of anonymous functions, which cannot be annotated in Python (although this may change in the future). This becomes especially problematic when building opaque spaces through the \code{using} method (Figure~\ref{fig:delphyne-strategy-example}). As a workaround, several functions and methods, such as \code{branch} and \code{using}, feature an optional argument \code{inner\_policy\_type} that allows the inner policy type to be explicitly specified when necessary.

\begin{figure}[t]
\begin{lcodebox*}
\begin{lstlisting}[style=delphyne]
@dataclass
class Join(Node):
    subs: Sequence[EmbeddedTree[Any, Any, Any]]

    def navigate(self) -> Navigation:
        ret = []
        for sub in self.subs:
            ret.append((yield sub))
        return tuple(ret)

def join[N: Node, P, T](
    subs: Sequence[StrategyComp[N, P, T]]) -> Strategy[N, P, Sequence[T]]:
    recv = yield spawn_node(Join, subs=subs)
    return recv.action
\end{lstlisting}
\end{lcodebox*}
\CodeFigSpaceBefore
\caption[Defining the Join effect]{Defining the \code{Join} Effect. Unlike the version defined in Figure~\ref{fig:derived-node-types-examples}, the \code{join} trigger can take a \emph{list} of computations as arguments (not just two), all with the same type.}\label{fig:delphyne-join-def}
\end{figure}

\subsubsection{Adding New Effects.}\label{ap:delphyne-new-effects} New effects can be defined by inheriting from the \code{Node} class and defining a corresponding trigger function. We show a definition of the \code{Join} effect in Figure~\ref{fig:delphyne-join-def}. Effect node types cannot be typed precisely (as in the Haskell embedding in Figure~\ref{fig:derived-node-types-examples}), since Python does not support GADTs or higher-kinded types, hence the use of \code{Any}. However, this only affects the internals of policy primitives that explicitly manipulate trees (see Appendix~\ref{ap:defining-new-search-algos-in-delphyne}) and does not affect strategy authors, since the trigger function \code{join} is precisely typed. Navigation functions are defined slightly differently than in Haskell: instead of taking a choice function as an argument, they return a Python generator that yields local spaces and receives elements from them (hence the use of \code{yield}). Node methods such as \code{spaces} and \code{map\_embedded} (see Appendix~\ref{ap:effect-def}) are automatically inherited from the \code{Node} parent class, where they are implemented using reflection by inspecting dataclass field annotations.

\begin{figure}[t]
\begin{lcodebox*}
\begin{lstlisting}[style=delphyne]
@search_policy
def par_dfs[P, T](
    tree: Tree[Branch | Fail, P, T], env: PolicyEnv, policy: P,
) -> StreamGen[T]:
    match tree.node:
        case Success(x): yield Solution(x)
        case Fail(): pass
        case Branch(cands):
            cands = yield from cands.stream(env, policy).all()
            yield from Stream.parallel(
                [par_dfs()(tree.child(a.tracked), env, policy) for a in cands])
\end{lstlisting}
\end{lcodebox*}
\CodeFigSpaceBefore{}
\caption[Defining a search policy in Delphyne]{Defining a Parallel Variant of Depth-First Search in Delphyne.}\label{fig:delphyne-par-dfs}
\end{figure}

\subsection{Embedding the Policy Language}\label{ap:python-embedding-policy}

On the side of the policy language, policy components (i.e., search and prompting policies, stream and tree transformers) can be given a precise external signature, enforcing their correct use and composition. However, their \emph{implementation} typically uses gradual typing since search trees are not fully typed. For example, the \code{Branch} class (representing branching nodes) has no parameter constraining the type of value being branched on or the type of the surrounding inner policy (see also Figure~\ref{fig:delphyne-join-def} for the definition of \code{Join}). In addition, locality (Section~\ref{sec:locality-references}) is only enforced at runtime. Writers of functions such as \code{dfs} must therefore be careful in ensuring that their implementation truly matches the advertised type. Overall, this is still a good trade-off since typical users are expected to spend much more effort writing strategies and assembling policy components than defining new effects and atomic policies (especially given a rich standard library).

\subsubsection{Type-Directed Policy Writing and Inner Policy Dictionaries.}\label{ap:inner-policy-dicts}
Inner policy types are typically defined as dataclasses in Delphyne, and policies are written by composing primitives (search algorithms, stream transformers, and tree transformers) in a type-directed process, as demonstrated in Section~\ref{sec:inner-policies-local-spaces}. This approach offers strong static type safety, but at the cost of verbosity: each strategy must be followed by an explicit inner policy type definition and each creation of an opaque space within a strategy requires passing a mapping from the ambient inner policy to a proper sub-policy, often in the form of an anonymous function (Figure~\ref{fig:delphyne-strategy-example}).

Delphyne offers an alternative that trades static type safety for concision: \emph{inner policy dictionaries}. An inner policy dictionary (with type \code{IPDict}) is simply a Python dictionary that maps strings denoting space tags (by default, the name of the query or sub-strategy inducing the space; see Appendix~\ref{ap:test-at-instruction}) to sub-policies. Strategies can elect to adopt \code{IPDict} as their associated inner policy type, in which case no additional inner policy type needs to be defined, nor any mapping passed to \code{using} (the ellipsis value \code{...} can be passed instead). The \code{guess} operator used in our third case study (Section~\ref{sec:universal-queries-case-study}) leverages inner policy dictionaries for concision (Figure~\ref{fig:universal-strategy}). Finally, note that the trade-off made here between static type safety and concision is not fundamental but stems from limitations of Python. In principle, macros could be used to automatically derive inner policy types and mappings from syntactic conventions. Although Python supports runtime code generation and evaluation, it does not do so in a way that is compatible with static type checking.

In any case, defining inner policy types and mappings is generally a small price to pay, and the resulting explicitness is valuable in its own right. It also offers additional expressiveness. For instance, an opaque space created inside a loop can have its associated policy indexed by the loop iteration number, in which case the corresponding inner policy type includes a function field.

\subsubsection{Defining New Search Algorithms.}\label{ap:defining-new-search-algos-in-delphyne}
Figure~\ref{fig:delphyne-par-dfs} shows how to define a search algorithm in Delphyne: a parallel variant \code{par\_dfs} of depth-first search. The \code{par\_dfs} function has a precise type signature but internally uses gradual typing. It returns a search stream defined via a Python generator. The \code{all} method collects all elements from a stream while forwarding resource-consumption messages (similar to \code{partial} in Figure~\ref{fig:stream-combinators}). The \code{Stream.parallel} combinator takes a list of search streams and returns a stream (similar to \code{parallel} in Figure~\ref{fig:stream-combinators}). As in our embedding of strategies in Python, we use Python coroutines (generators) instead of monads to define search streams.

\subsection{Editor and Tooling Support}\label{ap:editor-support}\label{ap:visualizing-trees}

Our proposed demonstration language (Section~\ref{sec:demonstration-language}) is explicitly designed to enable rich tooling for interactively writing and repairing demonstrations. We describe its implementation in Delphyne via a dedicated VSCode extension. The separation of strategies and policies also enables structured inspection of runs of oracular programs by visualizing the resulting traces (i.e., pruned trees containing all visited nodes; see Section~\ref{sec:locality-references}).

\subsubsection{Writing and Repairing Demonstrations}\label{ap:writing-repairing-demos}

\begin{figure}[tp]
\centering
\IncludeAppleScreenshot{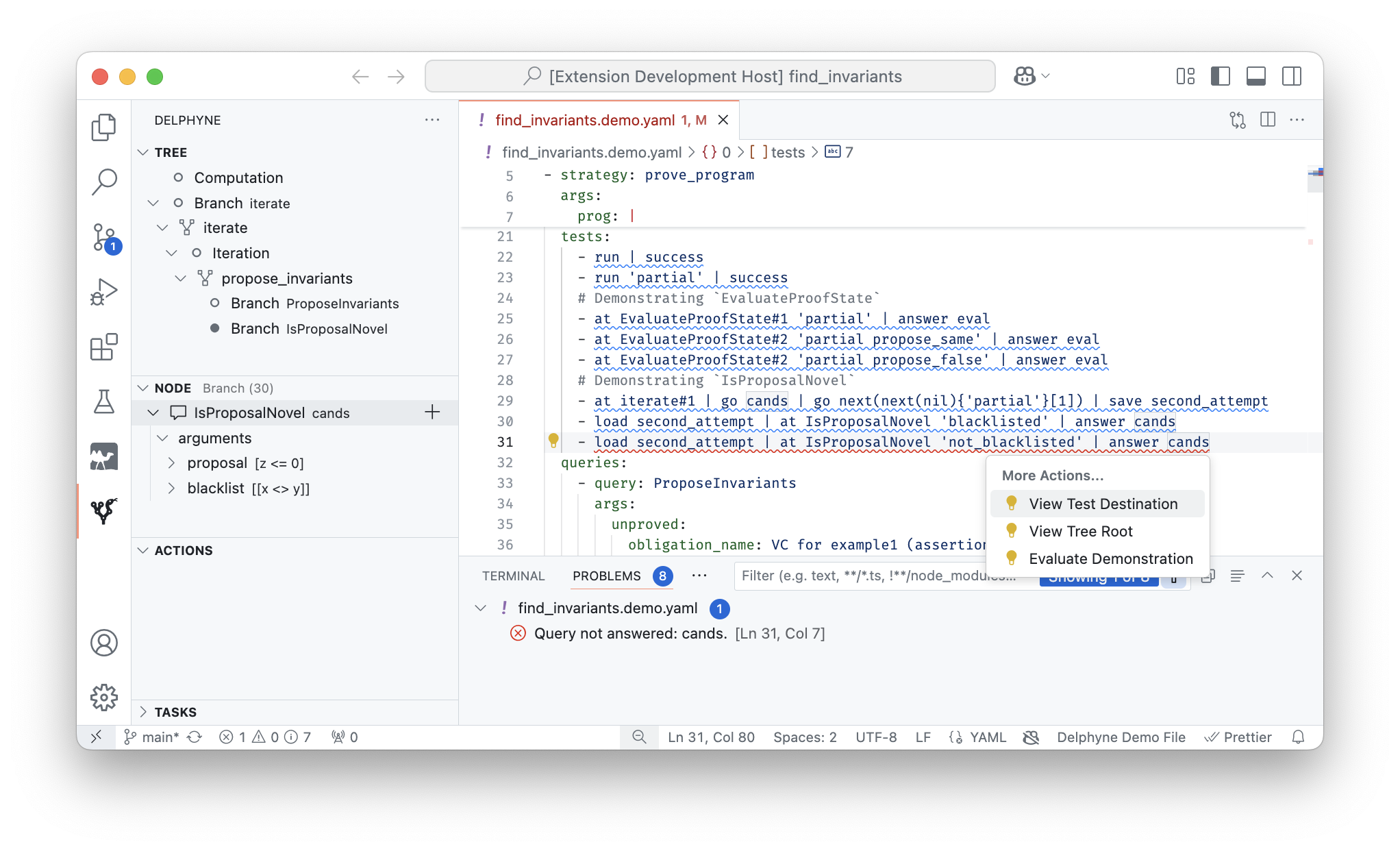}
\SmallSpaceBelowAppleScreenshot{}
\caption[Using the Delphyne VSCode extension to write a demonstration]{Using the Delphyne VSCode Extension to Write a Demonstration. Passing tests are underlined in blue while failing tests are underlined in red. Code actions allow inspecting the destination node of each test. Doing so for the failing test reveals that an \code{IsProposalNovel} query needs to be answered. This query can be added to the demonstration body by clicking on the \code{+} icon next to the query name. In general, the three views in the left pane allow inspecting arbitrary traces and describe one selected node at any  moment in time. The \code{Tree} view shows the position of this node in the trace, the \code{Node} view shows all spaces attached to it, and the \code{Actions} view shows all associated actions. The trace can be navigated by clicking on actions to access children, on spaces to access nested trees, or on ancestor nodes from the \code{Tree} view.}\label{fig:demo-screenshot}
\end{figure}

Figure~\ref{fig:demo-screenshot} illustrates the interactive writing of demonstrations using the Delphyne VSCode extension.
In Delphyne, demonstrations are written in YAML, making many standard YAML editing features available for free (collapsing, structural navigation, schema-based linting, anchor jumping). Demonstrations are gathered in demonstration files, and can be evaluated\footnote{Evaluating a demonstration consists in running its tests (Section~\ref{sec:demonstration-language}).} all together or individually. Users must \emph{explicitly} request the evaluation of demonstrations, since doing so may involve running expensive computations, but demonstrations can be evaluated concurrently. Once a demonstration has been evaluated, diagnostics are shown in the editor. When modifications to a demonstration file invalidate particular diagnostics, they automatically disappear, and the affected demonstration can be evaluated again.

A typical workflow for writing a demonstration is to start with an empty \code{queries} section and a single \code{run \!\!|\!\! success} test (Section~\ref{sec:demonstration-workflows}). Evaluating the demonstration causes the test to become \emph{stuck} at a given node. The user can then visualize this node in Delphyne’s \emph{Tree View} (Figure~\ref{fig:demo-screenshot}), along with the full trace generated by the test, and add the corresponding unanswered query to the demonstration body with a single click. An answer can then be written manually, possibly with accompanying explanations, after which the demonstration can be evaluated again. Alternatively, one can use the \emph{``Answer Query''} code action to open a pane showing the associated prompt and send it to LLMs, whose answers can be filtered, edited, and pasted back into the demonstration.

\begin{figure}
\centering
\IncludeAppleScreenshot{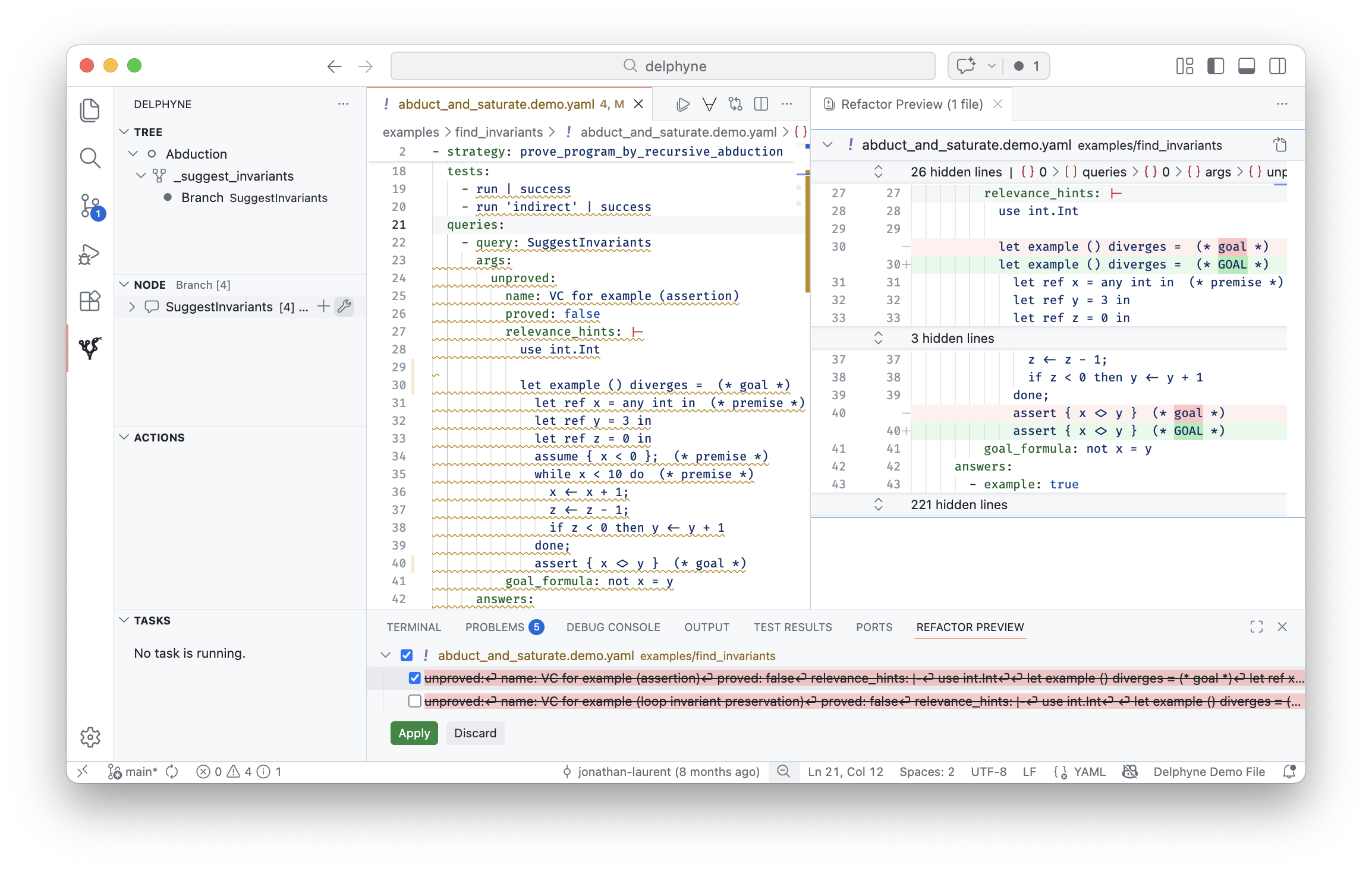}
\SpaceBelowAppleScreenshot{}
\caption[Repairing a broken demonstration interactively]{Repairing a Broken Demonstration Interactively. This screenshot illustrates a simple demonstration repair scenario that involves a program verification strategy that we developed in our first case study (Section~\ref{sec:loop-invariants-case-study}). In this scenario, a change to the rendering of proof obligations causes a demonstration to break, as some of its queries include proof obligations. For simplicity, we consider a minor change in which only the casing of the word \code{goal} is altered. Evaluating the demonstration after the change causes its tests to become stuck, and some queries to be diagnosed as \emph{unreachable}. Jumping to the node where the first test is stuck reveals the first unanswered query (\code{SuggestInvariants}). However, because a very similar unreachable query of the same type already exists in the demonstration body, Delphyne suggests a possible fix (revealed by clicking on the wrench icon): simply updating the existing query. The user can inspect the proposed patch, confirm that the change to the query does not require modifying the answer, and validate the patch.}\label{fig:demo-refactor-screenshot}
\end{figure}

When a strategy change breaks a demonstration (policy changes \emph{cannot} break demonstrations by construction), users can use Delphyne’s diagnostics and tree view to investigate and resolve the issue. In some cases, Delphyne can even suggest fixes, as demonstrated in Figure~\ref{fig:demo-refactor-screenshot}.

\subsubsection{Hybrid Workflows for Writing Demonstrations}\label{ap:implicit-answers-hybrid-workflows}

\begin{figure}
\centering
\IncludeAppleScreenshot{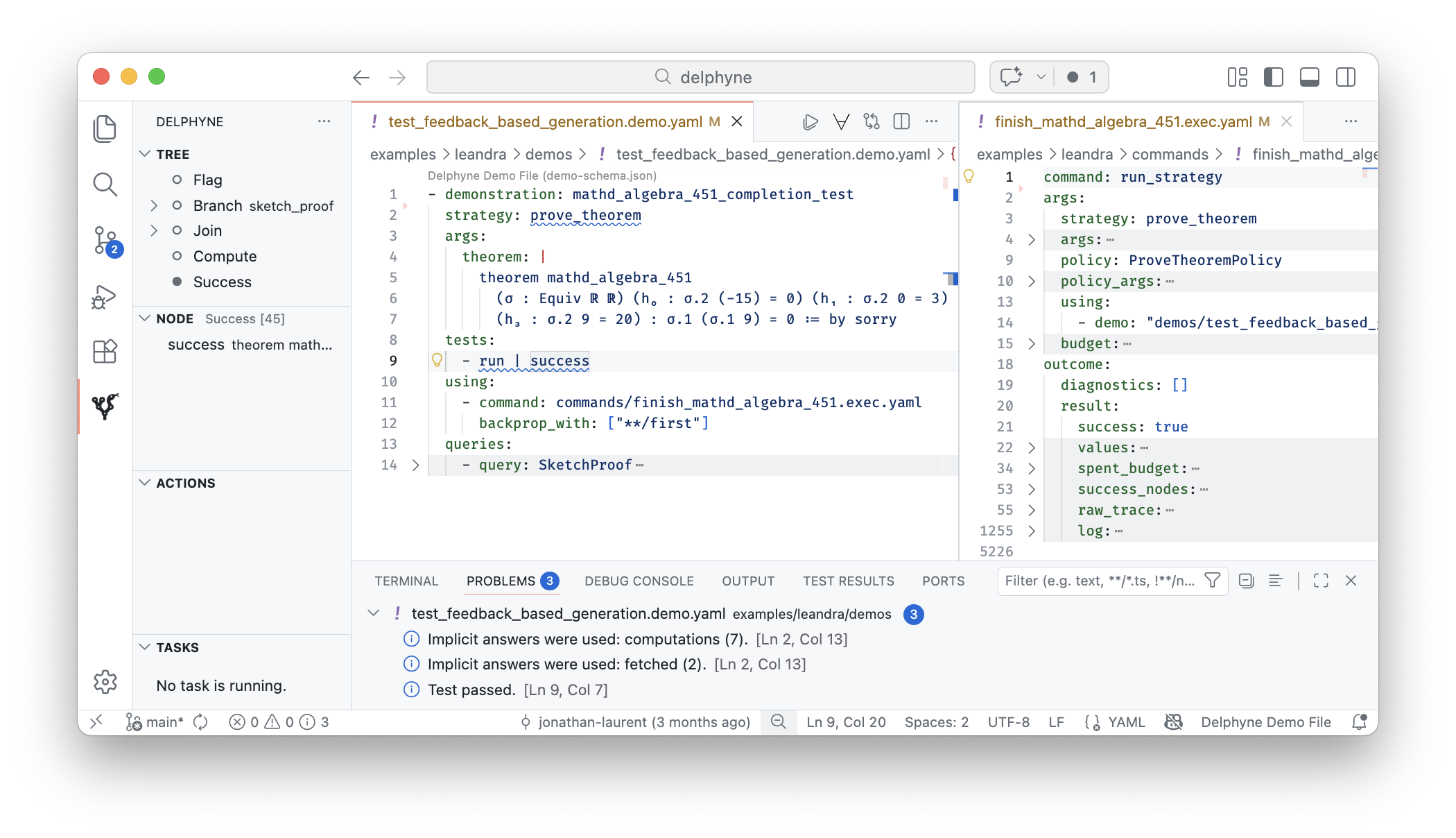}
\SpaceBelowAppleScreenshot{}
\caption[Hybrid writing of demonstrations]{Hybrid Writing of Demonstrations. This screenshot illustrates a concrete scenario of hybrid demonstration writing (Appendix~\ref{ap:implicit-answers-hybrid-workflows}) for the Lean theorem proving strategy developed in our second case study. As a reminder, the \code{prove\_theorem} strategy works by first generating a proof sketch and then filling in the remaining holes (Section~\ref{sec:lean-case-study}). In this scenario, the user first provides a proof sketch for a particular problem instance by answering the \code{ProofSketch} query (collapsed on Line~14). Then, they use an existing policy to try to finish the proof, as follows:
\begin{itinparaenum}
\item they use the \emph{``Run Strategy''} code action to create a template for a \emph{command file} specifying a run of the strategy,
\item they fill in policy and budget details in the template,
\item they launch the run, which ultimately results in an \code{outcome} section being added to the file that includes a trace (leading to the result shown in the right pane),
\item they add a \code{using} section to the demonstration referring to the command file (Lines~10--12), and finally
\item they re-evaluate the demonstration.
\end{itinparaenum}%
During re-evaluation, whenever unanswered queries are encountered, the demonstration interpreter tries to fetch answers from successful paths in the run's trace, which it treats as {implicit answers} (Appendix~\ref{ap:implicit-answers-hybrid-workflows}). Once the tests are passing and upon inspection, the user can use the \emph{``Add implicit answers (fetched)''} code action to explicitly add all implicit answers to the demonstration. Finally, note that the \code{backprop\_with} clause on Line~12 instructs the demonstration interpreter to fetch implicit answers using \emph{custom feedback hooks} defined by \code{prove\_theorem} for self-improvement (see Section~\ref{ap:self-improvement-and-reflection}). In particular, this ensures that answers leading to correct proofs of subgoals are identified as fetchable, even if other subgoals cannot be proved and the overall run fails.
}\label{fig:demo-hybrid-writing-screenshot}
\end{figure}

More advanced workflows are also possible, in which demonstrations can be partially specified with holes automatically filled by running external policies and extracting query answers from successful searches. This allows demonstration writers to retain tight control over some parts of the writing while outsourcing others, already tractable for LLMs, to automated generation, possibly followed by auditing and revision. Figure~\ref{fig:demo-hybrid-writing-screenshot} demonstrates such a scenario for our Lean theorem proving strategy (Section~\ref{sec:lean-case-study}). Such hybrid workflows are enabled by the following Delphyne features:

\begin{description}
    \item[Policy Overriding.] Given a set of query--answer pairs, policies can be wrapped so that whenever a query from the set is encountered, the specified prompting policy is bypassed and the associated answer is returned instead. This allows oracular programs to be run so that they follow partial demonstrations for all specified decisions, while performing normal search elsewhere. In Figure~\ref{fig:demo-hybrid-writing-screenshot}, this feature is enabled via the \code{using} clause shown in the right pane (part of the initial template generated by the ``Run Strategy'' code action).
    \item[Answer Fetching.] Given a set of query--answer pairs, the demonstration interpreter can be configured to use answers from this set whenever an unanswered query is encountered. Such a set can be sourced from other demonstrations, but also from traces produced by running oracular programs. By default, an answer in a trace is considered \emph{good} (and thus fetchable) if it occurs on a success path in the tree, but this behavior can be overridden by defining custom feedback hooks for self-improvement (see Appendix~\ref{ap:self-improvement-and-reflection}). In Figure~\ref{fig:demo-hybrid-writing-screenshot}, this feature is enabled via the \code{using} clause in the demonstration shown in the left pane.
    \item[Implicit Answers.] Whenever the demonstration interpreter obtains an answer to a query that does not appear in the demonstration's body, it remembers this answer as \emph{implicit}. Whenever a demonstration evaluation uses implicit answers, a diagnostic is emitted and the user is presented with the option of explicitly adding these answers to the demonstration's body. Implicit answers are useful in combination with {answer fetching}, but also in other contexts such as handling the \code{Compute} effect (Appendix~\ref{ap:compute-effect}). 
\end{description}

\subsubsection{Debugging Oracular Programs}\label{ap:debugging-visualizing}

Oracular programs can be debugged in different ways. First, from any run of an oracular program, a {trace} can be extracted that describes what parts of the search tree were visited (Section~\ref{sec:locality-references}). Traces can be serialized, unserialized, and visualized using Delphyne's {Tree View} (Figure~\ref{fig:demo-screenshot}). In addition, the \code{Message} effect from Delphyne's standard library can be used to attach debugging messages and metadata to trees.
Second, when running an oracular program, Delphyne allows the results of all LLM requests to be automatically cached, as well as the outcomes of impure, non-replicable computations performed via the \code{Compute} effect (Appendix~\ref{ap:compute-effect}). This allows any run to be easily and quickly replicated, possibly with a debugger attached to follow the execution of strategies and policies step by step.

\subsection{Performing Impure Computations in Strategies}\label{ap:compute-effect}

As explained in Appendix~\ref{ap:python-embedding-strategy}, strategy computations must be \emph{pure}, so that they always reify into the same tree. Thus, strategies cannot \emph{directly} call impure functions, including external tools such as automated theorem provers with wall-time timeouts (which may not always return the same result on a given input). Fortunately, Delphyne allows performing such computations within strategies by wrapping them within the \code{Compute} effect (Figure~\ref{fig:compute-trigger}).

\begin{figure}
\begin{ccodebox*}
\begin{lstlisting}[style=delphyne,style=delphyne-underlined]
def compute[**A, T, P](
    f: Callable[A, T]) -> Callable[A, dp.Strategy[Compute, P, T]]: ...(@\medskip@)
res = yield from compute(call_z3_prover)(formula)  # Example
\end{lstlisting}
\end{ccodebox*}
\CodeFigSpaceBefore{}
\caption{Wrapping Impure Computations with the \code{Compute} Effect.}\label{fig:compute-trigger}
\end{figure}

Invoking \code{compute} is akin to branching on the answer of a special \code{\_\_Compute\_\_} query that specifies the computation to be performed, as a fully qualified function identifier obtained via reflection along with a serialized dictionary of arguments. Such a query is not to be answered by calling an LLM oracle, but rather by executing the computation.

In demonstrations, \code{\_\_Compute\_\_} queries are seamlessly handled through the \emph{implicit answer} mechanism introduced in Appendix~\ref{ap:implicit-answers-hybrid-workflows}. When such a query is encountered in a navigation test, the underlying computation is run in the background and the result is treated as an implicit answer. A diagnostic indicates when implicit answers are used in a test (Figure~\ref{fig:demo-hybrid-writing-screenshot}). The user can use the \emph{``Add implicit answers (computations)''} code action to explicitly add the results of all computations to the demonstration body by inserting \code{\_\_Compute\_\_} entries into its \code{queries} section. This ensures that the demonstration is replicable and can be evaluated again without rerunning the computations.

In policies, \code{Compute} nodes can be eliminated from a tree via an \code{elim\_compute} tree transformer, which returns a new tree that transparently executes computations whenever needed.

Beyond including impure computations in strategies, the \code{Compute} effect can also be used to include \emph{expensive} computations, ensuring their proper caching.

\subsection{Writing ReAct Agents}\label{ap:interact}

\begin{figure}
\begin{ccodebox*}
\begin{lstlisting}[style=delphyne]
def interact[P, A, B](
    step: Callable[[AnswerPrefix], Opaque[P, Response[A]],
    process: Callable[[A], Opaque[P, B | Error]],
    tools: Mapping[type, Callable[[Any], Opaque[P, Any]]]
) -> Strategy[Branch, P, B]: ...
\end{lstlisting}
\end{ccodebox*}
\CodeFigSpaceBefore
\caption[Signature of the standard ``interact'' strategy]{Simplified Signature for the Standard \code{interact} Strategy. The \code{step} argument maps a conversation history to either a proposed answer or a tool call request, and can be implemented either via a query or a strategy (hence the \code{Opaque} return type). The \code{process} argument maps a proposed answer (of type \code{A}) to either a validated answer (of type \code{B}) or a feedback message (of type \code{Error}). The \code{tools} argument is a dictionary mapping types that specify tool interfaces to their implementations as oracular programs.}\label{fig:interact-signature}
\vspace{0.5em}
\end{figure}

\noindent Oracular programming is naturally suited for building \emph{vertical} LLM pipelines, where traditional computations orchestrate calls to LLMs. Importantly, it also encompasses the dual \emph{horizontal} paradigm, where an LLM orchestrates traditional computations via tool calls. These two approaches make fundamentally different trade-offs (see Section~\ref{sec:react-comparison} for a complete discussion), and oracular programming allows integrating both (e.g., giving a horizontal agent access to tools that are themselves implemented as oracular programs, either horizontal or vertical).

The Delphyne standard library includes a higher-order strategy named \code{interact} that generalizes ReAct-style~\cite{yao2022react} horizontal agents. A simplified type signature is provided in Figure~\ref{fig:interact-signature}. The \code{interact} strategy repeatedly invokes a query or sub-strategy (\code{step} argument) to try to produce a valid answer to a question, resending the full conversation history each time. Another query or strategy is tasked with evaluating answers and providing feedback when an answer is rejected (\code{process} argument). In addition, the first query or strategy may return tool call requests instead of answer proposals, with each available tool implemented as its own query or strategy (\code{tools} argument). The \code{interact} strategy is used in our case studies (Section~\ref{sec:case-studies}) to implement a baseline agent for invariant synthesis and multiple components of our Lean theorem-proving agent, which integrates vertical and horizontal oracular programs in a nested fashion.

Although \code{interact} allows implementing basic ReAct agents, it is significantly more general. Not only can all its arguments be implemented as arbitrary oracular programs (including the \code{step} function that generates each additional conversation message), it also induces a tree that branches at every step of the conversation. This tree can itself be explored via arbitrary policies. For example, although standard ReAct agents explore only a single child at each level, it is also possible to \emph{clone} the conversation state and explore continuations in parallel.

\section{Additional Discussion Items}\label{ap:extra-discussion-items}

This appendix elaborates on specific items from our Discussion and Related Work sections.

\subsection{Limitations of Tool Composition Through Model Contexts}\label{ap:limitations-tool-composition-through-context}
A standard framework for agentic design~\cite{yao2022react} gives a model access to a variety of tools (e.g., through MCP servers). Tool calls are composed by letting the model issue an initial call, add the result to the conversation context, and use it to produce subsequent calls. Despite great flexibility, this approach is not only expensive but also raises security concerns, since tool outputs may contain sensitive data or injected prompts~\cite{debenedetti2025defeating}. For this reason, the field is increasingly moving away from MCP servers and toward having agents compose tool calls by generating traditional programs, which can be analyzed and executed in sandboxed environments. In some cases, these programs can themselves issue calls to LLMs~\cite{zhang2025recursive,debenedetti2025defeating}, presenting a clear opportunity for oracular programming.

\subsection{Cognitive Overhead of Separating Strategies and Policies}\label{ap:unclear-strategy-policy-boundary} There are often multiple ways to split a program’s logic between strategies and policies, with subtle trade-offs. While powerful, this flexibility also introduces cognitive overhead. For example, consider our abduction-based invariant synthesis strategy (Section~\ref{sec:loop-invariants-case-study}). We chose to shift as much of our program's logic as possible to the policy side through the use of the \code{Abduction} effect, so as to enable the greatest variety of search algorithms to be used. However, a more elementary strategy can also be written that only uses \code{branch} and \code{ensure}, hardcoding the structure of recursive abduction calls into the strategy. Doing so still leaves plenty of room for policies to implement diverse search algorithms, but abstracts away key information about the relationship between branches---information that our saturation-based, \code{Abduction}-aware policy exploits.

\subsection{Comparison with LLM Orchestration Frameworks}\label{ap:llm-programming-extra}

A comparison with DSPy~\cite{khattab2024dspy} is provided in Section~\ref{sec:related-work}. We provide additional comparisons with LangGraph~\cite{langgraph} and Mellea~\cite{mellea_docs} below.

\subsubsection{Comparison with LangGraph} LangGraph~\cite{langgraph} provides graph-based orchestration abstractions for LLM agents and tool calls. Its runtime offers useful capabilities such as scheduling, observability, replay, and support for human-in-the-loop interactions.\footnote{Delphyne also provides strong observability and replay facilities (Appendix~\ref{ap:debugging-visualizing}). Integrating human oracles is an interesting avenue for future work.} However, its graph abstractions restrict how control flow can be expressed, reproducing only a limited subset of standard programming constructs in an ad-hoc manner (e.g., routing operators used in place of conditionals). By contrast, Delphyne leverages the full expressive power of Python to express control logic directly. LangGraph does offer a functional API that allows graphs to be constructed dynamically from Python code, but doing so reduces the benefits provided by its runtime. Delphyne remains unique in allowing the full separation of core and search logic and in offering a demonstration language.

\subsubsection{Comparison with Mellea} Mellea~\cite{mellea_docs} emphasizes treating LLM requests as unreliable, nondeterministic operations whose outcomes must be rigorously validated, and provides a rich library of utilities and prompting templates for doing so. However, LLM queries are fundamentally treated as guarded function calls, whereas Delphyne goes further by explicitly reifying strategies into trees amenable to advanced search. Mellea also revisits a key idea from object-oriented programming in the context of LLM programming by introducing the concept of an \code{MObject}, which packages data with LLM-powered operations. Future work may integrate this idea into Delphyne.

\section{Advanced Search for Loop Invariant Discovery}\label{ap:invariants-case-study}\label{ap:invariants}

This appendix provides details on our invariant synthesis case study (Section~\ref{sec:loop-invariants-case-study}). Appendix~\ref{ap:inv:protocol} elaborates on the experimental protocol, while Appendix~\ref{ap:inv-sys-prompt} shows the system prompts used for both the baseline and abduction-based agents. Appendix~\ref{ap:abduction-effect} shows the navigation function of the \code{Abduction} effect and proves it conditionally invertible. Finally, Appendix~\ref{ap:invariant-strategies-policies} provides code highlights for all strategies and policies.

\subsection{Experimental Protocol Details}\label{ap:inv:protocol}

Sensitive policy parameters were tuned for all agents, selecting those that led to the highest number of solved Code2Inv problems (using the average inference cost to break ties). Table~\ref{tab:model-pricing} provides details on model pricing.  %

\subsubsection{Tuning of the Abduction-Based Agent} The policy parameters (Figure~\ref{fig:inv-saturation-policy}) of our abduction-based agent were chosen as follows. Both temperatures of $1.5$ and $1.7$ were tried (a high temperature is clearly desirable since this agent relies on generating a lot of diverse candidates).
The \code{num\_suggestion\_completions} parameter was set to 8, aiming to balance the cost of LLM input and output tokens. In addition, the \code{max\_abduction\_depth} parameter was set to 2 (performing more than 2 nested abduction steps seems unnecessary for single-loop invariant generation). Finally, values of 4 and 8 were tried for the \code{max\_requests\_per\_attempt} parameter (i.e., the frequency at which all established facts are forgotten).

\subsubsection{Tuning of Baselines} The policy parameters (Figure~\ref{fig:inv-baseline-policy}) of our baselines were set as follows. For large/costly models, the temperature was set at 1 for o3 (no other value is supported) and $0.7$ for 4o (standard value used in related work for balancing reasoning ability and diversity~\cite{autospec-wen2024enchanting,loopy-kamath2023finding}). Values of 0, 1 and 3 were tried for the \code{max\_feedback\_cycles} parameter (which is particularly sensitive for large models given a relatively small budget). For the \emph{small} 4o-mini model, the \code{max\_feedback\_cycles} was set to 3 (much less important since a very high number of attempts can be performed from scratch in any case) but temperature values of 0.7, 1, and 1.5 were evaluated.

\begin{table}[t]
  \centering
  \begin{tabular}{lrrr}
    \toprule
    Model \qquad & Input & Cached Input & Output \\
    \midrule
    {gpt-4o} & \$2.50 & \$1.25 & \$10.00 \\
    {gpt-4o-mini} & \$0.15 & \$0.075 & \$0.60 \\
    {o3} & \$2.00 & \$0.50 & \$8.00 \\
    \bottomrule
  \end{tabular}
  \vspace{1em}
  \caption[API pricing for the invariant synthesis experiment]{OpenAI API Pricing (dollars per 1M tokens). Reported costs in our invariant synthesis experiment are computed based on these prices (Table~\ref{tab:code2inv}).}
  \label{tab:model-pricing}
\end{table}

\subsection{Complete System Prompts}\label{ap:inv-sys-prompt} 

\subsubsection{System Prompt Used By The Abduction-Based Strategy} We show below the system prompt used by the \code{SuggestInvariants} query from our abduction-based strategy (Figure~\ref{fig:inv-saturation}).

\medskip

\noindent \textbf{System Prompt.} Your goal is to prove the correctness of WhyML programs using the Why3 theorem prover. To do so, you must annotate programs with loop invariants in such a way that all assertions in the program can be proved automatically via the weakest-precondition calculus.

We only allow invariants that feature logical combinations of \textbf{linear} arithmetic expressions. In particular, we only accept multiplication by constant numerical literals. Thus, we accept invariants such as \code{x + y > 0} or \code{a > 0 || 2*x + y < n} but not \code{x * y > 0} or \code{x >= y * (y + 1) / 2}. I insist, and this is \textbf{very important}, you must \textbf{not} propose invariants involving complex arithmetic such as \code{2 * x >= y * (y + 1)}.

I will show you an annotated program in which Why3 did not manage to prove a particular assertion or invariant. Your task is to suggest a list of new invariant candidates that may unlock the proof.
In order to help you diagnose the problem, I added comments to the WhyML program indicating what assertion or invariant failed to be proved (\code{GOAL}) and what parts of the program provide relevant assumptions (\code{premise}). The name of the failing proof obligation also provides a hint about the nature of the problem, which is either:

\begin{itemize}
    \item \code{assertion}: the final assertion is not implied by the invariants
    \item \code{loop invariant init}: an invariant does not hold initially
    \item \code{loop invariant preservation}: an invariant cannot be proved to be preserved by the loop body (it may not be preserved or an additional invariant may be needed to complete the proof)
\end{itemize}

Please suggest new invariant candidates. Each invariant candidate must be obtained by following one of the \emph{tricks} discussed below. A trick can be used multiple times. Do not suggest candidates that are already established invariants. Answer as a JSON object representing a list of \code{(trick\_name, suggested\_invariant)} pairs and \textbf{nothing else}. Examples are provided that include additional explanations for clarity. Do not include such explanations in your answer.

\paragraph{Notes on Why3}

The \code{any T} construct generates an arbitrary object of type \code{T}. In particular, a loop whose guard is \code{any bool} can run for an arbitrary number of times.

\paragraph{Tricks}

Each trick is identified by a unique name. For each trick, we discuss \emph{when} it is applicable and \emph{what} the corresponding recipe is.

\begin{description}
\item[propose\_post] If the final assertion fails to prove but appears to hold through the whole program, propose it as an invariant.

\item[monotone] Whenever a variable \code{x} is only incremented (resp. decremented), propose invariant \code{x >= c} (resp. \code{x <= c}) for \code{c} some numerical constant.

\item[linear] Whenever a linear equality or inequality between variables appears to hold throughout the program (e.g. \code{x - y >= 0}, \code{3*x + 2*y = 1}...), consider proposing it as an invariant.

\item[abduct\_post] If the final assertion fails to prove, look for a missing assumption that implies it when assuming all established invariants (along with the negation of the loop guard). Propose this assumption as a new invariant candidate.

\item[abduct\_inv] If an invariant cannot be proved to be preserved, look for a missing assumption and propose it as a new invariant.

\item[strengthen\_inv] If an invariant cannot be proved to be preserved, consider making it stronger (e.g proposing \code{x < y} as a replacement for \code{x <> y}).

\item[guard\_inv] If proving an invariant \code{inv} requires assuming a global assumption \code{assum} that is only made \emph{after} the loop, consider proposing \code{assum -> inv} as an invariant instead.

\item[true\_or\_continue] If a property \code{P} always holds after the loop but cannot be proved as an invariant because \code{P} does not hold initially, consider proposing \code{P || loop\_guard} as an invariant instead.

\item[cond\_guard] If proving the preservation of an invariant candidate requires proving that a specific branch in the code cannot be taken (or is always taken), consider proposing an invariant that establishes this fact.

\item[relax\_loop\_guard] Suppose the loop guard is an inequality such as \code{expr < c} with \code{expr} an expression and \code{c} a constant. Then, if quantity \code{expr} cannot increase more than a constant amount \code{d} at each iteration, consider proposing \code{expr < c + d} as an invariant (and similarly for \code{<=}, \code{>} and \code{>=}).
\end{description}

\subsubsection{System Prompt Used for the Baseline Strategy}

The \code{AnnotateWithInvs} query from our baseline strategy (Figure~\ref{fig:inv-baseline}) uses the same system prompt as \code{SuggestInvariants} from the abduction-based strategy, except for the following differences:
\begin{itemize}
  \item Instead of generating a list of invariant candidates labeled with trick names, the model is instructed to output a copy of the original program with added invariant annotations.
  \item The same tricks are presented for abducting invariant candidates from Why3 feedback, but they are not attributed names and following them is not presented as mandatory.
\end{itemize}

\noindent All the rest is identical: the short introduction about Why3, the instruction to only generate linear invariants, and the explanations for interpreting Why3's feedback. The full system prompt (along with all code for the case studies) is provided in the supplementary material.

\subsection{An Effect for Recursive Abduction}\label{ap:abduction-effect}

\begin{figure}
\begin{lcodebox*}[left=\DoubleDigitLn]
\begin{lstlisting}[style=delphyne,style=withNumbers]
@dataclass
class Abduction(dp.Node):
    prove: ...  # types omitted for brevity
    suggest: ...
    search_equivalent: ...
    redundant: ...

    def navigate(self):

        def aux(fact):
            res = yield self.prove([], fact)
            status, payload = res[0], res[1]
            if status.value == "proved":
                return [(fact, payload)]
            elif status.value == "disproved":
                return []
            else:
                assert status.value == "feedback"
                feedback = payload
                suggestions = yield self.suggest(feedback)
                proved = []
                for s in suggestions:
                    extra = yield from aux(s)
                    proved.extend(extra)
                res = yield self.prove(proved, fact)
                status, payload = res[0], res[1]
                if status.value == "proved":
                    proved.append((fact, payload))
                return _remove_duplicates(
                    proved, by=lambda x: drop_refs(x[0]))

        proved = yield from aux(None)
        main_proof = _find_assoc(proved, None)
        assert main_proof is not None, "Failed to prove the main goal" (@\label{line:abduction-nav-assert}@)
        return main_proof
\end{lstlisting}
\end{lcodebox*}%
\CodeFigSpaceBefore{}
\caption[Navigation function for the ``abduction'' effect]{Navigation Function for \code{Abduction} nodes (Figure~\ref{fig:abduction-eff}). This navigation function first attempts to prove the main goal. If that fails, it recursively proves each abducted suggestion in order before trying again, while maintaining a global list of proved facts. If the main goal cannot be proved at the end of the recursion, an error is raised (Line~\ref{line:abduction-nav-assert}), providing feedback to demonstration writers. This navigation function is conditionally invertible, as explained in Appendix~\ref{ap:abduction-effect}.
}\label{fig:abduction-navigate}
\end{figure}

\noindent Delphyne's standard library defines a custom \code{Abduction} effect for recursive abduction, whose signature is shown in Figure~\ref{fig:abduction-eff}. We show the associated navigation function in Figure~\ref{fig:abduction-navigate}. It is \emph{invertible} (Section~\ref{sec:demo-completeness}) {if} {\it(i)} all proofs of a same fact are indistinguishable (proof irrelevance), {\it(ii)} the \code{suggest} function is \emph{surjective} in the sense that it can output any valid fact on any input, and {\it(iii)} the \code{prove} function is \emph{monotonic} in the sense that for any prefix \code{fs} of \code{fs'}, if there exists a proof element in \code{prove} \code{fs} \code{f}, then there exists a proof element in \code{prove} \code{fs'} \code{f}. A proof sketch proceeds as follows. From the reference of any action (i.e., a local value representing a proof of the main goal), one can extract a directed acyclic graph representing the dependencies between all involved auxiliary facts, where each fact depends on all other facts used to establish it via \code{prove}. The \code{navigate} function can then produce a proof of the main goal by having \code{suggest} output all dependencies of each fact that fails to be proved. Monotonicity is important, since \code{navigate} may attempt to prove each fact while assuming more auxiliary facts than its original dependencies.

\subsection{Details on Strategies and Policies}\label{ap:invariant-strategies-policies}

Figure~\ref{fig:inv-baseline} shows the strategy used by our baseline ReAct agents, while Figure~\ref{fig:inv-baseline-policy} shows the associated policy. Our abduction-based strategy is shown in Figure~\ref{fig:inv-saturation}, with its two associated policies in Figures~\ref{fig:inv-par-abduction-policy} and~\ref{fig:inv-saturation-policy}. All strategies and policies are fully typed. In policies, the \code{@} operator composes policies with streams or tree transformers, while the \code{\&} operator pairs a search policy parameterized by an inner policy (e.g., \code{dfs}) with a concrete inner policy. In addition, policies usually do not include an explicit parameter for the global budget limit, since the Delphyne experiment launcher provides facilities for automatically composing policies with the \code{with\_budget} stream transformer.

\begin{figure}
\begin{lcodebox*}[left=\DoubleDigitLn]
\begin{lstlisting}[style=delphyne,style=withNumbers]
@strategy
def prove_program_interactive(
    prog: why3.File,
) -> Strategy[Branch, dp.PromptingPolicy, why3.File]:
    annotated = yield from dp.interact(
        step=lambda prefix, _:
            AnnotateWithInvs(prog, prefix).using(dp.ambient_pp),
        process=lambda invs, _:
            check_invariants(prog, invs).using(dp.just_compute))
    return annotated

@strategy
def check_invariants(
    prog: why3.File, invariants: Sequence[why3.Formula]
) -> dp.Strategy[Compute, None, why3.File | dp.Error]:
    annotated = why3.add_invariants(prog, invariants)
    feedback = yield from dp.compute(why3.check)(prog, annotated)
    if feedback.success:
        return annotated
    feedback.obligations = [
        o for o in feedback.obligations if not o.proved]
    return dp.Error(label="feedback", meta=feedback)

@dataclass
class AnnotateWithInvs(dp.Query[dp.Response[Sequence[why3.Formula], Never]]):
    prog: why3.File
    prefix: dp.AnswerPrefix
    __parser__ = dp.last_code_block.yaml.response
\end{lstlisting}
\end{lcodebox*}%
\CodeFigSpaceBefore{}
\caption[Baseline strategy for finding loop invariants]{Baseline Strategy for Invariant Synthesis. The \code{interact} function is a higher order, standard library strategy that allows implementing ReAct agents (Appendix~\ref{ap:interact}). The \code{prove\_program\_interactive} strategy directly takes a prompting policy as its inner policy, since it only issues a single query. The \code{check\_invariants} strategy only uses the \code{Compute} effect (Appendix~\ref{ap:compute-effect}) and so does not need an inner policy. The \code{AnnotateWithInvs} query has a \code{prefix} field that is used to store past conversation history. Its answer type is wrapped in \code{Response}, allowing \code{interact} to access such history (the \code{Never} argument means that no tool call is allowed). System and instance prompts are stored in external Jinja2 template files.}\label{fig:inv-baseline}
\end{figure}

\begin{figure}
\begin{lcodebox*}%
\begin{lstlisting}[style=delphyne]
def prove_program_interactive_policy(
    model_name: str,
    temperature: float | None = None,
    max_feedback_cycles: int = 3
):
    model = dp.standard_model(model_name)
    sp = dp.loop() @ dp.dfs(max_depth=max_feedback_cycles+1)
    pp = dp.few_shot(model, temperature=temperature, max_requests=1)
    return sp & pp
\end{lstlisting}
\end{lcodebox*}%
\CodeFigSpaceBefore{}
\caption[Policy for the invariant synthesis ReAct baseline]{Policy Associated with the Baseline Strategy for Invariant Synthesis (Figure~\ref{fig:inv-baseline}). This policy repeatedly starts a new conversation with at most \code{max\_feedback\_cycles} rounds of feedback. See Appendix~\ref{ap:invariant-strategies-policies} for more explanations on the syntax.}\label{fig:inv-baseline-policy}
\end{figure}

\begin{figure}
\begin{lcodebox*}[left=\DoubleDigitLn]
\begin{lstlisting}[style=delphyne,style=withNumbers]
@strategy
def prove_program_by_recursive_abduction(
    prog: why3.File,
) -> Strategy[dp.Abduction, ProveProgIP, why3.File]:
    invs = yield from dp.abduction(
        prove=lambda proved, goal:
            _prove_goal(prog, proved, goal)
                .using(lambda p: p.prove),
        suggest=lambda feedback:
            _suggest_invariants(feedback)
                .using(lambda p: p.suggest),
        search_equivalent=lambda facts, fml:
            _search_equivalent(facts, fml)
                .using(lambda p: p.search_equivalent),
        redundant=lambda proved, fml:
            _is_redundant(proved, fml)
                .using(lambda p: p.is_redundant))
    return why3.add_invariants(prog, invs)

@strategy
def _suggest_invariants(
    unproved: Sequence[why3.Obligation],
) -> Strategy[Branch | Fail, dp.PromptingPolicy, Sequence[Formula]]:
    assert len(unproved) > 0
    # We focus on the first unproved obligation
    answer = yield from dp.branch(
        SuggestInvariants(unproved[0]).using(lambda p: p))
    return [s.invariant for s in answer.suggestions]

@dataclass
class SuggestInvariants(dp.Query[Sequence[InvariantSuggestions]]):
    unproved: why3.Obligation

@dataclass
class InvariantSuggestion:
    trick: str
    invariant: str

\end{lstlisting}
\end{lcodebox*}%
\CodeFigSpaceBefore{}
\caption[Abduction-based strategy for finding invariants]{An Abduction-Based Strategy for Invariant Synthesis. See Figures~\ref{fig:abduction-eff} and~\ref{fig:abduction-navigate} for the definition of the \code{abduction} effect, and Figures~\ref{fig:inv-par-abduction-policy} and~\ref{fig:inv-saturation-policy} for associated policies.}\label{fig:inv-saturation}
\end{figure}

\begin{figure}
\begin{lcodebox*}%
\begin{lstlisting}[style=delphyne]
def prove_program_by_repeated_recursive_abduction(
    model_name: str,
    max_suggestions: int | None = 2, max_abduction_depth: int = 2,
    num_parallel: int = 3, temperature: float | None = None,
):
    ip = ProveProgIP.make(
        dp.few_shot(
            dp.standard_model(model_name),
            temperature=temperature, max_requests=1))
    sp = dp.abduct_recursively(
        max_suggestions=max_suggestions, max_depth=max_abduction_depth)
    return dp.loop() @ dp.parallel([sp for _ in range(num_parallel)]) & ip
\end{lstlisting}
\end{lcodebox*}%
\CodeFigSpaceBefore{}
\caption[Simple parallel policy for abduction-based invariant synthesis]{A Simple Parallel Policy for Abduction-Based Invariant Synthesis. From the last line, a fixed number of concurrent attempts are repeatedly performed in parallel. The \code{abduct\_recursively} policy handles \code{Abduction} nodes using the same logic implemented by their navigation function (Figure~\ref{fig:abduction-navigate}), by recursively calling itself on abduction candidates while maintaining a global list of proved facts. The \code{ProveProgIP.make} function creates an inner policy for \code{prove\_prog} from a prompting policy for suggesting invariants (other optional arguments are available to configure prover timeouts). See Appendix~\ref{ap:invariant-strategies-policies} for more explanations on the syntax.}\label{fig:inv-par-abduction-policy}
\end{figure}

\begin{figure}
\begin{lcodebox*}%
\begin{lstlisting}[style=delphyne]
def prove_program_by_saturation(
    model_name: str,
    num_suggestion_completions: int = 4, max_abduction_depth: int = 2,
    max_requests_per_attempt: int = 4, temperature: float | None = None
):
    ip = ProveProgIP.make(
        dp.few_shot(
            dp.standard_model(model_name), temperature=temperature,
            num_concurrent=num_suggestion_completions, max_requests=1))
    sp = dp.abduct_and_saturate(
        max_rollout_depth=max_abduction_depth+1,
        max_proved=64, max_candidates=64, remember_disproved=False,
        max_raw_suggestions_per_step=8*num_suggestion_completions)
    per_attempt = dp.BudgetLimit({dp.NUM_REQUESTS: max_requests_per_attempt})
    return dp.with_budget(per_attempt) @ dp.loop() @ sp & ip
\end{lstlisting}
\end{lcodebox*}%
\CodeFigSpaceBefore{}
\caption[Saturation-based policy for abduction-based invariant synthesis]{A Saturation-Based Policy for Abduction-Based Invariant Synthesis. This policy was used in our Code2Inv experiment. All hyperparameters for \code{abduct\_and\_saturate} past the first one were set to limit the computational cost of propagating facts, which does not involve LLM queries. See Appendix~\ref{ap:inv:protocol} for details on how other parameters were tuned, and Appendix~\ref{ap:invariant-strategies-policies} for more explanations on the syntax.}\label{fig:inv-saturation-policy}
\end{figure}

\section{Self-Improvement of a Lean Prover Agent}\label{ap:lean-case-study}

This appendix provides details on our Lean theorem-proving case study (Section~\ref{sec:lean-case-study}). Appendix~\ref{ap:lean-example} presents an example scenario where a proof is discovered with our proposed strategy, and Appendix~\ref{ap:lean-learned-advice} shows examples of learned advice. Appendix~\ref{ap:self-improvement-and-reflection} describes self-improvement and the \code{Feedback} effect, and Appendix~\ref{ap:lean-strategy-policy-details} highlights aspects of the involved strategies and policies with concrete code examples.

\subsection{Example of a Proof Discovery Scenario}\label{ap:lean-example} We provide an example scenario of finding a Lean proof, reconstructed by browsing an actual search trace using the Delphyne VSCode extension (Appendix~\ref{ap:debugging-visualizing}), for the following theorem:
\begin{lcodebox*}%
\begin{lstlisting}[style=haskell]
theorem algebra_sqineq_unitcircatbpabsamblt1
  (a b : Real)
  (h0 : a ^ 2 + b ^ 2 = 1) :
  a * b + abs (a - b) <= 1 := by sorry
\end{lstlisting}%
\end{lcodebox*}%
\noindent Our strategy produces this sketch (after some search):
\begin{lcodebox*}%
\begin{lstlisting}[style=haskell]
have eq_sq : (a - b) ^ 2 = (1 : Real) - 2 * (a * b) := by sorry
have abs_eq_sqrt : abs (a - b) = Real.sqrt ((1 : Real) - 2 * (a * b)) := by sorry
have ab_le_half : a * b <= (1 : Real) / 2 := by sorry
have rhs_nonneg : 0 <= (1 : Real) - a * b := by sorry
have square_ineq : (1 : Real) - 2 * (a * b) <= (1 - a * b) ^ 2 := by sorry
\end{lstlisting}%
\end{lcodebox*}%
\noindent It is then tasked with producing proofs for all subgoals that do not close directly with \code{grind}, including the second one above. In doing so, it hallucinates a Mathlib theorem that does not exist (\haskellCode{Real.sqrt_eq_abs}). After receiving an error from Lean, it calls the \code{FindTheorem} tool:
\begin{lcodebox*}%
\begin{lstlisting}[style=haskell]
description: "lemma saying |x| = sqrt (x^2) for real x"
\end{lstlisting}%
\end{lcodebox*}%
\noindent This tool is implemented via a dedicated strategy, which itself uses Loogle to find good matches. After several unsuccessful Loogle requests yielding no relevant results (e.g., \haskellCode{Real.sqrt, abs, "abs_eq"}), it discovers a promising match using the query \haskellCode{"eq_abs", Real.sqrt, "sq"} and returns the corresponding documentation entry. The caller then manages to close the subgoal with:
\begin{lcodebox*}%
\begin{lstlisting}[style=haskell]
rw [Real.sqrt_sq_eq_abs (a - b) |> Eq.symm, eq_sq]\end{lstlisting}%
\end{lcodebox*}%

\subsection{Example of Learned Advice}\label{ap:lean-learned-advice}

This section presents examples of \emph{learned advice}, automatically generated via search reflection on the MiniF2F validation set and subsequently incorporated into the prompt. A complete list is available in the supplementary material.

\subsubsection{Advice for Proving Subgoals}

\begin{itemize}
\item \textbf{Use $\leftrightarrow$ lemmas via \code{.mp}/\code{.mpr} (or \code{.1}/\code{.2}), not as functions.} Many important lemmas return an equivalence $A \leftrightarrow B$. To use them, first instantiate any parameters to obtain the $\leftrightarrow$, then pick a direction with \code{.mp} ($A \to B$) or \code{.mpr} ($B \to A$); \code{.1} and \code{.2} are equivalent shorthands. Don't try to ``apply the lemma to a hypothesis'' like a function. This pattern shows up everywhere (divisibility, order facts, \code{Real.sqrt}/log lemmas, set membership equivalences). It both fixes type errors and makes the intended direction explicit.
\item \textbf{Control casts: work in one type, then come back by injectivity.} Avoid mixing $\mathbb{N}$, $\mathbb{Z}$, and $\mathbb{R}$ mid-proof. Cast early to a single target type and do all algebra there; when you need to return to $\mathbb{N}$, rewrite to a single cast (e.g., \code{$\leftarrow$Nat.cast\_add}/\code{$\leftarrow$Nat.cast\_mul}) and use \code{Nat.cast\_injective} on the resulting equality. Normalize casts before ring-like tactics with \code{simp [Nat.cast\_add, $\dots$]}.
For congruences, stay in $\mathbb{N}$ if your goal is \code{Nat.ModEq}; introducing $\mathbb{Z}$ gratuitously causes type mismatches without adding power.

\end{itemize}  

\subsubsection{Advice for Generating Proof Sketches}

\begin{itemize}
\item \textbf{Avoid $\mathbb{N}$ subtraction: rewrite to an additive invariant and induct on that.} Natural number subtraction is partial and awkward in Lean. When a goal has terms like $a^{n+1} - (n+2)$, recast it as an additive equality and prove the cleaner statement by induction. For instance, define \code{v n := u n + (n + c)} so the recurrence on \code{u} turns into a simple multiplicative or additive recurrence on \code{v}, which is easy to solve by \code{Nat.rec}. This both simplifies algebra (no case splits from \code{n - 1}) and exposes the right invariant: prove a closed form for \code{v}, then convert back to \code{u} by rearranging. The same idea works with helper sequences tailored to your recurrence so that the induction step is tautological.
\item \textbf{Prepare prerequisites before using \code{sqrt}/\code{log}/\code{floor}/\code{ceil}, then use their characterization lemmas.}
For square roots, first extract $0 \le x$ or $0 \le a$, then use theorems \haskellCode{Real.sqrt_eq_iff_sq_eq} and \haskellCode{sqr_sqrt} to move cleanly between $x^2 = a$ and $x = \sqrt{a}$.  
When bounding roots, use \haskellCode{Real.lt_sqrt} and \haskellCode{Real.sqrt_lt} to reduce inequalities to squared ones, supplying the required nonnegativity hypotheses.
For logs, isolate side conditions in advance (\haskellCode{0 < base}, \haskellCode{base != 1}, \haskellCode{0 < argument}).  
Use \haskellCode{Real.log_rpow} to take logs of powers, rewrite the equality into a cancellable form, and cancel with \haskellCode{mul_right_cancel0} using \code{log != 0} (from \haskellCode{Real.log_pos_iff}).
For floor/ceil, sandwich the number between consecutive integers and finish with \haskellCode{Int.floor_eq_iff} or \haskellCode{Int.ceil_eq_iff}.  
When square roots are involved, first obtain the bounds using the sqrt inequalities above, then apply the floor/ceil characterization.
\end{itemize}

\subsection{Self-Improving Oracular Programs}\label{ap:self-improvement-and-reflection}

\begin{figure}[t]
\centering
\includegraphics[width=0.65\textwidth]{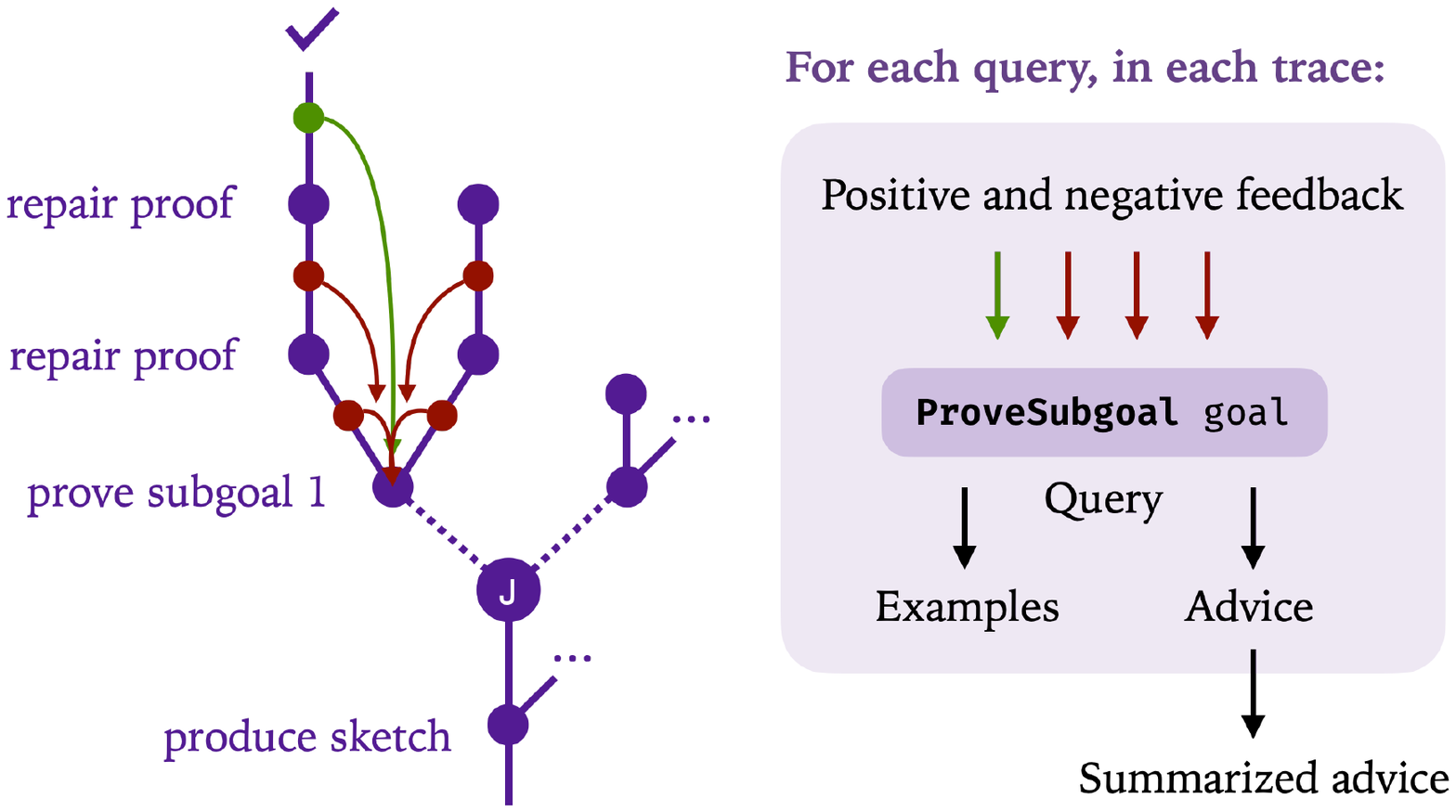}
\vspace{-0.8cm}
\caption[Custom feedback propagation]{Custom Feedback Propagation for our Lean Theorem Proving Strategy. Small green and red dots correspond to custom \code{Feedback} nodes (Appendix~\ref{ap:self-improvement-and-reflection}), while \code{J} denotes a \code{Join} node. Positive feedback is issued for each proved subgoal (green arrows), while Lean errors result in negative feedback (red arrows). Importantly, these custom feedback nodes can be specified by adding only four lines of code to our theorem-proving strategy (Figure~\ref{fig:leandra-strategy}). For each query in each trace, the received feedback is used to generate retrievable examples and also generalized into reusable advice, which is then aggregated and added to prompts.
}\label{fig:lean-feedback-backprop}
\end{figure}

Every oracular program is automatically equipped with a natural self-improvement mechanism: each time a problem is solved through search, examples of correct decisions can be extracted along the tree path leading to success, which can then be used to improve future executions. 
Oracular programming is particularly amenable to self-improvement, since the separation of strategies and policies ensures that \emph{any} run of an oracular program can be reified into a structured {trace} object (Section~\ref{sec:locality-references}), which can itself be analyzed in a process that we call \emph{search reflection}. The nested tree structure of traces carries rich information for learning. For example, answers on success paths can be automatically labeled as good, and the number of unsuccessful alternative paths explored at each decision level provides information about how \emph{challenging} the decision was and how valuable an extracted example would be. In addition, traces can incorporate further information by allowing strategies to encode domain-specific insights through custom feedback nodes.

\subsubsection{Emitting Feedback Messages.} Traces can be enriched with additional feedback nodes, refining and customizing the coarse self-improvement mechanism described above. Delphyne's standard library\footnote{This also demonstrates the power of oracular programming as a \emph{foundational} framework: all self-improvement utilities can be defined as library components that leverage Delphyne's extensible effect system.} defines a \code{Feedback} node type for this purpose, together with two trigger functions, \code{emit\_feedback} and \code{backprop\_feedback}, that induce such nodes.

Strategies can use \code{emit\_feedback} to provide custom hindsight feedback about any \emph{decision} made in a search tree. By \emph{decision}, we refer to the selection of an element from an opaque space. For any decision, four types of feedback messages can be emitted:
\code{GoodValue}, indicating that the decision was good;
\code{BetterValue}, specifying a better decision that could have been made;
\code{BadValue}, indicating that the decision was bad; and
\code{BadValueAlso}, specifying an alternative decision that would also have been bad. All messages can carry explanations as metadata, and the \code{BetterValue} and \code{BadValueAlso} messages additionally include alternative values.

For example, after successfully proving a subgoal, our Lean theorem-proving strategy emits a \code{GoodValue} message (Figure~\ref{fig:leandra-strategy}, Lines~\ref{line:leandra-emit-partial:start}--\ref{line:leandra-emit-partial:end}), registering a partial success in the event that the surrounding proof is not completed. In addition, the standard \code{interact} strategy (Appendix~\ref{ap:interact}), which is used several times in our Lean theorem-proving strategy, can be configured to emit a \code{BadValue} message when a proof sketch or tactic is rejected by Lean, and to emit a \code{BetterValue} message when a successful sketch or tactic is found after several repair iterations, indicating that such a proposal could have been made from the start (Figure~\ref{fig:leandra-strategy}, Lines~\ref{line:leandra-sketch-feedback} and~\ref{line:leandra-prove-subgoal-feedback}).

\subsubsection{Propagating Feedback in Traces.} Feedback messages can be automatically propagated through traces until they reach queries. Messages concerning the answer to a query stop there, while those concerning the outcome of a strategy are processed by feedback backpropagation handlers. Strategies can define such handlers via the \code{backprop\_feedback} function. These handlers receive a message about the final strategy outcome and return messages about inner decision points. When no custom handler is defined, the default backpropagation handler only recognizes the \code{GoodValue} message and automatically forwards it to all inner decision points.

The standard \code{interact} strategy (Appendix~\ref{ap:interact}) defines two different backpropagation handlers, which can be selected externally. For example, upon receiving a \code{GoodValue} message about its final outcome, it can either forward this message to the \emph{last} call to \code{step} that produced it, or convert it into a \code{BetterValue} message for the \emph{first} call to \code{step}.

\subsubsection{Leveraging Feedback.} A standard way to perform search reflection is to selectively enable a number of feedback emitters and handlers, propagate feedback, and gather it at the level of each query (Figure~\ref{fig:lean-feedback-backprop}). In our Lean case study, positive feedback received at each query is used to generate few-shot examples that are later selected via retrieval. Examples are extracted with priority from queries that also received the greatest amount of negative feedback, indicating that significant search was required to reach a good decision. In addition, for each query with both positive and negative feedback, an LLM is tasked with generalizing this feedback into a short piece of reusable advice. All feedback generated across a large number of problem instances is then sent to another LLM, which aggregates it into a small number of maximally useful and relevant advice items to be added to prompts (see Appendix~\ref{ap:lean-learned-advice} for examples).

Custom feedback nodes are leveraged by the demonstration interpreter in hybrid demonstration writing workflows (Appendix~\ref{ap:implicit-answers-hybrid-workflows}). When adding \code{using} clauses to demonstrations to fetch implicit answers from a trace (Figure~\ref{fig:demo-hybrid-writing-screenshot}), feedback is propagated and one fetchable answer is generated for each \code{GoodValue} or \code{BetterValue} message reaching a query. The \code{backprop\_with} argument can be used to customize this behavior by selectively enabling emitters and backpropagation handlers.

\subsection{Details on Strategies and Policies}\label{ap:lean-strategy-policy-details}

We provide sketches for our theorem-proving strategy and policy.

\subsubsection{Strategy Overview.}\label{ap:lean-strategy-overview} We provide a sketch for our Lean theorem-proving strategy in Figure~\ref{fig:leandra-strategy}. An example run is discussed in Appendix~\ref{ap:lean-example}. At the top level, the \code{prove\_theorem} strategy first generates a sketch via the \code{sketch\_proof} sub-strategy and then concurrently fills every hole in the proof via the \code{fill\_hole} sub-strategy, using the \code{Join} effect (Line~\ref{line:leandra-join}). Both \code{sketch\_proof} and \code{fill\_hole} use the standard \code{interact} strategy (Appendix~\ref{ap:interact}) to let oracles orchestrate calls to Lean  (Lines~\ref{line:leandra-sketch-interact} and~\ref{line:prove-subgoal-interact}). In addition, \code{fill\_hole} has access to an additional tool for finding auxiliary theorems in Mathlib, implemented via a custom \code{find\_theorem} strategy (Lines~\ref{line:leandra-prove-subgoals-tools:start}-\ref{line:leandra-prove-subgoals-tools:end}) that internally uses \code{interact} to orchestrate Loogle requests (the wrapping in the standard \code{nofail} combinator ensures that failures of the \code{find\_theorem} strategy do not result in failures of \code{fill\_hole}, but instead cause an empty tool response to be returned).

Four lines of strategy code were added to customize self-improvement feedback (Appendix~\ref{ap:self-improvement-and-reflection}). In \code{fill\_hole}, Lines~\ref{line:leandra-emit-partial:start} and~\ref{line:leandra-emit-partial:end} ensure that feedback is obtained from successful proofs of subgoals, even when the surrounding proof fails, by emitting hindsight feedback about the prior outcome of \code{prove\_subgoal}. This outcome originates from Line~\ref{line:leandra-fill-hole-branch-with-ref}, where \code{branch} is called with argument \code{with\_ref=True} so that a reference to the corresponding decision is returned and can later be used for targeting feedback. In addition, the \code{produce\_feedback=True} argument is passed to \code{iterate} (Lines~\ref{line:leandra-sketch-feedback} and~\ref{line:leandra-prove-subgoal-feedback}), so as to instruct it to {\it(i)} emit negative feedback when an answer is not validated by \code{process} and {\it(ii)} backpropagate positive feedback to the initial call to \code{step} when a good answer is generated after several rounds of feedback (the \code{unprocess} argument allows converting a final validated answer into the unvalidated answer that should, in hindsight, have been returned by the initial call to \code{step}; it should yield the identity function when left-composed with \code{process}).

\subsubsection{Policy Overview.}\label{ap:lean-policy-overview} We provide a sketch of the policy associated with our Lean theorem-proving strategy in Figure~\ref{fig:leandra-policy}. Its structure mirrors that of the associated strategy (Figure~\ref{fig:leandra-strategy}), and it is defined using Delphyne's \code{PolicyRecord} pattern, in which each sub-policy is represented as a serializable dataclass with an \code{instantiate\_with} method that produces an actual policy from its record representation. We provide more details on the implementation of \code{SketchProofPolicy} (Lines~\ref{line:sketch-proof-instantiate:start}--\ref{line:sketch-proof-instantiate:end}): to produce a proof sketch, it makes a series of attempts, each time using a different random procedure for selecting few-shot examples (Figure~\ref{fig:leandra-example-selector}). Such randomization improves the robustness of our policy, since different problem instances may benefit from different example-selection policies (e.g., prioritizing relevance or diversity in different settings).

\subsubsection{Example Selection Procedure.}\label{ap:lean-custom-example-selector} We define the aforementioned example selection procedure in Figure~\ref{fig:leandra-example-selector}, combining primitives from the Delphyne standard library. The \code{ExampleSelector} dataclass can be instantiated into a stream of example selectors, which is used in the implementation of \code{SketchProofPolicy} (Figure~\ref{fig:leandra-policy}, Lines~\ref{line:sketch-proof-instantiate:start}--\ref{line:sketch-proof-instantiate:end}). Each example selector in the stream first selects a number of high-quality, human-written examples from the main demonstration file, then uses MMR retrieval~\cite{carbonell1998use} with relevance weight $\lambda$ to select $n_s$ examples originating from self-improvement runs, and finally randomly samples $n_i$ examples from them. The parameters $n_s$, $n_i$, and $\lambda$ are initially set to fixed values and are then repeatedly and independently sampled from finite lists of choices (see the definition of \code{Randomized}).

\begin{figure}
\begin{lcodebox*}[left=\DoubleDigitLn]
\begin{lstlisting}[style=delphyne,style=withNumbers]
@strategy
def prove_theorem(theorem) :
    sketch, goals = yield from dp.branch(sketch_proof(theorem).using(...))
    subproofs = yield from dp.join( (@\label{line:leandra-join}@)
        [fill_hole(theorem, sketch, i, goal)
        for i, goal in enumerate(goals)])
    return fill_sketch(theorem, sketch, subproofs)

@strategy
def sketch_proof(theorem):
    sketch_and_goals = yield from dp.interact( (@\label{line:leandra-sketch-interact}@)
        step=lambda prefix, _:
            SketchProof(theorem, prefix).using(...),
        process=lambda sketch, _:
            check_sketch(theorem, sketch).using(...),
        # Instruct 'interact' to issue feedback
        produce_feedback=True, unprocess=lambda sg: sg[0]) (@\label{line:leandra-sketch-feedback}@)
    return sketch_and_goals

@strategy
def fill_hole(theorem, sketch, hole_index, goal):
    proof, proof_ref = yield from dp.branch( (@\label{line:leandra-fill-hole-branch-with-ref}@)
        prove_subgoal(theorem, sketch, hole_index, goal).using(...),
        return_ref=True)
    # We can issue partial feedback even if the surrounding proof fails
    yield from dp.emit_feedback("subproof", [ (@\label{line:leandra-emit-partial:start}@)
        dp.send(dp.GoodValue(), proof_ref)]) (@\label{line:leandra-emit-partial:end}@)
    return proof

@strategy
def prove_subgoal(theorem, sketch, hole_index, goal):
    proof = yield from dp.interact( (@\label{line:prove-subgoal-interact}@)
        step=lambda prefix, _:
            ProveGoal(goal, prefix).using(...),
        process=lambda proof, _:
            process_proof(theorem, sketch, hole_index, proof).using(...),
        tools={TheoremRequest: lambda call: (@\label{line:leandra-prove-subgoals-tools:start}@)
            dp.nofail(find_theorem(call).using(...), default=None)}, (@\label{line:leandra-prove-subgoals-tools:end}@)
        produce_feedback=True, unprocess=lambda proof: proof) (@\label{line:leandra-prove-subgoal-feedback}@)
    return proof
\end{lstlisting}
\end{lcodebox*}%
\CodeFigSpaceBefore{}
\caption[A strategy for proving theorems in Lean]{A Strategy for Proving Theorems in Lean. See Appendix~\ref{ap:lean-strategy-overview} for explanations.}\label{fig:leandra-strategy}
\end{figure}

\begin{figure}
\begin{lcodebox*}[left=\DoubleDigitLn]
\begin{lstlisting}[style=delphyne,style=withNumbers]
@dataclass
class ProveTheoremPolicy(dp.PolicyRecord[...]):
    sketch: SketchProofPolicy | None = None
    subgoal: ProveSubgoalPolicy | None = None
    
    def instantiate_with(self, env: dp.PolicyEnv): ...

@dataclass
class SketchProofPolicy(dp.PolicyRecord[...]]):
    model_name: str = "gpt-5"
    effort: dp.ReasoningEffort = "medium"
    examples: ExampleSelectorStream | None = None
    max_full_attempts: int = 4
    max_feedback_rounds_per_attempt: int = 4
    lean_timeout: float = 8.0

    def instantiate_with(self, env: dp.PolicyEnv): (@\label{line:sketch-proof-instantiate:start}@)
        examples = self.examples or ExampleSelectorStream()
        selectors = itertools.islice(
            examples.instantiate(env.random), self.max_full_attempts)
        return dp.sequence((... for sel in selectors)) (@\label{line:sketch-proof-instantiate:end}@)

@dataclass
class ProveSubgoalPolicy(dp.PolicyRecord[...]):
    model_name: str = "gpt-5-mini"
    effort: dp.ReasoningEffort = "low"
    max_full_attempts: int = 3
    max_feedback_rounds_per_attempt: int = 3
    max_requests_per_attempt: int = 7
    examples: ExampleSelectorStream | None = None
    lean_timeout: float = 8.0
    find_theorem: FindTheoremPolicy | None = None
    ...

@dataclass
class FindTheoremPolicy(dp.PolicyRecord[...]):
    model_name: str = "gpt-5-mini"
    effort: dp.ReasoningEffort = "low"
    max_requests: int = 4
    ...
\end{lstlisting}
\end{lcodebox*}%
\CodeFigSpaceBefore{}
\caption[A policy for proving theorems in Lean]{Policy Sketch for the Lean Theorem-Proving Strategy from Figure~\ref{fig:leandra-strategy} (Appendix~\ref{ap:lean-policy-overview}).}\label{fig:leandra-policy}
\end{figure}

\begin{figure}
\begin{lcodebox*}[left=\DoubleDigitLn]
\begin{lstlisting}[style=delphyne,style=withNumbers]
@dataclass
class ExampleSelectorStream:
    """
    Specification for a stream of example selectors that include a certain
    amount of permanent examples along with extra ones selected using MMR.
    """
    model_name: dp.StandardOpenAIEmbeddingModel = "text-embedding-3-large"
    num_selected: Randomized[int] = Randomized(5, [5])
    num_included: Randomized[int] = Randomized(5, [5, 15])
    mmr_lambda: Randomized[float] = Randomized(0.5, [0.3, 0.5, 0.7])

    def instantiate(self, rng: random.Random):
        nss = self.num_selected.stream(rng)
        nis = self.num_included.stream(rng)
        mls = self.mmr_lambda.stream(rng)
        for ns, ni, ml in zip(nss, nis, mls):
            permanent = dp.all_examples.such_that( (@\label{line:example-selector-stream-permanent}@)
                lambda ex: ex.demo_file is not None
                and MAIN_DEMO_FILE in ex.demo_file.stem)
            extra = dp.maximum_marginally_relevant( (@\label{line:example-selector-stream-mmr}@)
                k=ns, lambda_param=ml, model_name=self.model_name
            ).random(ni)
            # We want the closest examples returned by MMR to appear last
            yield (extra + permanent).reverse()

@dataclass
class Randomized[T]:
    """
    Specification for a hyperparameter that is first picked
    deterministically and then randomly.
    """
    first: T
    then: Sequence[T]

    def stream(self, rng: random.Random) -> Iterable[T]:
        yield self.first
        while True:
            yield rng.choice(self.then)
\end{lstlisting}
\end{lcodebox*}%
\CodeFigSpaceBefore{}
\caption{Example Selection Procedure for the Policy from Figure~\ref{fig:leandra-policy}. See Appendix~\ref{ap:lean-custom-example-selector} for explanations.}\label{fig:leandra-example-selector}
\end{figure}

\section{Universal Queries}\label{ap:universal-queries-case-study}

\begin{figure}[t]
\begin{lcodebox*}[left=\DoubleDigitLn]
\begin{lstlisting}[style=delphyne,style=withNumbers]
import sympy as sp
import delphyne as dp 
from delphyne import Branch, Fail, IPDict, Strategy, strategy

@strategy
def find_param_value(expr: str) -> Strategy[Branch | Fail, IPDict, int]:
    """
    Find an integer `n` that makes a given math expression nonnegative
    for all real `x`. Prove that the resulting expression is nonnegative
    by rewriting it into an equivalent form.
    """
    x, n = sp.Symbol("x", real=True), sp.Symbol("n")
    symbs = {"x": x, "n": n}
    try:
        n_val = yield from dp.guess(int, using=[expr])
        expr_sp = sp.parse_expr(expr, symbs).subs({n: n_val})
        equiv = yield from dp.guess(str, using=[str(expr_sp)])
        equiv_sp = sp.parse_expr(equiv, symbs)
        equivalent = (expr_sp - equiv_sp).simplify() == 0
        yield from dp.ensure(equivalent, "not_equivalent")
        yield from dp.ensure(equiv_sp.is_nonnegative, "not_nonneg")
        return n_val
    except Exception as e:
        yield from dp.fail("sympy_error", message=str(e))
\end{lstlisting}
\end{lcodebox*}%
\CodeFigSpaceBefore{}
\caption{A Strategy that Leverages Universal Queries. The task is to find a value for an integer parameter $n$ that makes a given mathematical expression nonnegative for all values of $x$. For example, given expression $x^2-2x+n$, $n=-1$ is an incorrect answer (take $x=0$), but $n = 1$ is a correct answer since $x^2 - 2x + 1 = (x-1)^2$. The strategy proceeds in two steps. It first guesses a value for $n$, using the value of the \code{expr} local variable as context (\code{guess} allows providing a custom set of local variables as context rather than dumping the full stack). Then, after performing the substitution, it tries to rewrite the resulting expression, in such a way that SymPy can prove it nonnegative through basic interval arithmetic. For concision, \code{guess} is designed to work with {inner policy dictionaries} (\code{IPDict}; see Appendix~\ref{ap:inner-policy-dicts}). Examples of associated policies are shown in Figure~\ref{fig:universal-policy}.}\label{fig:universal-strategy}
\end{figure}

\begin{figure}
\begin{lcodebox*}%
\begin{lstlisting}[style=delphyne]
def serial_policy():
    model = dp.standard_model("gpt-5-mini")
    return dp.dfs() & {
        "n_val": dp.few_shot(model),
        "equiv": dp.take(2) @ dp.few_shot(model)}

def parallel_policy():
    model = dp.standard_model("gpt-5-mini")
    return dp.loop() @ dp.par_dfs() & {
        "n_val": dp.few_shot(model, max_requests=1, num_completions=3),
        "equiv": dp.few_shot(model, max_requests=1, num_completions=2)}
\end{lstlisting}%
\end{lcodebox*}%
\CodeFigSpaceBefore{}
\caption{Two Policies for the Strategy from Figure~\ref{fig:universal-strategy}. The first policy uses sequential depth-first search, making at most two proof attempts for every parameter guess and sampling choices using gpt-5-mini. The second policy uses a parallel variant of dfs, where multiple completions are repeatedly sampled and the resulting branches explored in parallel (see definition of \code{par\_dfs} in Figure~\ref{fig:delphyne-par-dfs}).}\label{fig:universal-policy}
\end{figure}

\begin{figure}[t]
\begin{displayCode*}[left=4mm,right=4mm,top=4mm,bottom=4mm]
I am executing a program that contains nondeterministic assignments along with assertions (e.g., in the form of \code{ensure} and \code{fail} statements). I am stuck at one of these nondeterministic assignments and your goal is to generate an assigned value, in such a way that the program can go on and not fail any assertion. Specifically, I'll give you three pieces of information:

\begin{itemize}
\item A nondeterministic program.
\item The name of the variable being assigned at the program location where I am stuck.
\item Some values for a number of local variables.
\end{itemize}

You must generate a correct value to assign. The expected type of this value is indicated inside the nondeterministic assignment operator. Terminate your answer with a code block (delimited by triple backquotes) that contains a YAML object of the requested type.
\end{displayCode*}
\CodeFigSpaceBefore{}
\caption[System prompt for universal queries]{System Prompt for Universal Queries.}\label{fig:universal-query-system-prompt}
\end{figure}

\noindent This appendix demonstrates a short oracular program that uses \emph{universal queries} (Section~\ref{sec:universal-queries-case-study}) for maximal concision and solves a simple symbolic mathematics problem. The full strategy is presented in Figure~\ref{fig:universal-strategy}, and two policies are presented in Figure~\ref{fig:universal-policy}. Delphyne's default system prompt for universal queries is shown in Figure~\ref{fig:universal-query-system-prompt}. The full code for this example, along with the results of successful executions, is available in the supplementary material.

\stopcontents[appendices]

\end{document}